\documentclass[submission, LectureNotes]{SciPost}
\usepackage{graphicx}
\usepackage{amsmath}
\usepackage{amssymb}
\usepackage{esint}
\usepackage{adjustbox}

\newcommand{\bd}{\begin{displaymath}}
\newcommand{\ed}{\end{displaymath}}
\renewcommand{\vec}[1]{{\bf #1}}

\begin{document}

\begin{flushright}
QMUL-PH-19-25
\end{flushright}

\begin{center}{\Large \textbf{
Aspects of High Energy Scattering
}}\end{center}

\begin{center}
C. D. White\textsuperscript{1*}
\end{center}

\begin{center}
{\bf 1} Centre for Research in String Theory, School of Physics and
Astronomy,\\ Queen Mary University of London, 327 Mile End Road, London
E1 4NS, UK\\
* christopher.white@qmul.ac.uk
\end{center}

\begin{center}
\today
\end{center}


\section*{Abstract}
{\bf
Scattering amplitudes in quantum field theories are of widespread
interest, due to a large number of theoretical and phenomenological
applications. Much is known about the possible behaviour of
amplitudes, that is independent of the details of the underlying
theory. This knowledge is often neglected in modern QFT courses, and
the aim of these notes - aimed at graduate students - is to redress
this. We review the possible singularities that amplitudes can have,
before examining the generic behaviour that can arise in the
high-energy limit. Finally, we illustrate the results using examples
from QCD and gravity. }

\vspace{10pt}
\noindent\rule{\textwidth}{1pt}
\tableofcontents\thispagestyle{fancy}
\noindent\rule{\textwidth}{1pt}
\vspace{10pt}

\section{Motivation}
\label{sec:motivation}

Much research in contemporary theoretical high energy physics involves
scattering amplitudes (see
e.g. refs.~\cite{Elvang:2013cua,Henn:2014yza} for recent reviews),
which are themselves related to the probabilities for interactions
between various objects to occur. Usually, this is within the context
of a specific field or string theory, and we might therefore be
interested in the following questions:
\begin{itemize}
\item {\it What physical behaviour occurs in a given theory?} This
  might be a known or proposed theory at the LHC, for example, in
  which case we would want to know what to look for in experiments. In
  more formal theories (such as ${\cal N}=4$ Super-Yang-Mills), we
  might be trying to find out what types of behaviour are possible,
  and how this is similar or different to more applied theories. In
  gravity, we might be testing the limits of our understanding
  (e.g. by probing scattering above the Planck scale).
\item {\it What mathematical structures can amplitudes contain?}
  Recent work has revealed interesting connections between amplitudes
  and pure mathematics (e.g. number theory, special functions,
  abstract algebra)~\cite{Duhr:2014woa}. Knowing the types of
  mathematical object or function that generically occur in amplitudes
  can help us to address open mathematical problems, or provide
  shortcuts to calculating amplitudes in the first place (see
  e.g.~\cite{Caron-Huot:2016owq,Dixon:2016nkn,Almelid:2017qju,Chicherin:2017dob,Henn:2018cdp,Dixon:2014xca}).
\item {\it Are there special kinematic regimes where calculations
  become simpler?} It might be possible to gain insights into
  amplitudes at all orders in perturbation theory, for special regimes
  of energy, transverse momentum etc. Such ideas have great practical
  importance. For example, we must often sum certain contributions to
  all orders in perturbation theory in order to obtain meaningful
  results for comparison to
  data~\cite{Luisoni:2015xha,Becher:2014oda}. More formally, special
  kinematic regimes have a key role to play in elucidating the
  complete behaviour of a theory, overlapping with the first two
  questions.
\end{itemize}
Lately, people have also become interested in how different theories
can be related to each other. For example, string and field theories
can be related by taking certain limits, or through the AdS / CFT
correspondence~\cite{Maldacena:1997re}. There are certain string
theories, such as {\it ambitwistor string
  theory}~\cite{Mason:2005kn,Casali:2015vta,Geyer:2014fka}, that
encode the behaviour of field theories in a string-like
language. Furthermore, there are relationships between different types
of field theory. Chief amongst these is perhaps the {\it double
  copy}~\cite{Bern:2008qj,Bern:2010ue,Bern:2010yg} which, together
with similar relationships, relates scattering
amplitudes~\cite{Bern:2019prr} and classical
solutions~\cite{Monteiro:2014cda,Luna:2015paa,Luna:2016due,White:2016jzc,Luna:2016hge,DeSmet:2017rve,Bahjat-Abbas:2017htu,Luna:2017dtq,Berman:2018hwd,Bahjat-Abbas:2018vgo,CarrilloGonzalez:2019gof,Anastasiou:2018rdx,LopesCardoso:2018xes,Cardoso:2016ngt,Anastasiou:2014qba,Goldberger:2016iau,Goldberger:2017frp,Goldberger:2017vcg,Goldberger:2017ogt}
in a wide variety of theories including biadjoint scalar, non-abelian
gauge and gravity theories, with and without supersymmetry. With this
in mind, we might ponder the following:
\begin{itemize}
\item {\it Can we find common languages, that make e.g. QCD and
  gravity look the same?} Even though the physics in different
  theories can vary greatly, it might be possible to interpret this
  physics through a common calculational procedure, that itself makes
  clear why the physical behaviour is forced to be different.
\item {\it Are there generic behaviours, that any physical theory has
  to obey?} If we know that certain terms {\it must} appear in
  perturbation theory, we are no longer surprised when they do, and
  indeed know to be on the look-out for them!
\end{itemize}
The aim of these lectures is to explore these issues within a
particular context, namely the high energy, or {\it Regge
  limit}~\footnote{Classic reviews of the Regge limit can be found
  in~\cite{Collins:1977jy,Eden:1966dnq,Forshaw:1997dc}. Reference~\cite{DelDuca:2018nsu}
  brings these nicely up to date.}, to be defined more carefully in
what follows. This is important in QCD, given that present-day
collider experiments (e.g. the LHC) probe kinematic regimes in which
additional contributions from the high energy limit - that go beyond
fixed orders in perturbation theory - are believed to be
important~\cite{Andersen:2019yzo,Andersen:2017sht,Andersen:2017kfc,Andersen:2012gk,Andersen:2011hs,Andersen:2011zd,Andersen:2009he,Andersen:2009nu,Andersen:2008gc,Andersen:2008ue,Jung:2010si,White:2006yh,White:2006xv,Ball:2017otu,Bonvini:2016wki,Altarelli:2008aj,Altarelli:2003hk,Altarelli:2001ji,Altarelli:1999vw,Ciafaloni:2007gf,Ciafaloni:2006yk,Ciafaloni:2003rd}.
In gravity, the Regge limit is relevant for the scattering of black
holes, or other high energy objects. This is of interest due to the
recent discovery of gravitational waves by LIGO, but also for studies
which aim to explore the possible existence and structure of a
gravitational S-matrix if we go above the Planck scale (see
e.g. ref.~\cite{Giddings:2010pp}).

In fact, study of the Regge limit goes back to the
1960s~\footnote{Strictly, speaking, Regge's original paper was in
  1959, although this was in a non-relativistic context!}. This was
part of a large body of work that attempted to construct and
understand scattering amplitudes from general principles alone, and
which is loosely known as {\it S-matrix theory}. The programme was
particularly important for strong interactions, given that the
relevant field equations (QCD) were not yet known. Nor was it clear
that there would ever be a perturbative description of strong
interactions, given that asymptotic freedom was not known about. Some
of the conclusions of S-matrix theory remain highly useful and
relevant. For example, one can show that at high energy, amplitudes
have similar overall behaviour, regardless of the underlying
theory. This allows us to interpret modern results, clarify their
structure, and also provides important cross-checks for new
calculations.

The structure of these notes is as follows. In
section~\ref{sec:Smatrix}, we will review the basics of S-matrix
theory, by reviewing the definition of the S-matrix, and outlining the
assumptions that go into its characterisation. We will then examine a
particular consequence of these assumptions, namely that singularities
in the complex energy plane are related to physical bound states and
thresholds. In section~\ref{sec:Regge}, we look at the Regge limit,
and deduce that amplitudes must generically grow in a power-like way
as the energy becomes large, where this growth can be associated with
the exchange of an infinite family of bound states. In
section~\ref{sec:examples}, we illustrate this with examples in QCD
and gravity. Although the behaviour in those two theories will be very
different, the calculations themselves will be very similar, and both
agree with the generic predictions that come from S-matrix theory.

\section{Scattering and the S-matrix}
\label{sec:Smatrix}

\subsection{The S-matrix}
\label{sec:Sdef}

\begin{figure}
\begin{center}
\scalebox{0.5}{\includegraphics{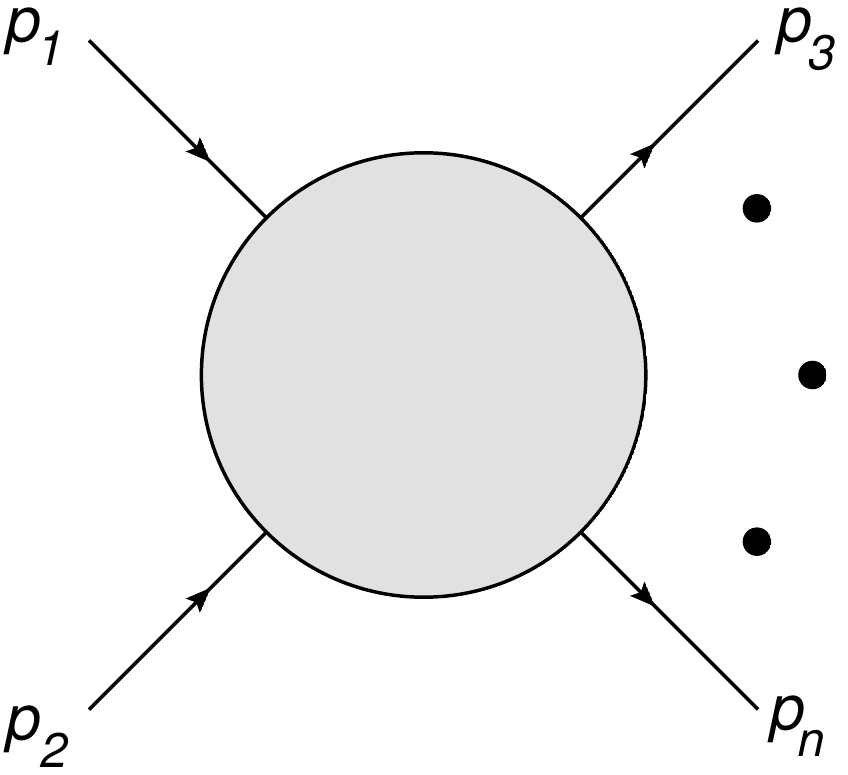}}
\caption{A scattering process, involving particles with 4-momenta
  $\{p_i\}$.}
\label{fig:scattering}
\end{center}
\end{figure}

In this section, we review the definition of the S-matrix. This
material will overlap with what you may have seen in previous courses
on quantum field theory. However, it is important to bear in mind that
we are trying to be as general as possible here, without assuming any
particular underlying theory. We will use the term ``particle'' to
describe those objects which are colliding with each other, but we
should not take this to mean that these particles are in any way
associated with fields. Furthermore, much of what we say would also
apply to other types of colliding object, such as strings or branes
(the latter, of course, were not known about in the 1960s!).

We have to start somewhere, so let us assume that we can talk about a
scattering process such as that shown in figure~\ref{fig:scattering},
which we can split into three different regimes. At very early times
($t\rightarrow-\infty$), we have well-defined incoming particles, here
with 4-momenta $p_1$ and $p_2$. At very late times
($t\rightarrow\infty$), we have a number of well-defined outgoing
particles, with 4-momenta $\{p_3,\ldots,p_n\}$. In between these two
extremes, some sort of interaction happens, and we would like to know
how to systematically describe it. First, however, we should clarify
what we mean by ``well-defined'' incoming and outgoing particles. This
typically means that the interaction is short-range, so that the
incoming and outgoing particles can be considered as free. We can then
expect to describe all possible sets of incoming particles by some set
of states $\{|n;{\rm in}\rangle\}$, and outgoing particles by a
different set of states $\{|m;{\rm out}\rangle\}$. Here the integers
$n$ and $m$ label the numbers of particles, which is not itself
sufficient to fully characterise each state, although this will be a
useful short-hand notation in what follows. To fully characterise the
states, we need to know the momenta of the particles, and also any
additional quantum numbers such as charge and spin. Here and
throughout, we shall ignore the complications of spin etc., and simply
label states by the 3-momenta of the particles e.g.
\begin{displaymath}
|n;{\rm in}\rangle\equiv |\vec{p}_1,\vec{p}_2,\ldots,\vec{p}_n;{\rm in}
\rangle,
\end{displaymath}
(similarly for outgoing particle states). Note that the 3-momenta are
indeed sufficient: if the incoming / outgoing particles are free, then
they must obey the relativistic relation
\begin{displaymath}
p_i^2=E_i^2-\vec{p}_i^2=m_i^2,
\end{displaymath}
where $E_i$ and $m_i$ are the energy and mass respectively. We can
choose to normalise these states however we like, given that in any
formula for probabilities to transition between states, we must divide
by the normalisation. As you will have seen in your field theory
courses, it is conventional to normalise one-particle states according
to the convention
\begin{equation}
\Big\langle \vec{p}_i;\begin{array}{c}{\rm in}\\{\rm out}\end{array}
\Big|\vec{p}_j;
\begin{array}{c}{\rm in}\\{\rm out}\end{array}
\Big\rangle=(2\pi)^3 2E_i\delta^3(\vec{p}_i-\vec{p}_j).
\label{statenorm}
\end{equation}
Here the delta function indicates that states with different momenta
are orthogonal (i.e. particles with different momenta are physically
different!). The element of choice affects the prefactors, which here
contain an overall power of $(2\pi)$, and the energy, to be explained
shortly. We have also indicated in eq.~(\ref{statenorm}) that the same
normalisation criterion holds either for incoming or outgoing particle
states. It is straightforward to generalise eq.~(\ref{statenorm}) to
two arbitrary multiparticle states:
\begin{equation}
\Big\langle \vec{p}_1,\vec{p}_2,\ldots,\vec{p}_n;
\begin{array}{c}{\rm in}\\{\rm out}\end{array}
\Big|\vec{p}'_1,\vec{p}'_2,\ldots \vec{p}'_m;
\begin{array}{c}{\rm in}\\{\rm out}\end{array}
\Big\rangle=\delta_{nm}\prod_{i=1}^n
(2\pi)^3 2E_i\delta^3(\vec{p}_i-\vec{p}'_i).
\label{statenorm2}
\end{equation}
Here we see in particular that states with different numbers of
particles are physically distinct, and thus orthogonal. We will assume
that the states are {\it complete}, namely that any state in the
(Fock) space we are working in can be represented as a superposition
of the possible particle states. More formally, this is expressed by
the completeness relation
\begin{equation}
\sum \Big| n;
\begin{array}{c}{\rm in}\\{\rm out}\end{array}
\Big\rangle \Big\langle n;
\begin{array}{c}{\rm in}\\{\rm out}\end{array}
\Big|=1,
\label{completeness}
\end{equation}
where the right-hand side denotes the identity operator, and the sum
on the left-hand side is over all possible particle numbers, and also
their momenta. More fully, this can be written out as
\begin{equation}
\sum_{n=0}^\infty \left(\prod_{i=1}^n\int
\frac{d^3\vec{p}_i}{(2\pi)^3 2E_i}\right)
\Big|\vec{p}_1,\ldots,\vec{p}_n;
\begin{array}{c}{\rm in}\\{\rm out}\end{array}
\Big\rangle \Big\langle \vec{p}_1,\ldots,\vec{p}_n;
\begin{array}{c}{\rm in}\\{\rm out}\end{array}
\Big|=1.
\label{completeness2}
\end{equation}
The denominator in the integral over the particle momenta arises from
the normalisation of the states in eq.~(\ref{statenorm}), as can be
easily checked by acting on a particular particle state with
eq.~(\ref{completeness2}). The reason for this normalisation is that
the measure of the momentum integral can be shown to be Lorentz
invariant by itself, which is very convenient when constructing
formulae for cross-sections etc. (see your QFT courses!). If we assume
that the set of possible outgoing particles is the same as the set of
possible incoming ones, it must be true that the in and out states
defined above form a basis of the {\it same} free theory. Thus, they
must be relatable. To make this concrete, we can define the operator
\begin{equation}
\hat{S}=\sum_m |m;{\rm in}\rangle\langle m;{\rm out}|,
\label{Sdef}
\end{equation}
and orthogonality of the basis states then implies that
\begin{equation}
{\hat S}|n;{\rm out}\rangle=|n;{\rm in}\rangle,
\end{equation}
so that the $\hat {S}$ operator ``turns out states into in
states''. We can further show that
\begin{equation}
\langle n;{\rm in}|\hat{S}|m;{\rm in}\rangle
=\langle n;{\rm out}|m;{\rm in}\rangle
=\langle n;{\rm out}|\hat{S}|m;{\rm out}\rangle,
\label{Sinout}
\end{equation}
namely that matrix elements of the $\hat{S}$ operator do not depend on
whether we take the in or out states as our basis. We can arrive at a
physical interpretation of $\hat{S}$ by looking at the middle of
eq.~(\ref{Sinout}), and remembering that the probability to transition
from some initial state $|m;{\rm in}\rangle$ to some final state
$|n;{\rm out}\rangle$ is given, via the usual rules of quantum
mechanics, by
\begin{equation}
P_{nm}\propto |\langle n;{\rm out}|m;{\rm in}\rangle|^2.
\label{Pnmdef}
\end{equation}
Thus, squared matrix elements of the $\hat{S}$ operator are related to
the probability for a given set of particles to scatter to
another. For this reason, $\hat{S}$ is called the {\it scattering
  operator}, and scattering probabilities are related to {\it elements
  of the S-matrix}
\begin{equation}
S_{nm}=\Big\langle n;
\begin{array}{c}{\rm in}\\{\rm out}\end{array}
\Big|\hat{S}\Big|m;
\begin{array}{c}{\rm in}\\{\rm out}\end{array}
\Big\rangle.
\label{Smatrix}
\end{equation}
Given that it relates in and out states, it must encode all properties
of the interaction. If we are able to write down a complete set of
particle states, and can also give all possible elements of the
S-matrix, we have completely described what a given theory can do, at
least when it comes to scattering. Given that eq.~(\ref{Smatrix}) does
not care whether we use the in or out states as our basis (as long as
we choose the same states on each side of the matrix element!), we can
omit the explicit in / out notation in what follows, unless otherwise
stated.

\subsection{Properties of the S-matrix}
\label{sec:properties}

The previous section reviewed material that you are probably familiar
with from your quantum field theory courses. However, we stress again
that we have at no point assumed a particular underlying theory -
merely the existence of quantum particles that can scatter. In doing
so, and in saying that the in and out states could be related to each
other, we have assumed the {\it superposition principle} of quantum
mechanics. Furthermore, we assumed that the incoming and outgoing
particles were described by states in a free (non-interacting) theory,
so that the interactions themselves were {\it short-range}. By making
these and more properties explicit, we can examine the consequences
for the S-matrix in a systematic way. We will see that we can surmise
quite a lot about what scattering looks like on generic grounds,
subject to a very reasonable set of assumptions.

As well as superposition and short range interactions, we will further
assume that our theory is consistent with special relativity, so that
the S-matrix must be {\it Lorentz invariant}~\footnote{Strictly
  speaking, we mean Poincar\'{e} invariant, as we will assume momentum
  conservation throughout. We have stated Lorentz invariance above,
  however, as this is the explicit postulate that is usually written
  in books on this subject.}. In practice, this means that elements of
the S-matrix must depend only on scalar products of 4-vectors. Another
important property follows from completeness of the states. From the
definition of the scattering operator in eq.~(\ref{Sdef}), one finds
\begin{equation}
\hat{S}^\dag=\sum_{m} |m;{\rm out}\rangle\langle m;{\rm in}|,
\label{Sdag}
\end{equation}
and one may then use eq.~(\ref{completeness}) to show that
\begin{equation}
\hat{S}^\dag \hat{S}=\hat{S}\hat{S}^\dag=1.
\label{unitarity}
\end{equation}
In words, we say that the S-matrix is {\it unitary}, but we stress
again that this is merely a consequence of the fact that we assumed
that our particle states are complete: given any two complete bases of
a Fock space, any transformation that relates them must be
unitary. Furthermore, the physical interpretation of unitarity is that
probability is conserved in scattering processes. This is again a
statement about the completeness of the states: if there was a
complete set of in states, for example, but {\it not} a complete set
of out states, it would be possible for information to ``get lost''
when we scatter our particles. This would then show up as a failure to
conserve probability, although it is difficult to see how such a
theory makes any sense at all, let alone how such a consequence could
ever be testable. 

Let us write the unitarity condition in more detail. We can first
rewrite the measure for momentum integration that we encountered in
eq.~(\ref{completeness2}), using the well-known identity (see
e.g. ref.~\cite{Schwartz:2013pla})
\begin{equation}
\int\frac{d^3\vec{p}_i}{(2\pi)^3 2E_i}
=\int\frac{d^4 p_i}{(2\pi)^3}\delta^{(+)}(p_i^2-m_i^2).
\label{measure}
\end{equation}
Here we have introduced
\begin{equation}
\delta^{(+)}(p_i^2-m_i^2)=\delta(p_i^2-m_i^2)\theta(p_i^0),
\label{delta+}
\end{equation}
where $\delta(x)$ and $\theta(x)$ are the Dirac delta and Heaviside
functions respectively. Physically, eq.~(\ref{delta+}) tells us that
the particle whose momentum we are summing over must be on-shell
(i.e. obey the relativistic energy-momentum relation $p_i^2=m_i^2$),
and have positive energy ($p_i^0>0$). Using eq.~(\ref{measure}), the
unitarity condition of eq.~(\ref{unitarity}), sandwiched between given
initial and final states with momenta $\{p_i\}$ and $\{p'_i\}$
respectively, can be written as
\begin{align}
&\quad\sum_{n=0}^\infty \left(\prod_{i=1}^n\frac{d^4 q_i}{(2\pi)^3}
\delta^{(+)}(q_i^2-m_i^2)\right)
\langle \vec{p}'_1,\ldots,\vec{p}'_{m'}|\hat{S}|\vec{q}_1,\ldots,\vec{q}_n
\rangle\langle \vec{q}_1,\ldots,\vec{q}_n|\hat{S}^\dag|\vec{p}_1,
\ldots,\vec{p}_m\rangle\notag\\
&=
\langle\vec{p}'_1,\ldots,\vec{p}'_m|\vec{p}_1,\ldots,\vec{p}_m\rangle,
\label{unitarity2}
\end{align}
where we have inserted a complete set of states with 4-momenta
$\{q_i\}$. Equation~(\ref{unitarity2}) is going to be enormously
powerful in what follows, but there is no getting around the fact that
it is quite horrible to look at. It is thus useful to represent it
graphically, and a nice notation has been developed in
ref.~\cite{PhysRev.135.B745}. Consider, for example, the case $m=m'=2$
(i.e. two incoming and two outgoing particles). Given that we have
\begin{displaymath}
\langle \vec{p}'_1,\vec{p}'_2|\vec{p}_1,\vec{p}_2\rangle\sim
\delta^{(3)}(\vec{p}_1-\vec{p}'_1)
\delta^{(3)}(\vec{p}_2-\vec{p}'_2),
\end{displaymath}
we can draw this as
\begin{equation}
\langle \vec{p}'_1,\vec{p}'_2|\vec{p}_1,\vec{p}_2\rangle\equiv
\raisebox{-10pt}{\scalebox{0.4}{\includegraphics{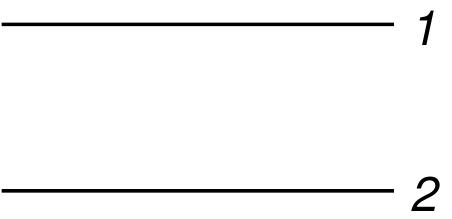}}}
\qquad,
\label{deltafig}
\end{equation}
where the horizontal lines indicate that the momenta of the first and
second particles remain unchanged. Likewise, we can draw an S-matrix
element with two incoming particles as follows:
\begin{equation}
\langle \vec{p}'_1,\vec{p}_2|\hat{S}|\vec{q}_1,\ldots \vec{q}_n\rangle
\equiv 
\raisebox{-10pt}{\scalebox{0.4}{\includegraphics{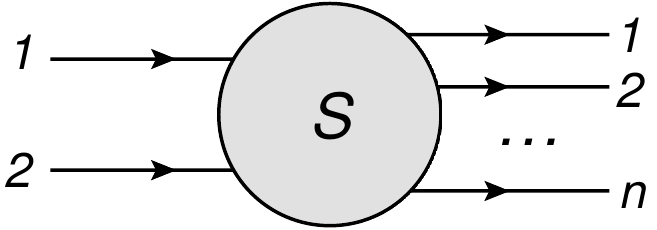}}}
\qquad.
\label{Smatrixfig}
\end{equation}
Next, we can introduce {\it internal lines} in our diagrams, by
identifying
\begin{equation}
\raisebox{0pt}{\scalebox{0.4}{\includegraphics{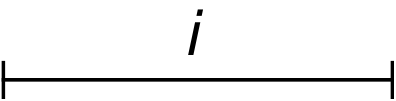}}}
\equiv \int\frac{d^4 p_i}{(2\pi)^3}\delta^{(+)}(p_i^2-m_i^2).
\label{internalline}
\end{equation}
Using this notation, the condition of eq.~(\ref{unitarity2}) can be
drawn as 
\begin{align}
&\raisebox{-9pt}{\scalebox{0.35}{\includegraphics{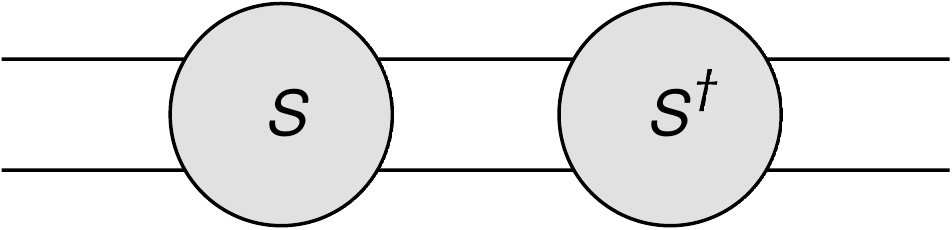}}}
\quad+\quad
\raisebox{-9pt}{\scalebox{0.35}{\includegraphics{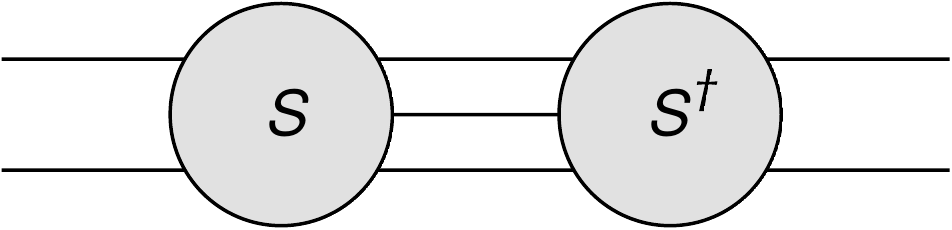}}}
\quad+\quad
\raisebox{-9pt}{\scalebox{0.35}{\includegraphics{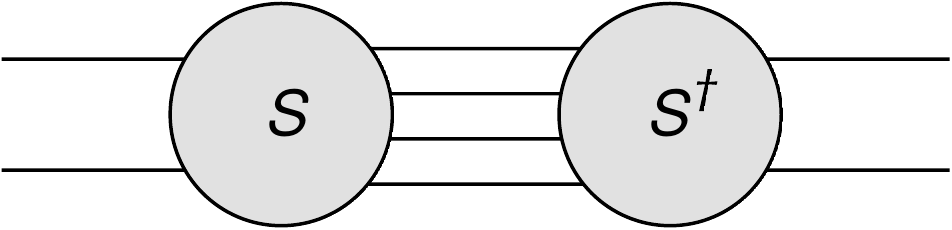}}}
\quad +\quad \ldots
\notag\\
& =\raisebox{-5pt}{\scalebox{0.35}{\includegraphics{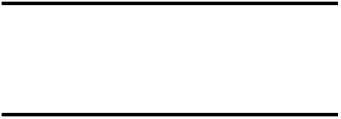}}}\quad.
\label{unitarityeq1}
\end{align}
Here the first line represents the left-hand side of
eq.~(\ref{unitarity2}), with $m=m'=2$. The left and right-hand sides
of each graph correspond to the two particles with momenta
$\{\vec{p}_i\}$ and $\{\vec{p}'_i\}$ respectively, and the sum over
all intermediate states with momenta $\{\vec{q}_i\}$ is represented by
including an increasing number of internal lines. Each of the latter
is associated with an integral as in eq.~(\ref{internalline}). On the
right-hand side of eq.~(\ref{unitarityeq1}), we have two plain lines
representing the overlap between two 2-particle states, consistent
with eqs.~(\ref{unitarity2},
\ref{deltafig}). Equation~(\ref{unitarityeq1}) is only one of an
infinite number of {\it unitarity equations}, given that we can draw
such a pictorial relation for all possible values of $m$ and $m'$.

We can go further than this, by noting that the fact that we assumed
interactions were short-range implies that there is a significant
probability that particles will not interact at all. It therefore
makes sense to write S-matrix elements in a form that explicitly
separates out the contributions where nothing, or something,
happens. More specifically, let us take a given initial state
$|i\rangle$ and final state $|f\rangle$, and write the corresponding
S-matrix element as
\begin{equation}
S_{fi}=\delta_{fi}+i(2\pi)^4\delta^{(4)}(P_f-P_i){\cal
  A}_{fi}.
\label{Sif}
\end{equation}
Here $\delta_{fi}$ schematically represents the contribution in which
the final state is the same as the initial state, which is represented
graphically as in eq.~(\ref{deltafig}). For the second term, we have
introduced conventional factors of $i$ and $(2\pi)^4$, and extracted an
overall delta function which tells us that the momenta of the initial
and final states, $P_i$ and $P_f$ respectively, must be equal. This is
a consequence of the Poincar\'{e} invariance that we assumed
earlier. Finally, the quantity ${\cal A}_{fi}$ (defined for given
initial and final states) encodes the contribution in which there is a
genuine interaction between the incoming and outgoing particles, and
is known as the {\it scattering amplitude}. Similar to the graphical
notation introduced for the S-matrix above, we may draw eq.~(\ref{Sif})
as~\cite{PhysRev.135.B745}
\begin{equation}
\raisebox{-9pt}{\scalebox{0.4}{\includegraphics{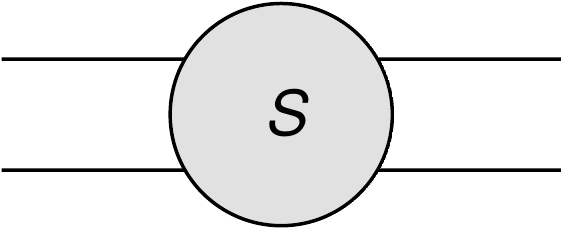}}}
\quad=\quad
\raisebox{-2.5pt}{\scalebox{0.4}{\includegraphics{Ueq4.pdf}}}
\quad+\quad
\raisebox{-9pt}{\scalebox{0.4}{\includegraphics{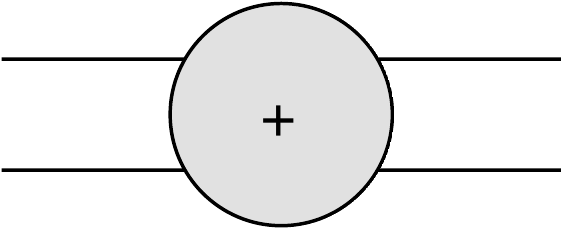}}}
\quad, 
\label{ampfig1}
\end{equation}
where the notation in the second term represents everything in the
second term on the right-hand side of eq.~(\ref{Sif}). By taking the
complex conjugate of eq.~(\ref{Sif}), we find
\begin{equation}
S^\dag_{fi}=\delta_{fi}-i(2\pi)^4\delta^{(4)}(P_f-P_i){\cal
  A}^\dag_{fi},
\label{Sif2}
\end{equation}
which, similarly to the above, we may draw as
\begin{equation}
\raisebox{-9pt}{\scalebox{0.4}{\includegraphics{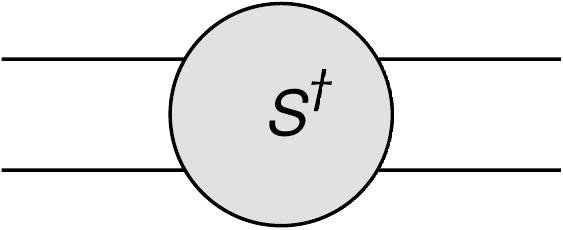}}}
\quad=\quad
\raisebox{-2.5pt}{\scalebox{0.4}{\includegraphics{Ueq4.pdf}}}
\quad-\quad
\raisebox{-9pt}{\scalebox{0.4}{\includegraphics{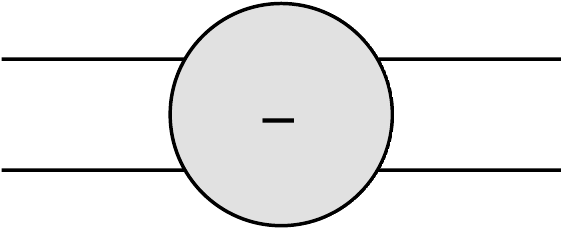}}}
\quad. 
\label{ampfig2}
\end{equation}
The pictures are known as {\it bubble diagrams} in the literature and
old textbooks~\cite{Eden:1966dnq,Collins:1977jy}, and we follow the
notation of ref.~\cite{PhysRev.135.B745} in letting the plus and minus
blobs denote the scattering amplitude and its complex conjugate
respectively, together with the momentum-conserving delta function and
factor of $i(2\pi)^4$. 

Equation~(\ref{ampfig1}) shows that the S-matrix for $2\rightarrow 2$
scattering decomposes into a trivial term containing two disconnected
lines, and a contribution in which the incoming and outgoing particles
interact. We may consider a similar concept for any number of incoming
and outgoing particles $m$ and $m'$, where the general idea is that
one must sum over all possible subsets of interacting particles in the
initial and final states. Examples include the following:
\begin{align}
\raisebox{-10pt}{\scalebox{0.4}{\includegraphics{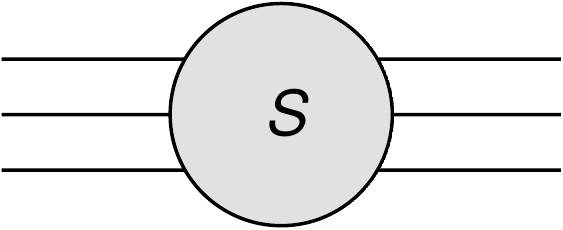}}}
\quad&=\quad
\raisebox{-4pt}{\scalebox{0.4}{\includegraphics{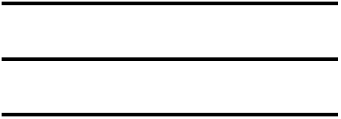}}}
\quad+\quad
\sum 
\raisebox{-7pt}{\scalebox{0.4}{\includegraphics{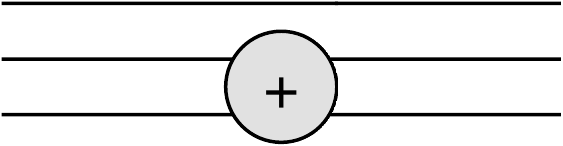}}}
\quad+\quad
\raisebox{-10pt}{\scalebox{0.4}{\includegraphics{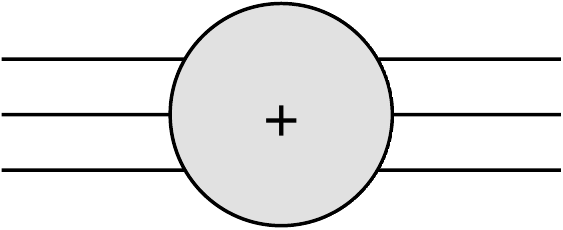}}}
\quad
\label{S3bubble}
\end{align}
for $m=m'=3$. Here, the plus bubble notation denotes an amplitude for
the interaction of the relevant number of particles entering or
leaving the bubble, and the sum in the second term on the right-hand
side is over all possible combinations of particles, and thus all
interacting subsets. A more non-trivial example for $m=m'=4$ is 
\begin{align}
\raisebox{-10pt}{\scalebox{0.4}{\includegraphics{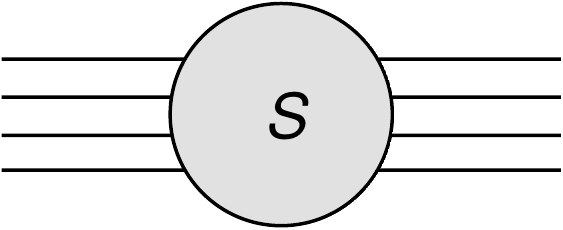}}}
\quad&=\quad
\raisebox{-4pt}{\scalebox{0.4}{\includegraphics{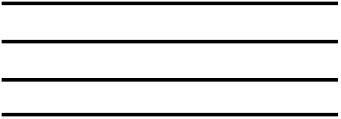}}}
\quad+\quad
\sum 
\raisebox{-10pt}{\scalebox{0.4}{\includegraphics{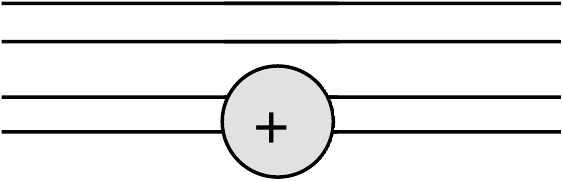}}}
\quad+\quad
\sum
\raisebox{-8pt}{\scalebox{0.4}{\includegraphics{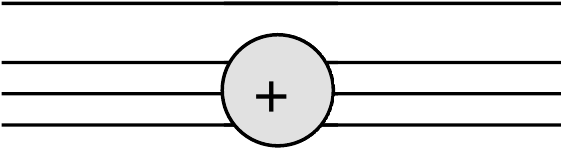}}}\notag\\
&\quad+\quad
\sum
\raisebox{-12pt}{\scalebox{0.4}{\includegraphics{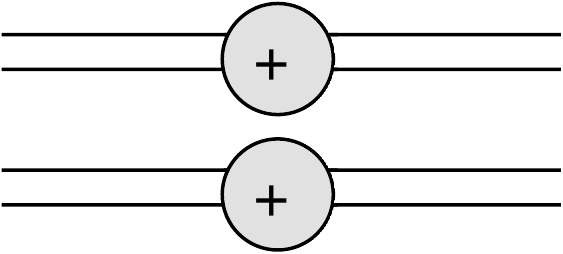}}}
\quad+\quad
\raisebox{-10pt}{\scalebox{0.4}{\includegraphics{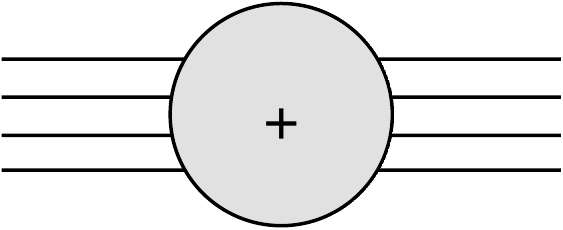}}}
\quad,\label{S4bubble} 
\end{align}
and clearly demonstrates how the combinatorial complexity starts to
increase with increasing particle number. The above diagrams show the
connectedness structure of a given S-matrix element, and illustrate an
important property known as {\it cluster decomposition}, namely that
widely-separated particles should interact independently of each
other. This is clearly related to the idea that experiments on
opposite sides of the universe should not influence each other,
although is not as strong a requirement as strict locality (i.e. the
idea that objects can only influence things {\it immediately} next to
them). Nevertheless, cluster decomposition is also a consequence of
our assuming that interactions were suitably short-range above. Going
beyond S-matrix theory, cluster decomposition plays an important role
in arguing that quantum field theory is the unique result of unifying
special relativity and quantum mechanics, as stressed in some QFT
textbooks~\cite{Weinberg:1995mt,Schwartz:2013pla}~\footnote{See
  ref.~\cite{Veltman:1994wz} for another QFT book with a useful
  discussion of the implications of unitarity.}. Independently of
fields, the combination of unitarity, Lorentz invariance and cluster
decomposition has important consequences for the structure of the
S-matrix, as we will see. Before moving on, however, we should make
explicit another assumption that is often left implicit in the
literature on this subject, namely that we are assuming that all of
our particle states are massive. This is clearly at odds with some of
the physical theories underlying the reality that we appear to live
in: electromagnetism and gravity, for example, are apparently carried
by massless particles. The assumption is valid, however, for the
strong interactions studied in the 1960s, which are presently
understood to be described by QCD. The latter theory indeed has a
so-called {\it mass gap}, such that the lowest energy states which
carry the interaction at sufficiently long distance are (massive)
mesons. We will return to this point in section~\ref{sec:examples}.

\subsection{Analyticity}
\label{sec:analytic}

Lorentz invariance implies that the scattering amplitude ${\cal A}$
(for a given initial and final state) is a function of Lorentz
scalars. Given the set of particle momenta $\{p_i\}$ (which may be in-
or outgoing) and corresponding masses $m_i$, this means that the
amplitude can only depend on the set of invariant variables
$\{p_i\cdot p_j, m_i^2\}$. Furthermore, it is a complex function,
given that S-matrix elements are in general complex. The physical
reason for this is straightforward: particles have wave-like
properties in quantum mechanics, and thus any mathematical object that
describes their scattering must be able to keep track of phase
information. This is precisely what complex numbers are for!

\begin{figure}
\begin{center}
\scalebox{0.6}{\includegraphics{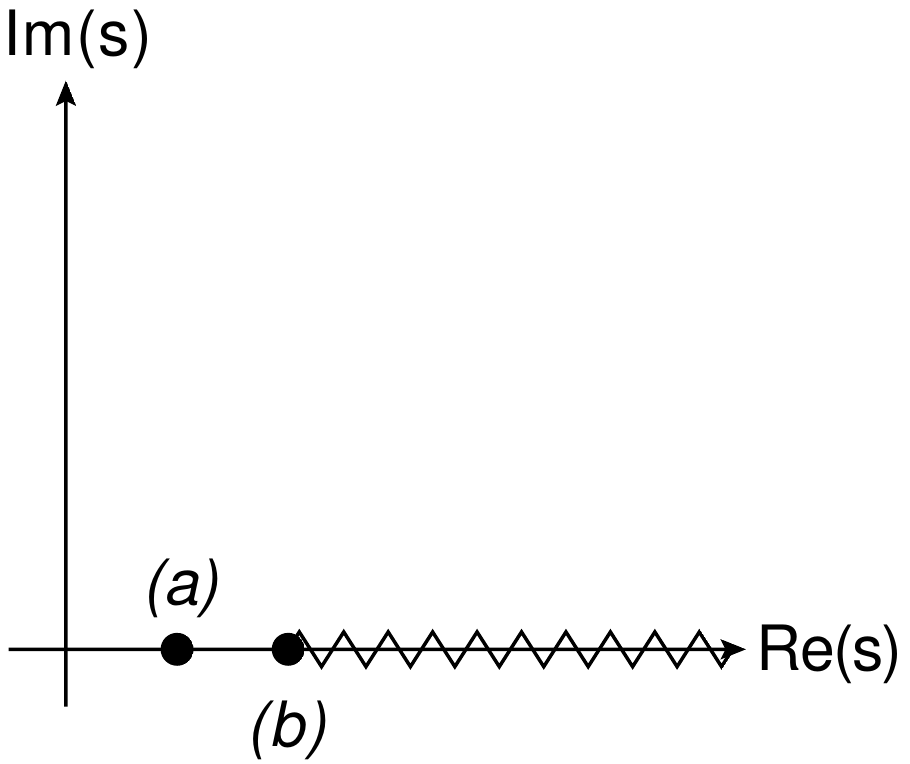}}
\caption{Example singularity structure of the complex scattering
  amplitude ${\cal A}(s)$ in the complex $s$ plane, showing (a) a
  pole; (b) a branch point.}
\label{fig:discontinuities}
\end{center}
\end{figure}

So far, then, we see that the scattering amplitude ${\cal A}$ is a
complex function of apparently real variables: dot products of
4-momenta (which may be positive or negative), and masses. However,
the unitarity and cluster decomposition properties that we introduced
in the previous section turn out to imply the following: {\it
  scattering amplitudes are the real-boundary values of analytic
  functions}. This is a somewhat cryptic phrase, so let's decode it
carefully. Let us first consider the squared centre of mass energy,
which is conventionally denoted by $s$:
\begin{equation}
s=\left(\sum_{i}p_i\right)^2,
\label{sdef}
\end{equation}
where the sum is over all incoming particles $i$. Ordinarily, $s$ is a
real number, but let us now imagine extending it to be complex. We can
then extend the function ${\cal A}(s)$ into the entire complex
plane~\footnote{In general, the amplitude will be a function of more
  invariants than just $s$. The present discussion then generalises.},
such that its value on the real axis ${\rm Im}(s)=0$ gives the
physical amplitude we are seeking. What type of function might this
be? We may think of insisting that this function be bounded (i.e. not
infinite) and differentiable (i.e. {\it holomorphic}) in the whole
complex plane. However, {\it Liouville's theorem} of complex analysis
tells us that any such function can only be a constant, which clearly
does not lead to any interesting scattering! Instead, we conclude that
the complex function ${\cal A}(s)$ must have singularities, leading to
{\it poles} and {\it branch cuts} in the complex $s$ plane. Away from
these singularities, though, it will be analytic (i.e. expandable in a
power series). An example singularity structure is shown in
figure~\ref{fig:discontinuities}, and reveals the presence of poles
and cuts on the real axis itself. The branch cut is needed because
crossing the real axis will be associated with a discontinuity, so
that ${\cal A}(s)$ becomes multi-valued. The cut reminds us of this,
and we must then specify which side of the cut corresponds to the
physical amplitude. This turns out to be the upper half plane of
$s\in{\mathbb C}$, so that the physical amplitude is the value of
${\cal A}(s)$ on the real axis, as approached from above. This is the
``real-boundary value'' referred to in the above sentence.

Where do the singularities come from? In other words, is there a {\it
  physical} explanation for the {\it mathematical} properties of our
particular complex function ${\cal A}(s)$? Indeed there is, and we can
establish the following two relations:
\begin{enumerate}
\item[(a)] Cuts arise from {\it multiparticle thresholds}. That is, if
  it is only kinematically allowed to make $N$ final states particles
  above a certain (squared) centre of mass energy $s_N$, we call
  $\sqrt{s_N}$ the {\it threshold energy}, and there is a branch point
  of ${\cal A}(s)$ at $s_N$.
\item[(b)] Poles are associated with {\it (bound) single particle
  states}. If there is a single particle state in the theory - which
  may be a fundamental or composite particle of mass $M$ - then ${\cal
    A}(s)$ has a pole at $s=M^2$.
\end{enumerate}
We can now see why we expect to find poles and cuts on the real axis
e.g. we certainly encounter particle thresholds for physical values of
the centre of mass energy, corresponding to ${\rm Im}(s)=0$! By
property (a), this will lead to a cut on the real axis, as shown in
figure~\ref{fig:discontinuities}. However, neither (a) or (b) is
obvious, so let us try to sketch how one might go about proving them. 

To derive (a), let us assume we know all possible (multi-)particle
states in our theory, and let us take a value of $\sqrt{s}$ such that
we can only have two particles in the final state (i.e. we are below
the three particle threshold). We may then simplify the unitarity
equation of eq.~(\ref{unitarityeq1}) to simply read
\begin{align}
&\raisebox{-9pt}{\scalebox{0.35}{\includegraphics{Ueq1.pdf}}}
\quad=\quad
\raisebox{-5pt}{\scalebox{0.35}{\includegraphics{Ueq4.pdf}}}\quad,
\label{uni1}
\end{align}
given that intermediate states with three or more particles (all of
which must be on-shell) are no longer allowed to contribute. We can
then implement our cluster decomposition property, by replacing each
S-matrix element with the amplitude notation of eqs.~(\ref{ampfig1},
\ref{ampfig2}):
\begin{equation}
\left(
\raisebox{-3.5pt}{\scalebox{0.35}{\includegraphics{Ueq4.pdf}}}
\quad+\quad
\raisebox{-9pt}{\scalebox{0.35}{\includegraphics{plusnotation.pdf}}}\right)
\left(
\raisebox{-3.5pt}{\scalebox{0.35}{\includegraphics{Ueq4.pdf}}}
\quad-\quad
\raisebox{-9pt}{\scalebox{0.35}{\includegraphics{minusnotation.pdf}}}\right)
\quad=\quad
\raisebox{-3.5pt}{\scalebox{0.35}{\includegraphics{Ueq4.pdf}}}.
\label{uni2}
\end{equation}
Remembering what all the symbols mean, we can multiply this out and
simplify to get
\begin{equation}
\raisebox{-9pt}{\scalebox{0.35}{\includegraphics{plusnotation.pdf}}}
\quad-\quad
\raisebox{-9pt}{\scalebox{0.35}{\includegraphics{minusnotation.pdf}}}
\quad=\quad 
\raisebox{-9pt}{\scalebox{0.35}{\includegraphics{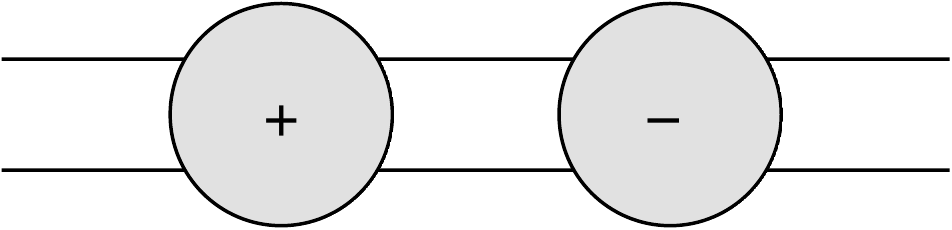}}}
\label{uni3}
\end{equation}
which, translated back into algebra, means the following:
\begin{align}
i(2\pi)^4&\delta^{(4)}(p'_1+p'_2-p_1-p_2)\left({\cal A}-{\cal A}^\dag\right)
=
\int\frac{d^4 q_1}{(2\pi)^3}\delta^{(+)}(q_1^2-m_1^2)i(2\pi)^4
\delta^{(4)}(p_1+p_2-q_1-q_2){\cal A}\notag\\
&\times \int\frac{d^4 q_2}{(2\pi)^3}\delta^{(+)}(q_2^2-m_2^2)i(2\pi)^4
\delta^{(4)}(q_1+q_2-p'_1-p'_2)(-{\cal A}^\dag).
\label{unieq1}
\end{align}
Here we have chosen $\{p_i\}$ and $\{p'_i\}$ to be the incoming and
outgoing momenta respectively, and $\{q_i\}$ to be the momenta of the
intermediate particles. On the right-hand side, we must include the
integrals over the intermediate particle momenta (with the on-shell
condition included), and the four-dimensional delta functions
enforcing momentum conservation for each sequential part of the
diagram: from the initial state to the intermediate state, and from
the intermediate state to the final state. Every four-dimensional
delta function is accompanied by a factor of $i(2\pi)^4$ according to
the conventions outlined above. Finally, each plus or minus bubble
corresponds to a factor of ${\cal A}$ or ${\cal A}^\dag$
respectively. We can clearly simplify eq.~(\ref{unieq1})
considerably. For example, we can rearrange the numerical constants,
and also combine the two four-dimensional delta functions on the
right-hand side, by using up one of the $q_i$ integrals. Following
e.g. ref~\cite{Eden:1966dnq}, we can then choose to rewrite
eq.~(\ref{unieq1}) as
\begin{align}
i(2\pi)^4&\delta^{(4)}(p'_1+p'_2-p_1-p_2)\left({\cal A}-{\cal A}^\dag\right)
=
i(2\pi)^4\delta^{(4)}(p'_1+p'_2-p_1-p_2)\notag\\
&\times i\int\frac{d^4 k}{(2\pi)^4}
[2\pi i\delta^{(+)}(k^2-m^2)]\,[2\pi i\delta^{(+)}({k'}^2-{m'}^2)]\,
{\cal A}\,{\cal A}^\dag,
\label{unieq2}
\end{align}
where we have carried out the $q_2$ integral and defined $k=q_1$,
$k'=p_1+p_2-k$, so that $k$ and $k'$ are now the 4-momenta in the
intermediate state, with masses $m$ and $m'$ respectively. Finally, we
may cancel out a common factor to give
\begin{align}
{\cal A}-{\cal A}^\dag
=i\int\frac{d^4 k}{(2\pi)^4}
[2\pi i\delta^{(+)}(k^2-m^2)]\,[2\pi i\delta^{(+)}({k'}^2-{m'}^2)]\,
{\cal A}\,{\cal A}^\dag.
\label{unieqres}
\end{align}
This has a pleasing form, and indeed one can show that one reproduces
eq.~(\ref{unieqres}) by adopting a new set of rules for our plus and
minus bubble graphs, which tell us how to translate diagrams featuring
multiple bubbles:
\begin{enumerate}
\item[(i)] Each loop in a plus / minus bubble graph carries a momentum
  integration
  \begin{displaymath}
    i\int\frac{d^4 k}{(2\pi)^4}
  \end{displaymath}
  over a {\it loop momentum} $k$.
\item[(ii)] Each internal line of a graph with momentum $k$ gives a
  factor
  \begin{displaymath}
    2\pi i\delta^{(+)}(k^2-m^2).
  \end{displaymath}
\item[(iii)] Each plus (minus) bubble carries a factor ${\cal A}$
  $(-{\cal A}^\dag)$, as we already defined above.
\end{enumerate}
It is relatively straightforward to check that these rules apply for
any general bubble diagram containing plus and minus bubbles, and any
number of loops. They follow as a direct consequence of the rules for
S-matrix diagrams, and the definition of the plus and minus bubbles
(see e.g. ref.~\cite{Eden:1966dnq} for a full derivation, or the
original work of ref.~\cite{PhysRev.135.B745}). Returning to the
present case, we see that the right-hand side of eq.~(\ref{unieqres})
is pure imaginary if we are above the two particle threshold (as we
have indeed assumed): the factors of $i$ inside the integrand cancel
each other out to give $-1$, and the amplitude combined with its
complex conjugate will be a positive real number. Another way to see
this is to note that the left-hand side of eq.~(\ref{unieqres}) is
\begin{displaymath}
{\cal A}-{\cal A}^\dag=2i{\rm Im}({\cal A}),
\end{displaymath}
so that eq.~(\ref{unieqres}) implies
\begin{align}
2{\rm Im}({\cal A})
=\int\frac{d^4 k}{(2\pi)^4}
[2\pi i\delta^{(+)}(k^2-m^2)]\,[2\pi i\delta^{(+)}({k'}^2-{m'}^2)]\,
|{\cal A}|^2\neq 0,
\label{ImA}
\end{align}
where the right-hand side is {\it real}. We thus conclude that the
amplitude ${\cal A}$ has a non-zero imaginary part above the
two-particle threshold. Below the threshold, the right-hand side of
eq.~(\ref{ImA}) is replaced simply with zero, as it currently assumes
that intermediate states with two exchanged particles are
kinematically allowed. Hence, ${\cal A}\in{\mathbb R}$ on the real
axis of the complex $s$ plane, for values of $\sqrt{s}$ which are
below the two-particle threshold~\footnote{There may still be poles
  for some values of (real) $s$ below the threshold. However, these
  are by definition localised at single points, so that there is at
  least some non-zero region of the real $s$ axis for which ${\cal A}$
  is itself completely real.}. This information is useful, as it
allows to almost immediately conclude that not only does ${\cal A}$
have an imaginary part for values of $\sqrt{s}$ above the threshold,
but it must also be discontinuous across the real $s$ axis. To see
this, note that the {\it Schwarz reflection principle} of complex
analysis states that any function $f(s)$ which is real on some part of
the real $s$ axis satisfies
\begin{equation}
f^*(s)=f(s^*).
\label{Schwarz}
\end{equation}
In words: the function evaluated with the complex conjugate of its
argument, is the same as the complex conjugate of the function of the
original argument~\footnote{Depending on your disposition, this might
  be one of those rare cases in which the formula is much clearer than
  the accompanying words!}. Let us now consider our amplitude ${\cal
  A}(s)$ evaluated off the real axis, by shifting its argument into
the upper half plane i.e. we will look at ${\cal A}(s+i\epsilon)$,
$s,\epsilon\in{\mathbb R}$, with $\epsilon>0$. Equation~(\ref{ImA})
combined with the Schwarz reflection principle of eq.~(\ref{Schwarz})
then implies
\begin{displaymath}
{\rm Im}[{\cal A}(s+i\epsilon)]\propto {\cal A}(s+i\epsilon)
-{\cal A}(s-i\epsilon)\neq 0.
\end{displaymath}
If we now take $\epsilon\rightarrow 0^+$, we see that ${\cal A}$ must
indeed have a discontinuity across the real $s$ axis, which justifies
property (a) above: the function ${\cal A}(s)$ has a cut in the
complex $s$ plane, associated with the 2-particle threshold. The
branch point will be at the value of $s$ such that $\sqrt{s}$ is the
centre of mass energy at which two particles can be produced at rest
in the final state. Some further comments are in order:
\begin{itemize}
\item Here we considered only the two-particle threshold. However,
  similar arguments can be used to show that there will also be cuts
  associated with {\it any} $m$-particle threshold,
  $m\geq3$. Alternatively, we can simply take a single branch cut to
  the right along the real axis, stemming from the two-particle branch
  point, as this will cover all the other cuts.
\item Above we essentially defined the {\it physical} amplitude as a
  limit
  \begin{displaymath}
    {\cal A}(s)=\lim_{\epsilon\rightarrow
      0^+}{\cal A}(s+i\epsilon),\quad s\in{\mathbb R}.
  \end{displaymath}
This amounts to saying that the physical amplitude is given by
approaching the real axis of $s$ from the upper half plane, and you
may be wondering why we did not choose the lower half plane. The
answer is that the choice we have made can be (sort of) justified on
the grounds of causality, by looking at the scattering of general
wavepackets~\cite{Eden:1966dnq,Collins:1977jy}. However, such
arguments have never been made truly rigorous (although see
ref.~\cite{PhysRev.135.B1255} for an admirable attempt).
\item It is worthwhile noting that the prescription used here to
  define the physical amplitude turns out to agree with the
  consequences of the Feynman $i\epsilon$ prescription in quantum
  field theory. This provides some justification for our choice of the
  upper-half plane, but of course we are trying to construct arguments
  that are {\it independent} of any particular theory. 
\item You might be wondering if it is mysterious that, if we are
  working in QFT or in pure S-matrix theory, we always need some oddly
  pedantic prescription involving esoteric factors of $i\epsilon$. It
  is not mysterious at all - as soon as we surmise that the amplitude
  must be a complex function in general, and that it must also have
  discontinuities and therefore potentially be multi-valued, we are
  {\it obliged} to state how the physical value of the amplitude is to
  be obtained, which is bound to end up looking vaguely similar in
  different approaches.
\end{itemize}
Having sketched how to derive property (a) above, let us now turn to
property (b), whose proof is perhaps more difficult than you might
think it should be, but also rather fun. We start by stating that one
can carry out a similar exercise to the derivation of eq.~(\ref{uni3})
using the unitarity equation for $3\rightarrow 3$ scattering, where we
assume the centre of mass energy is below the four-particle
threshold. Combining this with the cluster decomposition of
eq.~(\ref{S3bubble}), one can derive the following~\footnote{Deriving
  eq.~(\ref{S3uni}) is an excellent exercise to see if you understand
  the bubble graph algebra. The full derivation can be found in
  ref.~\cite{Eden:1966dnq}.}:
\begin{align}
\raisebox{-9pt}{\scalebox{0.35}{\includegraphics{plus33.pdf}}}&
\quad-\quad
\raisebox{-9pt}{\scalebox{0.35}{\includegraphics{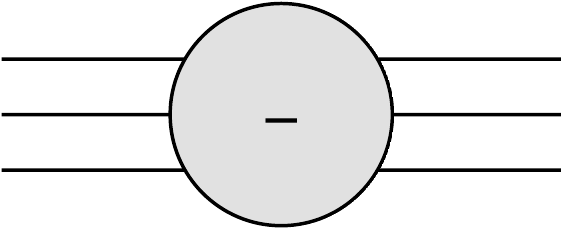}}}
\quad=\quad 
\raisebox{-9pt}{\scalebox{0.35}{\includegraphics{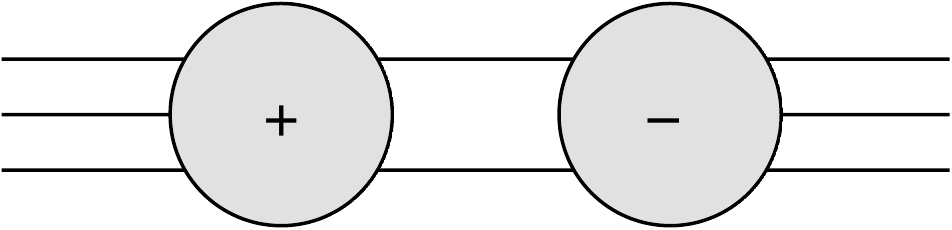}}}
\quad+\quad
\raisebox{-9pt}{\scalebox{0.35}{\includegraphics{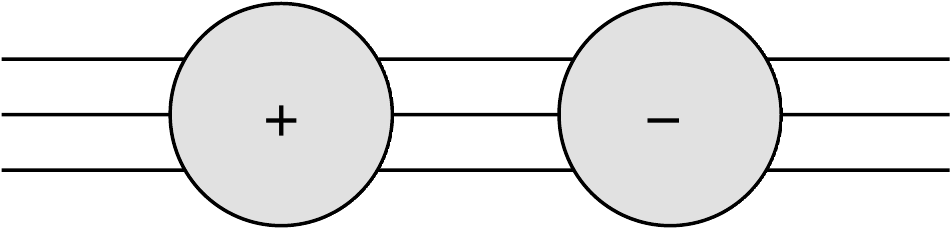}}}
\notag\\
&\quad+\quad
\sum
\raisebox{-9pt}{\scalebox{0.35}{\includegraphics{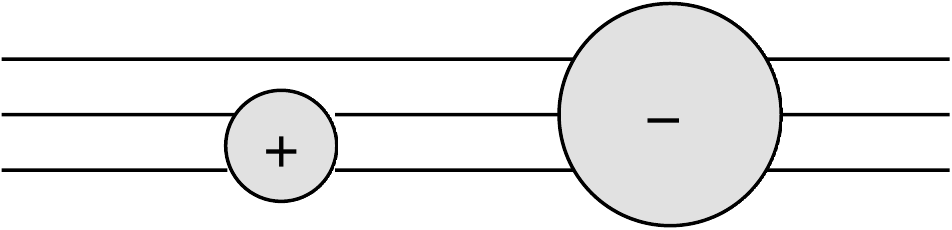}}}
\quad+\quad
\sum
\raisebox{-9pt}{\scalebox{0.35}{\includegraphics{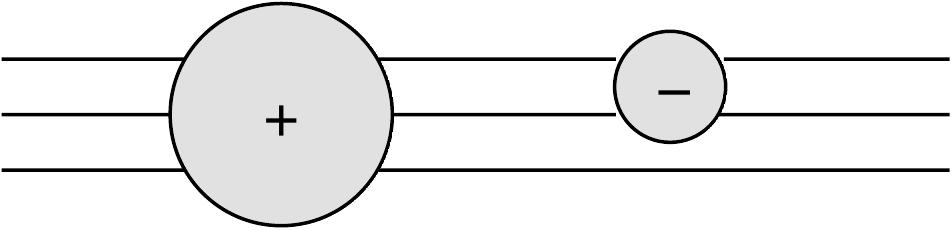}}}\notag\\
&\quad+\quad
\sum 
\raisebox{-12pt}{\scalebox{0.35}{\includegraphics{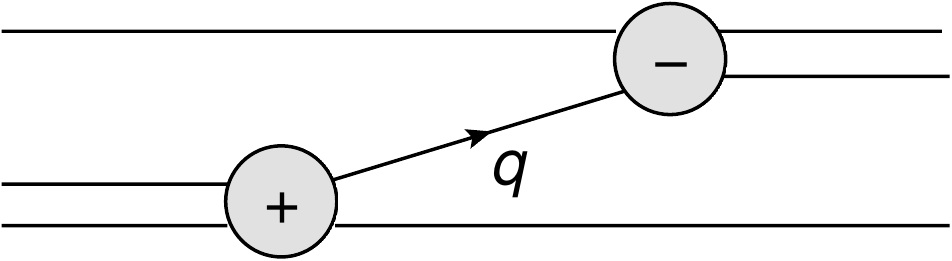}}}
\quad.
\label{S3uni}
\end{align}
There are many contributions on the right-hand side, but we will only
need to focus on the last one i.e. the sum of terms on the third
line. Every term in this sum has a nice physical interpretation: there
is a $2\rightarrow 2$ scattering process, one of whose particles
travels some distance and then interacts with another initial
particle, in a second $2\rightarrow 2$ scattering process. We have
labelled the 4-momentum of the exchanged particle by $q$ in
eq.~(\ref{S3uni}). Using our rules for such diagrams, this
contribution has the form
\begin{equation}
{\cal A}_{2\rightarrow 2}\,[2\pi i\delta^{(+)}(q^2-m^2)]\,
(-{\cal A}^\dag_{2\rightarrow 2}),
\label{ampfac}
\end{equation}
where ${\cal A}_{2\rightarrow 2}$ is the amplitude for $2\rightarrow
2$ scattering, as the notation suggests. The contribution in the third
line of eq.~(\ref{S3uni}) is thus singular for $q^2=m^2$, and in fact
this singularity has a nice physical interpretation: due to the
short-range interactions, it is infinitely more likely that the
$3\rightarrow 3$ scattering process occurs via two successive
$2\rightarrow 2$ scatterings, than any other possibility. One can
argue on very general grounds~\cite{PhysRev.135.B745} that the only
way eq.~(\ref{S3uni}) can be satisfied is if the $3\rightarrow 3$
scattering amplitude (i.e. the first term on the left-hand side of
eq.~(\ref{S3uni})) {\it also} has a contribution that is singular at
$q^2=m^2$, and which we may draw as
\begin{equation}
\raisebox{-9pt}{\scalebox{0.35}{\includegraphics{plus33.pdf}}}
\quad\xrightarrow{q^2\rightarrow m^2}\quad
\raisebox{-12pt}{\scalebox{0.35}{\includegraphics{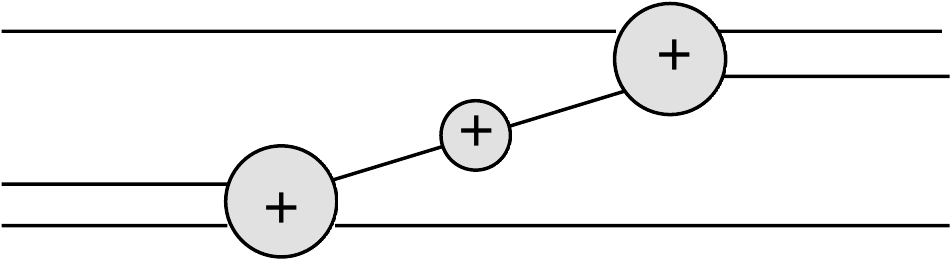}}}\quad,
\label{amp33lim}
\end{equation}
where we have introduced a dressed internal line factor
\begin{equation}
\raisebox{-8pt}{\scalebox{0.4}{\includegraphics{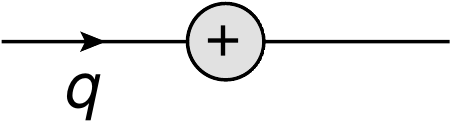}}}
\quad\equiv\quad
D^{(+)}(q^2,m^2),
\label{qline}
\end{equation}
and our task is to find the function $D^{(+)}(q^2,m^2)$. This will
then show us what kind of singularity an amplitude contains when a
single on-shell particle is exchanged. Given eq.~(\ref{qline}), we may
also write
\begin{equation}
\raisebox{-8pt}{\scalebox{0.4}{\includegraphics{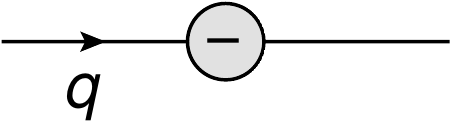}}}
\quad\equiv\quad
D^{(-)}(q^2,m^2)=[D^{(+)}]^*.
\label{qline2}
\end{equation}
That is, as for amplitudes, we use a minus bubble to represent the
complex conjugate of the quantity associated with a plus bubble. Armed
with these definitions, we may consider eq.~(\ref{S3uni}) in the limit
$q^2\rightarrow m^2$, keeping only those terms which are singular. The
result is
\begin{align}
&\raisebox{-12pt}{\scalebox{0.35}{\includegraphics{Dplus.pdf}}}
\quad-\quad
\raisebox{-12pt}{\scalebox{0.35}{\includegraphics{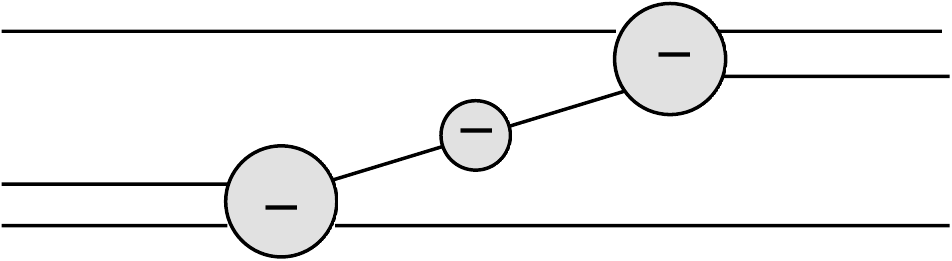}}}
\quad=\quad\notag\\
&\quad \raisebox{-12pt}{\scalebox{0.35}{\includegraphics{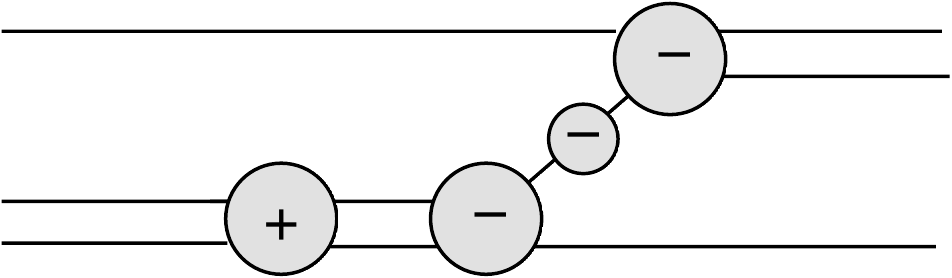}}}
\quad+\quad
\raisebox{-12pt}{\scalebox{0.35}{\includegraphics{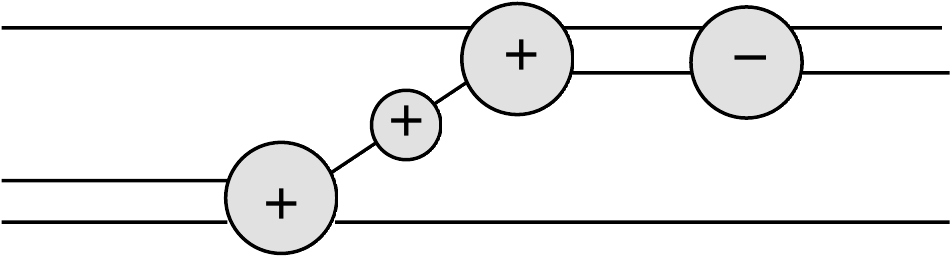}}}
\quad+\quad
\raisebox{-12pt}{\scalebox{0.35}{\includegraphics{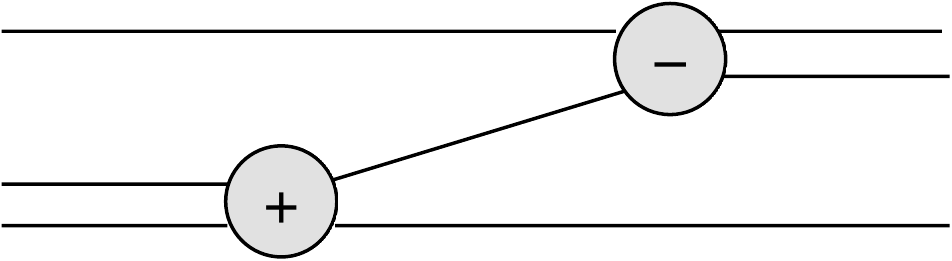}}}\quad.
\label{Deq}
\end{align}
To see where this comes from, note that on the left-hand side of
eq.~(\ref{S3uni}), each $3\rightarrow3$ scattering amplitude will be
dominated, by its singular piece as $q^2\rightarrow m^2$, which by
definition is given by eq.~(\ref{amp33lim}) and its complex
conjugate. On the right-hand side of eq.~(\ref{S3uni}), the only
singular terms (other than in the last line) come from taking the
limit of $q^2\rightarrow m^2$ in any $3\rightarrow3$ scattering
amplitudes that appear. There are two of these in the final term in
the first line, which give rise to the first two terms on the
right-hand side of eq.~(\ref{Deq}). If we are focusing on the exchange
of a single type of particle, only one term in the sum on the last
line of eq.~(\ref{S3uni}) contributes i.e. the term in which the
particle of interest is exchanged. 

We can now simplify eq.~(\ref{Deq}) as follows. First, we may use the
$2\rightarrow 2$ unitarity equation (eq.~(\ref{uni3})) on the lower
line of the first term on the right-hand side of eq.~(\ref{Deq}), so
that this becomes
\begin{equation}
\raisebox{-12pt}{\scalebox{0.35}{\includegraphics{Deq1.pdf}}}
\quad=\quad
\raisebox{-12pt}{\scalebox{0.35}{\includegraphics{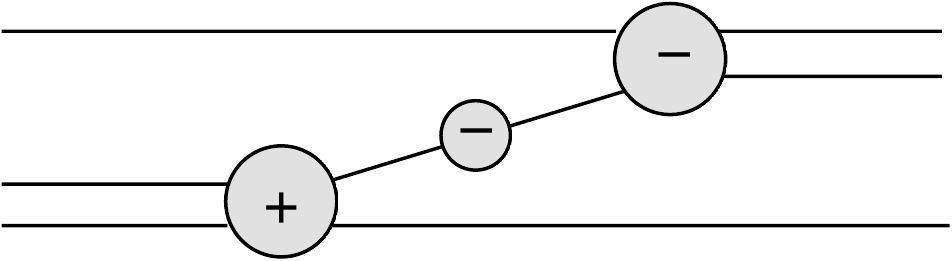}}}
\quad-\quad
\raisebox{-12pt}{\scalebox{0.35}{\includegraphics{Dminus.pdf}}}
\quad.
\label{Dpmm}
\end{equation}
Likewise, one has
\begin{equation}
\raisebox{-12pt}{\scalebox{0.35}{\includegraphics{Deq2.pdf}}}
\quad=\quad
\raisebox{-12pt}{\scalebox{0.35}{\includegraphics{Dplus.pdf}}}
\quad-\quad
\raisebox{-12pt}{\scalebox{0.35}{\includegraphics{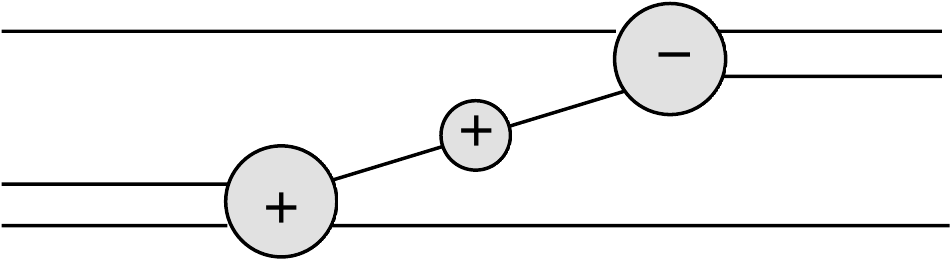}}}
\quad.
\label{Dppm}
\end{equation}
Substituting eqs.~(\ref{Dpmm}) and~(\ref{Dppm}) into eq.~(\ref{Deq}),
we find the simpler relation
\begin{equation}
\raisebox{-12pt}{\scalebox{0.35}{\includegraphics{Dppm.pdf}}}
\quad-\quad
\raisebox{-12pt}{\scalebox{0.35}{\includegraphics{Dpmm.pdf}}}
\quad=\quad
\raisebox{-12pt}{\scalebox{0.35}{\includegraphics{Deq3.pdf}}}\quad,
\label{Drelation}
\end{equation}
which when translated back into algebra yields
\begin{displaymath}
{\cal A}_{2\rightarrow 2}\,D^{(+)}(q^2,m^2)\,{\cal
  A}^\dag_{2\rightarrow 2} -{\cal A}_{2\rightarrow
  2}\,D^{(-)}(q^2,m^2)\,{\cal A}^\dag_{2\rightarrow 2}
={\cal A}_{2\rightarrow 2}\,[2\pi i\delta^{(+)}(q^2-m^2)]\,
{\cal A}^\dag_{2\rightarrow 2}.
\end{displaymath}
This shows why the above simplifications have been useful: we now
have the {\it same} factors of a $2\rightarrow 2$ amplitude and / or
its complex conjugate in each term on the left- and right-hand
sides. We can thus cancel them, to get 
\begin{equation}
D^{(+)}(q^2,m^2)-D^{(-)}(q^2,m^2)=2\pi i\delta^{(+)}(q^2-m^2).
\label{Drelation2}
\end{equation}
This has a unique solution that is regular in the upper half plane of
$q^2$:
\begin{equation}
D^{(+)}=-\frac{1}{q^2-m^2+i\epsilon}.
\label{D+sol}
\end{equation}
Here the $i\epsilon$ ensures that we are indeed in the upper half
plane, as was required from causality. Drawing everything together, we
have found that the $3\rightarrow 3$ scattering amplitude contains a
{\it pole} at $q^2=m^2$, where this corresponds to the exchange of a
single particle with 4-momentum $q$ and mass $m$. At this pole, the
amplitude {\it factorises} into scattering amplitudes on either side
of the exchanged particle, times a function (that of
eq.~(\ref{D+sol})) describing the exchange. Some further comments:
\begin{itemize}
\item Here the singularity is in $q^2$ (the exchanged momentum) which,
  by momentum conservation, will be equal to the invariant mass of all
  4-momenta on one side of the exchanged particle. The above argument
  generalises, for any number of momenta on either side of the
  exchange, provided we use the corresponding multiparticle unitarity
  equations.
\item The exchanged particle may be fundamental or composite. All that
  matters in the above argument is that it is some one-particle state
  of the theory.
\item The singularity structure we have derived is not a consequence
  of unitarity alone, but also of cluster decomposition: we relied
  crucially on the decomposition of the S-matrix into multiple
  connected pieces.
\item The results above agree with quantum field theory, as they must
  do given that we are being fully general
  here. Equation~(\ref{D+sol}) clearly matches the propagator of a
  scalar particle, if we identify the $i\epsilon$ required to ensure
  we are in the upper half plane of $q^2$ with the Feynman $i\epsilon$
  prescription used to implement causality in QFT. However, the
  factorisation of an amplitude into smaller amplitudes, times a
  divergent propagator-like function, also works if the exchanged
  state is described by a composite operator composed of many fields
  (e.g. a complicated bound state): see e.g. section 10.2 of
  ref.~\cite{Weinberg:1995mt}, or section 24.3 of
  ref.~\cite{Schwartz:2013pla}.
\end{itemize}
In summary, we have established that the S-matrix must be a complex
function of Lorentz scalars, where the latter are themselves
considered as complex variables. The amplitude will then have poles
and cuts associated with bound states and multiparticle thresholds
respectively. Why, though, is this useful? Firstly, we will see some
explicit examples of amplitudes having poles in the squared centre of
mass energy $s$ later on. We will then be able to directly interpret
these in terms of bound states! Secondly, knowing the singularity of
the structure of the amplitude ${\cal A}$ is a crucial preqrequisite
for being able to analytically continue this function. We will need to
do this later, when describing high energy behaviour. Furthermore, the
ability to continue a function to different {\it regions} of its
kinematic variables allows us to conclude that the amplitudes for
different scattering processes are in fact related. We explore this in
the following section.

\subsection{Physical regions and crossing}
\label{sec:crossing}

\begin{figure}
\begin{center}
\scalebox{0.6}{\includegraphics{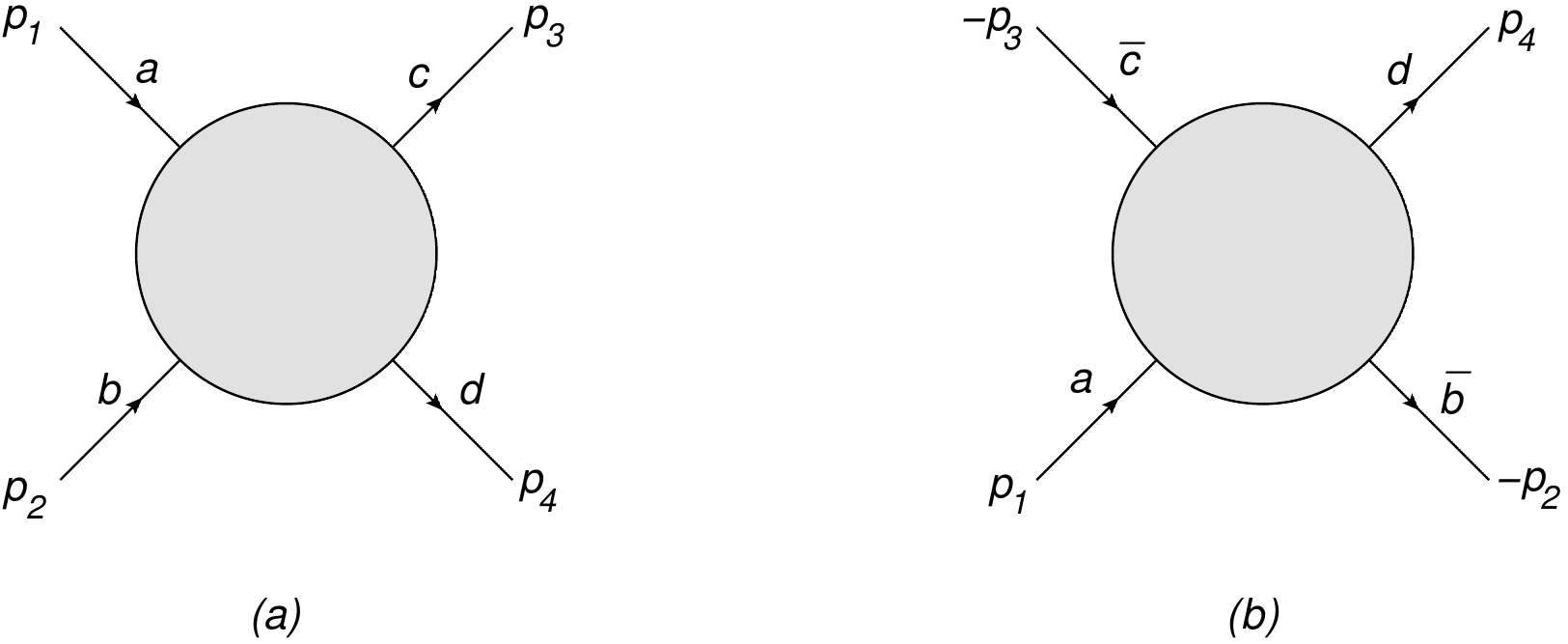}}
\caption{(a) A $2\rightarrow 2$ scattering process, where 4-momenta
  $\{p_i\}$ and particle species $\{a,b,c,d\}$ are labelled; (b)
  crossed process, in which (anti-)particle momenta have been
  exchanged.}
\label{fig:2to2}
\end{center}
\end{figure}

We have seen that amplitudes depend on Lorentz invariants, regarded as
complex variables. However, not all regions of the complex plane of
each Lorentz invariant are kinematically accessible. To illustrate
this, let us consider the $2\rightarrow2$ scattering process
\begin{equation}
a(p_1)+b(p_2)\rightarrow c(p_3)+d(p_4)
\label{2to2a}
\end{equation}
depicted in figure~\ref{fig:2to2}. 
For simplicity, we will consider a case in which all particles have
equal mass, so that 
\begin{displaymath}
p_i^2=m^2\quad \forall i. 
\end{displaymath}
For any $2\rightarrow 2$ process, it is conventional to define the
so-called {\it Mandelstam invariants}
\begin{equation}
s=(p_1+p_2)^2,\quad t=(p_1-p_3)^2,\quad u=(p_1-p_4)^2.
\label{mandies}
\end{equation}
We have seen the first of these already: it represents the squared
total energy in the centre of mass frame. Furthermore, the second
invariant $t$ has a nice interpretation as the square of the total
4-momentum exchanged between the incoming particles. The third
invariant $u$ is not independent from the others. Using the momentum
conservation condition
\begin{equation}
p_1+p_2=p_3+p_4,
\label{momcon}
\end{equation}
one may derive the fact that 
\begin{equation}
s+t+u=\sum_{i=1}^4 m_i^2=4m^2.
\label{stu}
\end{equation}
Let us now see what the region of the $(s,t,u)$ space is that
corresponds to physically allowed scattering. We can choose any frame
to examine this, given that each parameter is Lorentz invariant. Let
us then choose the centre of mass frame, in which one may parametrise
the incoming and outgoing momenta according to
\begin{equation}
p_1=(E,\vec{p}),\quad p_2=(E,-\vec{p}),\quad E^2-\vec{p}^2=m^2.
\label{incoming}
\end{equation}
and
\begin{equation}
p_3=(E',\vec{p}'),\quad p_4=(E',-\vec{p}'),\quad (E')^2-(\vec{p}')^2=m^2.
\label{outgoing}
\end{equation}
respectively. Energy conservation then implies that
$E=E'$. Furthermore, there will be some angle $\theta$ between the
3-momenta $\vec{p}$ and $\vec{p}'$ in general, so that we have
\begin{align}
s&=4E^2=4(|\vec{p}|^2+m^2),\notag\\
t&=-2|\vec{p}|^2(1-\cos\theta),\notag\\
u&=-2|\vec{p}|^2(1+\cos\theta).
\label{stuparam}
\end{align}
Physical scattering corresponds to $E\geq 0$ and $|\cos\theta|\leq 1$,
thus to the region
\begin{equation}
s\geq 4m^2,\quad t\leq 0,\quad u\leq 0.
\label{schannel}
\end{equation}
This is a highly restricted region of the total $(s,t,u)$
space. However, other regions of this space can also be given a
physical meaning. Imagine, for example, rotating the diagram of
figure~\ref{fig:2to2}(a) to get the diagram of
figure~\ref{fig:2to2}(b). One can easily see from rotating the diagram
that $p_1$ and $p_4$ remain incoming and outgoing momenta
respectively. However, $p_3$ switches from the final state to the
initial state. If we want the 4-momentum to be flowing {\it in} to the
scattering process, we must reverse the sign of $p_3$. Likewise, $p_2$
moves to the final state, and again must be reversed in sign if we
want the momentum to be flowing {\it out} of the process. In
figure~\ref{fig:2to2}(b), we have also allowed for the particle
species to change in general, if a particle moves from the initial to
the final state or vice versa i.e. we have replaced $(b,c)\rightarrow
(\bar{b},\bar{c})$. To see how they are related, consider that there
is some additive quantum number possessed by all the particles e.g. a
conserved charge $Q$ of some kind. The existence of the scattering
process in figure~\ref{fig:2to2}(a) then implies
\begin{equation}
Q_a+Q_b=Q_c+Q_d.
\label{chargecon1}
\end{equation}
Similarly, the process of figure~\ref{fig:2to2}(b) implies
\begin{equation}
Q_a+Q_{\bar{c}}=Q_{\bar{b}}+Q_d,
\label{chargecon2}
\end{equation}
and comparison of eqs.~(\ref{chargecon1}) and~(\ref{chargecon2})
implies~\footnote{Equation~(\ref{antiparticle}) is actually a stronger
  condition than that directly implied by comparing
  eq.~(\ref{chargecon1}) with eq.~(\ref{chargecon2}). However, one
  can consider swapping the particles from the final to the initial
  state (or vice versa) one by one, in which case
  eq.~(\ref{antiparticle}) does indeed follow.}
\begin{equation}
Q_{\bar{b}}=-Q_b,\quad Q_{\bar{c}}=-Q_c.
\label{antiparticle} 
\end{equation}
This argument will work for any such quantum number, and we rapidly
conclude that $\bar{b}$ and $\bar{c}$ must be the {\it antiparticles}
of $b$ and $c$. From the figure, we see that the new process
corresponds to the old one (i.e. it is the same amplitude, just
rotated) with the replacements $b\rightarrow \bar{c}$, $c\rightarrow
\bar{b}$, and $s\leftrightarrow t$. It will thus have a physical
region
\begin{equation}
t\geq 4m^2,\quad s\leq 0,\quad u\leq 0,
\label{tchannel}
\end{equation}
and we expect
\begin{equation}
{\cal A}_{ab\rightarrow cd}(s,t,u)=
{\cal A}_{a\bar{c}\rightarrow \bar{b}d}(t,s,u).
\label{crossingeq}
\end{equation}
Clearly this is a different region of $(s,t,u)$ space than that of
eq.~(\ref{schannel}), as might be expected given that our two
different physical regions are meant to correspond to physically
distinct scattering processes. Switching between processes that
correspond to different regions of the same amplitude function is
called {\it crossing}, and the fact that the amplitudes of
eq.~(\ref{crossingeq}) are equal is called {\it crossing
  symmetry}. However, we have to be very careful about whether this is
really true, despite the fact that the above arguments might look very
convincing! We have in fact made an assumption in
eq.~(\ref{crossingeq}), namely that one can safely analytically
continue the amplitude in one physical region into another (n.b. this
happens when we send $(s,t)\rightarrow (t,s)$). As we send $s$ or $t$
from one region of their respective complex planes to another, we
might encounter singularities, that disrupt the simple relationship of
eq.~(\ref{crossingeq}), in that we might not be able to interpret the
function we get in the new region as a scattering amplitude. However,
provided that the only singularities we encounter are the poles and
cuts we already found, and that we know amplitudes are meant to have,
then everything should be OK. In asserting crossing symmetry, we thus
assume that the only singularities in the entire complex plane of any
Mandelstam invariant are the poles and cuts associated with bound
states and thresholds. This is sometimes called {\it maximal
  analyticity of the first kind}~\cite{Collins:1977jy}, which is a
rather silly and fancy name, but at least makes clear that an
assumption is being made~\footnote{There is actually an even more
  subtle assumption being made, namely that swapping $s$ and $t$ takes
  us from the physical Riemann sheet of one complex amplitude to the
  physical sheet of another. It is best to simply say that crossing
  symmetry is {\it assumed} rather than {\it derived}. Perturbative
  amplitudes in QFT certainly respect it.}. Once we have crossing
symmetry, however, we can surmise the presence of cuts and poles in
the complex $t$ and $u$ planes, which are obtained from their
counterparts in the $s$ plane by the appropriate crossing
relationships.

\begin{figure}
\begin{center}
\scalebox{0.6}{\includegraphics{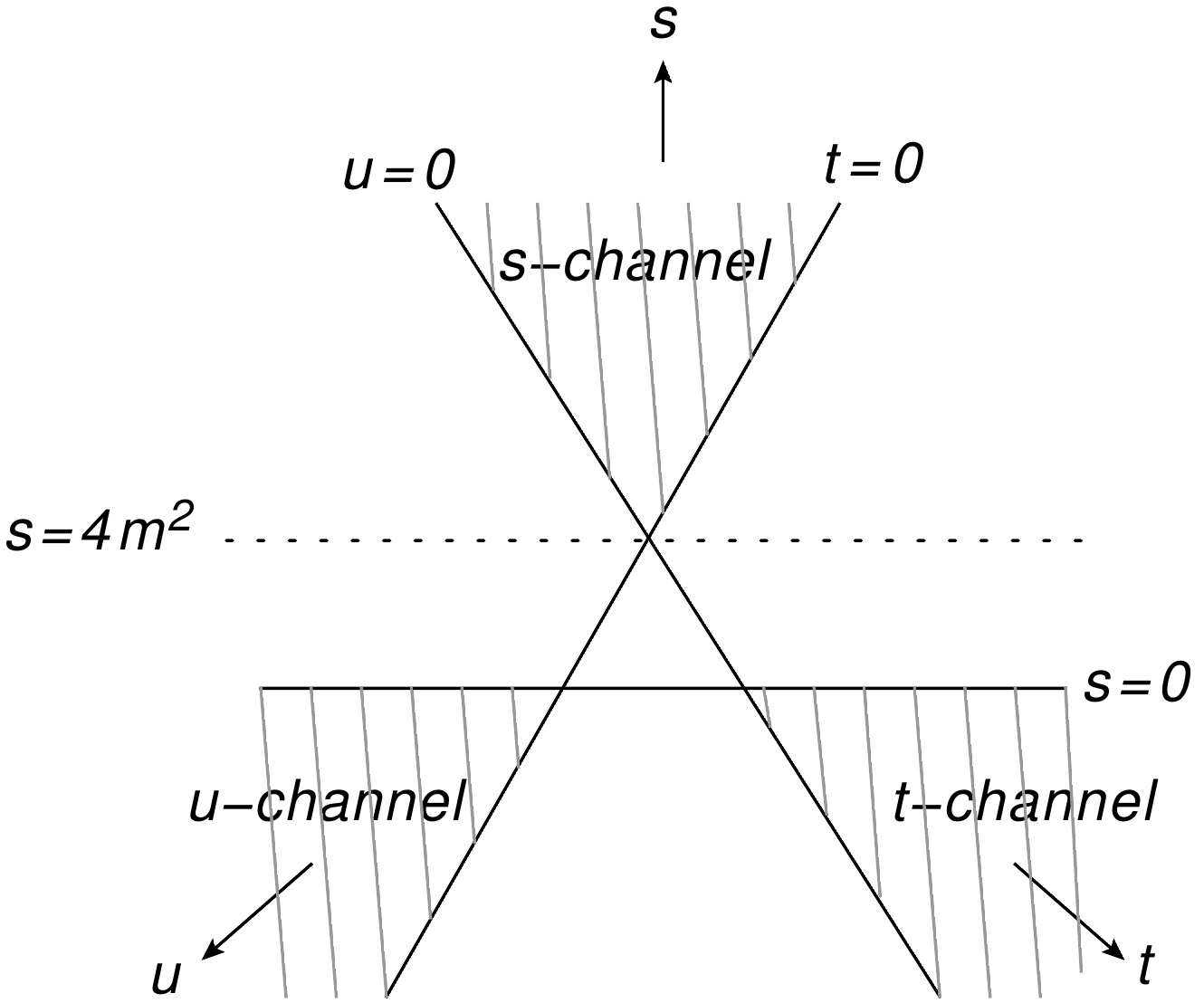}}
\caption{Mandelstam diagram showing the three physical regions of
  $2\rightarrow 2$ scattering, in the case where all particle masses
  are equal.}
\label{fig:mandelstam}
\end{center}
\end{figure}

Above, we interchanged $s$ and $t$ and obtained a different
process. We could also have performed a different crossing, and
obtained
\begin{equation}
{\cal A}_{ab\rightarrow cd}(s,t,u)=
{\cal A}_{\bar{d}b\rightarrow c\bar{a}}(u,t,s).
\label{crossingeq2}
\end{equation}
Here $u$ and $s$ have been interchanged, so we conclude from
eq.~(\ref{schannel}) that we have produced a new physical region,
defined by
\begin{equation}
u\geq 4m^2,\quad t\leq 0,\quad s\leq 0.
\label{uchannel}
\end{equation}
One way to draw all these possible physical regions is on a so-called
{\it Mandelstam diagram}, which has $(s,t,u)$ axes mutually aligned at
$60^\circ$ to one another. This is shown, for our example of equal
masses, in figure~\ref{fig:mandelstam}. 
The three physical regions we have identified in eqs.~(\ref{schannel},
\ref{tchannel}, \ref{uchannel}) are known as the ``$s$-channel'',
``$t$-channel'' and ``$u$-channel'' respectively, where the parameters
$(s,t,u)$ increase in the directions shown. The $s$-channel starts at
$s=4m^2$, as we found above, and we illustrate this on the
figure. Such figures are a tad obfuscating at first glance,
particularly given that they have long since fallen out of
fashion. However, they do provide a nice way to visualise all physical
regions at once. Note that the boundaries of the physical regions in
figure~\ref{fig:mandelstam} are particularly nice (i.e. straight
lines), due to our choice of equal masses. Things are more complicated
for general particles / masses, as you may like to investigate
yourself! In such cases, there will still be three physical regions,
and we can still talk about analytic continuation from one channel to
another.

\section{Regge Theory}
\label{sec:Regge}

\subsection{The Regge limit}
\label{sec:limit}

So far, we have studied general structures of amplitudes
(e.g. singularities, crossing symmetry) and how these arise from
general assumptions (e.g. superposition, Lorentz invariance,
unitarity, localised interactions). In this section, we will focus on
a particular kinematic region, and see what else we can learn. In
particular, we will study the high-energy or {\it Regge limit} of
$2\rightarrow 2$ scattering~\footnote{The use of the phrase {\it high
    energy} for this limit is something of an abuse of terminology,
  albeit one that is common throughout the literature. We will use the
  terms {\it high energy} and {\it Regge} interchangeably
  throughout.}. Defining invariants as in eq.~(\ref{mandies}), we can
formally define the Regge limit as
\begin{equation}
s\gg -t \gg m^2.
\label{Reggelim}
\end{equation}
Here we are working in the physical $s$-channel region of
eq.~(\ref{schannel}), for which $t$ will be negative. For simplicity,
we will continue to assume common particle masses $m$, but this will
not matter too much for what follows. Physically, the above limit
corresponds to the centre of mass energy being much larger than the
momentum exchanged by the scattering particles, which corresponds to
highly forward scattering, in which the colliding particles barely
glance off each other. Furthermore, the requirement of high energy
does not itself place a restriction on the relative ordering of the
(squared) momentum transfer $|t|$, and the squared particle masses
$m^2$. We have chosen a particular ordering above, but you might
sometimes see the Regge limit being defined with the alternative
choice:
\begin{displaymath}
s\gg m^2\gg -t.
\end{displaymath}
Which choice one makes ultimately depends on whether one wants to keep
track of mass information, including allowing a certain mass to become
large. For our present discussion, this will not be important, and we
will therefore stick with the limit of eq.~(\ref{Reggelim}).

Our reason for studying the high energy limit is that it has a number
of applications in non-abelian gauge theories and gravity. In collider
physics, for example, present-day experiments such as the LHC (and to
some extent the Tevatron and HERA beforehand) probe kinematic regions
in which additional contributions arising from the Regge limit become
relevant. This may affect the description of the quark and gluon
distributions within the
proton~\cite{White:2006yh,White:2006xv,Ball:2017otu,Bonvini:2016wki,Altarelli:2008aj,Altarelli:2003hk,Altarelli:2001ji,Altarelli:1999vw,Ciafaloni:2007gf,Ciafaloni:2006yk,Ciafaloni:2003rd},
for example, or the radiation profile of multiple
jets~\cite{Andersen:2019yzo,Andersen:2017sht,Andersen:2017kfc,Andersen:2012gk,Andersen:2011hs,Andersen:2011zd,Andersen:2009he,Andersen:2009nu,Andersen:2008gc,Andersen:2008ue,Jung:2010si,Chachamis:2017vgr,Chachamis:2015ico,Caporale:2015int,Caporale:2015vya}.

In gravity, the high energy limit is related to scattering
above the Planck scale (i.e. {\it transplanckian
  scattering})~\cite{Amati:1987wq, Muzinich:1987in, tHooft:1987vrq,
  Amati:1987uf,
  Amati:1990xe,Verlinde:1991iu,Amati:1992zb,Amati:1993tb,Giddings:2010pp,Kabat:1992tb,DAppollonio:2010krb,Akhoury:2013yua,Collado:2018isu,KoemansCollado:2019lnh,KoemansCollado:2019ggb},
and also to high energy collisions of black holes, which have become
extremely topical due to the recent discovery of gravitational waves
by LIGO.

We will see that it is possible to say quite a lot about the high
energy behaviour of scattering amplitudes, based solely on general
principles. The main idea involved is that large $s$ behaviour in the
$s$-channel is related to particle exchange in the $t$-channel. To make
sense of this statement, let us first note that it is possible to
write the amplitude in the physical $s$-channel as a sum of
contributions with definite angular momentum in the $t$-channel:
\begin{equation}
{\cal A}(s,t)=16\pi \sum_{l=0}^\infty (2l+1)a_l(t)P_l(z_t).
\label{partialwave}
\end{equation}
Here the overall factor of $16\pi$ is purely conventional, as is the
factor of $(2l+1)$ in each term (although it has a simple
interpretation as the total number of states with orbital angular
momentum $l$). The function $P_l(z_t)$ is a {\it Legendre polynomial}
involving the $t$-channel scattering angle $\theta_t$ via the argument
\begin{equation}
z_t=\cos\theta_t.
\label{ztdef}
\end{equation}
You may recall that the Legendre polynomials are used in the solution
of the hydrogen atom in quantum mechanics, where they arise
specifically as the eigenfunctions of orbital angular momentum. Here
the principle is similar: whatever the amplitude is, it must be
expandable in terms of a complete set of eigenfunctions involving the
$t$-channel scattering angle. These eigenfunctions are precisely the
Legendre polynomials. The final ingredient of each term in
eq.~(\ref{partialwave}) is $a_l(t)$, which tells us ``how much'' of
each Legendre polynomial we have. This will depend upon $t$ in
general, given that we have expressed the amplitude in terms of
definite angular momentum in the
$t$-channel. Equation~(\ref{partialwave}) is called a {\it partial
  wave expansion}, and the coefficients $\{a_l(t)\}$ are called {\it
  partial wave amplitudes}. You may already be familiar with this from
the study of scattering theory in non-relativistic quantum
mechanics. If not, we can surmise eq.~(\ref{partialwave}) on very
general grounds, as described above.

\begin{figure}
\begin{center}
\scalebox{0.5}{\includegraphics{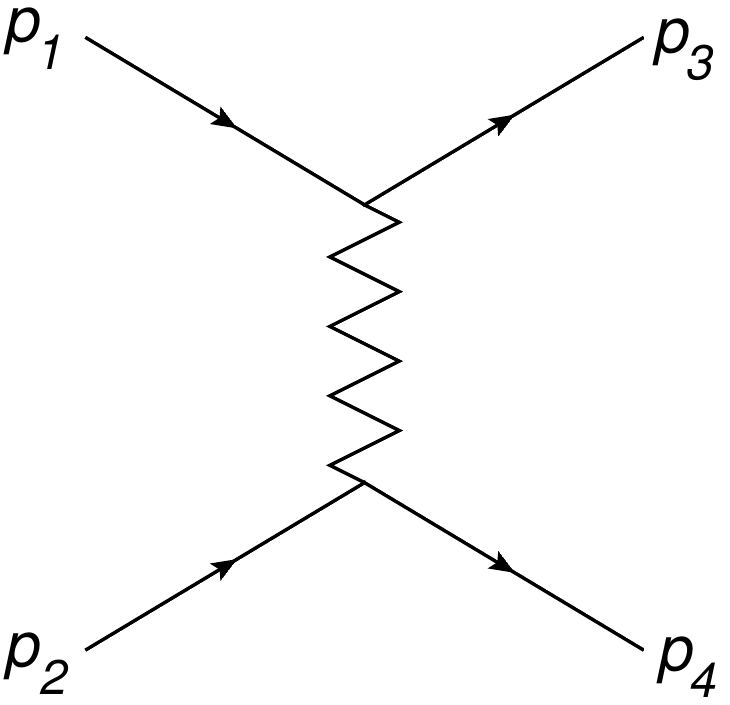}}
\caption{Exchange of a particle of spin $J$ in the $t$-channel of
  $2\rightarrow2$ scattering, where we consider the physical
  scattering (from left to right) in the $s$-channel. Strictly
  speaking, this is not a Feynman diagram, given that we are not
  necessarily in a QFT.}
\label{fig:tchannel}
\end{center}
\end{figure}

Equation~(\ref{partialwave}) is a bit unusual, in that we are talking
about the amplitude in the physical $s$-channel, but expressed as a
$t$-channel partial wave amplitude. This is because we are trying to
justify the statement made above, that exchanges in the $t$-channel
can be related to definite high energy behaviour in the
$s$-channel. To go further, we need the following formula for the
$t$-channel scattering angle in terms of the Mandelstam invariants $s$
and $t$:
\begin{equation}
\cos\theta_t=1+\frac{2s}{t-4m^2}.
\label{costhetat}
\end{equation}
To see where this comes from, note that the scattering angle in the
$s$-channel is given implicitly in eq.~(\ref{stuparam}), from which we
find
\begin{displaymath}
|\vec{p}|^2=-\frac{t}{(1-\cos\theta_s)}=\frac{s}{4}-m^2\quad
\Rightarrow\quad
\cos\theta_s=1+\frac{2t}{s-4m^2}.
\end{displaymath}
The result of eq.~(\ref{costhetat}) then follows from crossing
symmetry, and our $s$-channel amplitude becomes
\begin{equation}
{\cal A}(s,t)= 16\pi \sum_{l=0}^\infty (2l+1)a_l(t)P_l\left(
1+\frac{2s}{t-4m^2}\right).
\label{partialwave2}
\end{equation}
Now let us take the case where a single particle of definite spin $J$
is exchanged in the $t$-channel, as in figure~\ref{fig:tchannel}. Note
that this is not, strictly speaking, a Feynman diagram, as we are not
necessarily in a QFT. Even if we were, the exchanged particle may be
some complicated non-perturbative bound state, rather than one of the
fundamental particles in the theory. As we saw in
section~\ref{sec:Smatrix}, amplitudes are singular if a particle is
exchanged in a given channel. Thus, the expansion of
eq.~(\ref{partialwave2}) is dominated by a single term, namely the one
corresponding to $l=J$ (i.e. the angular momentum in the $t$-channel
must correspond to the exchanged spin):
\begin{displaymath}
{\cal A}(s,t)\rightarrow 16\pi (2J+1)a_J(t)P_J
\left(1+\frac{2s}{t-4m^2}\right).
\end{displaymath}
We can now take the Regge limit of eq.~(\ref{Reggelim}), using the
following known asymptotic behaviour of Legendre polynomials:
\begin{equation}
\lim_{z\rightarrow\infty} P_l(z)\sim z^l.
\end{equation}
We conclude that
\begin{equation}
{\cal A}(s,t)\sim \left(\frac{s}{t}\right)^J.
\label{ARegge}
\end{equation}
In words: {\it power-like growth of amplitudes with squared centre of
  mass energy $s$ is associated with exchange of (bound) particle
  states in the $t$-channel.} Our wording stresses that the exchanged
particles may be fundamental or composite.

The above argument is, of course, a simplification, given that it
relied on the exchange of only one particle state. In practice, many
particles may be exchanged in the $t$-channel, so that multiple
partial waves contribute. One must then consider the full series of
eq.~(\ref{partialwave2}), and here there is a problem: if we
na\"{i}vely try to carry out the sum of eq.~(\ref{partialwave2}) in
the physical $s$-channel region, it does not converge! The origin of
this problem is that the $t$-channel partial wave expansion, strictly
speaking, only applies in the physical region of $t$-channel
scattering, namely that of eq.~(\ref{tchannel}). However, we are
choosing to apply it in the physical $s$-channel region of
eq.~(\ref{schannel}), with nary a thought as to whether or not it has
been correctly analytically continued. There is a truly ingenious way
to get around this problem that was first proposed by Regge in
non-relativistic quantum mechanics~\cite{Regge:1959mz}. It involves
rewriting the series of eq.~(\ref{partialwave2}) as a complex
integral, that can then be made to give sensible results in the
physical $s$-channel region by deforming it. The upshot is that if a
theory contains a spectrum of states that can be exchanged in the
$t$-channel, then this indeed leads to generic high energy behaviour
in the $s$-channel.

\subsection{Complex angular momentum}
\label{sec:complexangmom}

\begin{figure}
\begin{center}
\scalebox{0.8}{\includegraphics{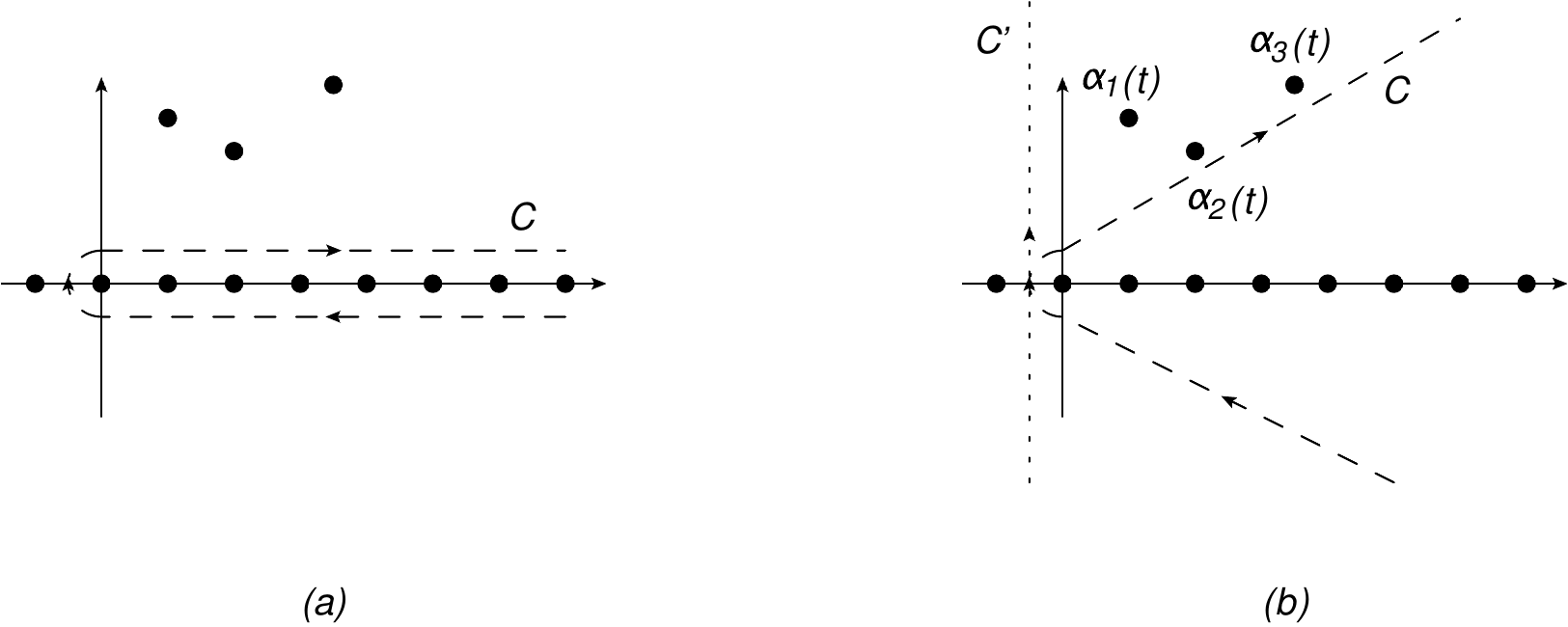}}
\caption{(a) Pole structure of the integrand in
  eq.~(\ref{Sommerfeld}), where poles in the real axis arise from
  zeroes of the $\sin$ function; (b) Deformation of the original
  integration contour $C$ (enclosing the real axis) to make a vertical
  contour $C'$, which is closed at infinity.}
\label{fig:Sommerfeld}
\end{center}
\end{figure}

Quite obviously, the angular momentum $l$ in eq.~(\ref{partialwave2})
must be a non-negative integer. Regge's idea was instead to let this
be a complex variable. This is not such a daft idea, given that we
have already let Lorentz invariants (formed from linear momenta)
become complex. How, though, do we achieve a complex $l$? The first
step is to replace the partial wave amplitudes in
eq.~(\ref{partialwave2}) as follows:
\begin{displaymath}
\{a_l(t)\}\rightarrow a(l,t),
\end{displaymath}
where on the right-hand side we have a complex function $a(l,t)$ of
$l,t\in{\mathbb C}$, whose only requirement is that it reproduces the
partial wave amplitudes if $l$ is a non-negative integer:
\begin{displaymath}
a(l,t)\rightarrow a_l(t),\quad l\in\{0,1,2,\ldots\}.
\end{displaymath}
Unfortunately, this does not look unique: na\"{i}vely, it seems we
have the freedom to replace
\begin{displaymath}
a(l,t)\rightarrow a(l,t)+f(l,t),
\end{displaymath}
where
\begin{displaymath}
f(l,t)=0,\quad l\in\{0,1,2,\ldots\}.
\end{displaymath}
If the procedure is not unique, how then can we be confident of
getting a definite answer for the $s$-channel scattering amplitude,
which certainly ought to be unique? The resolution lies in something
called {\it Carlson's theorem}, which says that {\it if two different
  analytic functions do not grow too fast at infinity, they cannot
  coincide at non-negative integers.} In other words, $a(l,t)$ is {\it
  unique}, provided it does not grow too fast. The precise criterion
that Carlson requires is that
\begin{displaymath}
a(l,t)< C e^{\pi|l|},\quad |l|\rightarrow\infty,
\end{displaymath}
where $C$ is an arbitrary constant. Unfortunately, this is not quite
satisfied! One can show that the partial wave amplitudes $\{a_l(t)\}$
have contributions that alternate in sign, so that
\begin{equation}
a_l(t)\sim (-1)^l=e^{-i\pi l},
\label{altsign}
\end{equation}
which clearly violates the requirements of Carlson's theorem on the
imaginary axis. We can rescue things by separating the odd and even
partial waves: each of these has contributions associated with the
{\it same} sign, which will not then behave like eq.~(\ref{altsign}),
and thus we can use Carlson's theorem to guarantee uniqueness of our
complexified odd / even partial wave expansions. The conventional way
to do this is to use the property of Legendre polynomials
\begin{equation}
P_l(-z)=(-1)^l P_l(z)
\label{Legendrereflect}
\end{equation}
to define 
\begin{equation}
{\cal A}^\pm (s,t)=8\pi \sum_{l=0}^\infty (2l+1)a^\pm_l(t)
[P_l(z_t)\pm P_l(-z_t)].
\label{Aplusminus}
\end{equation}
This defines the even (odd) amplitude ${\cal A}^+$ (${\cal A}^-$),
where the factor in the square brackets will pick out even (odd)
partial waves only for the upper (lower) sign. Furthermore, we have
defined the even and odd partial wave amplitudes $a^\pm(t)$, such that
\begin{displaymath}
a_l(t)=\begin{cases}a_l^+(t),\quad l\,\,\rm{even};\\
a_l^-(t),\quad l\,\,\rm{odd}.
\end{cases}
\end{displaymath}
We do not have to worry about defining $a_l^+(t)$ for $l$ odd (or
$a_l^-(t)$ for $l$ even), given that such terms will be killed off by
the combination of Legendre polynomials in eq.~(\ref{Aplusminus}).
The full amplitude is then given by
\begin{equation}
{\cal A}(s,t)={\cal A}^+(s,t)+{\cal A}^-(s,t).
\label{Asum}
\end{equation}
We now have two different partial wave expansions, involving two sets
of partial wave amplitudes $\{a_l^+(t)\}$ and $\{a_l^-(t)\}$, each of
which separately satisfies the requirements of Carlson's theorem. We
can thus define two {\it unique} functions $a^+(l,t)$ and $a^-(l,t)$
in the complex $l$ plane, such that $a^+(l,t)$ reduces to $a^+_l(t)$
for even $l$, and $a^-(l,t)$ to $a^-_l(t)$ for odd $l$. Next, we can
rewrite the expansions of eq.~(\ref{Aplusminus}) as complex
integrals~\footnote{In eq.~(\ref{Sommerfeld}), the Legendre
  polynomials are understood to be replaced by their analytic
  continuations for complex $l$, for which a known form in terms of
  hypergeometric functions exists.}:
\begin{equation}
{\cal A}^\pm(s,t)=-4\pi i\int_C dl (2l+1)a^\pm(l,t)
\frac{[P_l(-z_t)\pm P_l(z_t)]}{\sin(\pi l)},
\label{Sommerfeld}
\end{equation}
where $C$ is a contour that completely encloses the positive real
axis, as shown in figure~\ref{fig:Sommerfeld}(a). 
To justify this formula, note that the function $\sin(\pi l)$ in the
denominator has zeroes at all integer values of $l$. Thus, there is a
series of equally spaced poles of the integrand on the real $l$ axis,
at all values $l\in{\mathbb Z}$. Each of these is in fact a simple
pole, given that one has
\begin{displaymath}
\sin(\pi l)\rightarrow (-1)^n (l-n)\pi+{\cal O}[(l-n)^2],\quad 
n\in\mathbb{Z}. 
\end{displaymath}
This also allows to find the residue of each pole, using the general
result for simple poles
\begin{displaymath}
{\rm Res}_{z=z_0} f(z)=\lim_{z\rightarrow z_0} (z-z_0) f(z).
\end{displaymath}
Denoting the integrand of eq.~(\ref{Sommerfeld}) by $f(l)$, we then
have
\begin{align}
{\rm Res}_{l\rightarrow n\in{\mathbb Z}} f(l)
&=\lim_{l\rightarrow n}\frac{(-4\pi i)}{(-1)^n}
\frac{(l-n)}{(l-n)\pi}(2l+1) a^\pm (l,t)[P_l(-z_t)\pm P_l(z_t)]\notag\\
&=-\frac{4\pi i}{(-1)^n\pi}(2n+1)a^\pm (n,t)[P_n(-z_t)\pm P_n(z_t)].
\label{residue}
\end{align}
We can then carry out the complex integral of eq.~(\ref{Sommerfeld})
using {\it Cauchy's theorem}
\begin{equation}
\oint_C f(z) =2\pi i\sum_{\rm i} {\rm Res}_{z=z_i} f(z),
\label{Cauchy}
\end{equation}
where $\{z_i\}$ denotes the locations of all poles. In our present
case this gives
\begin{align}
{\cal A}^\pm(s,t)&=2\pi i\sum_{n=0}^\infty {\rm Res}_{l=n}f(l)\notag\\
&=8\pi \sum_{n=0}^\infty (2n+1)a^\pm_n(t)[P_n(z_t)\pm P_n(-z_t)],
\label{Aint}
\end{align}
where we have used eq.~(\ref{Legendrereflect}), and the fact that
$a^\pm(l,t)$ is the same as $a^\pm_l(t)$ for even or odd $l$
respectively. We have thus indeed shown that the complex integral
representation of eq.~(\ref{Sommerfeld}) reproduces the $t$-channel
even and odd partial wave expansions of
eq.~(\ref{Aplusminus}). Equation~(\ref{Sommerfeld}) has a name in the
literature: it is known as the {\it Sommerfeld-Watson transform}.

\subsection{Regge poles}
\label{sec:Reggepoles}

So far, we have made the claim that high energy behaviour in the
$s$-channel should be dictated by particle exchange in the
$t$-channel. This led us to write the amplitude in the $s$-channel as
a $t$-channel partial wave expansion, which however did not
converge. Thus, we then rewrote the expansion in terms of two highly
complicated complex integrals in the angular momentum plane, where our
ultimate aim is to find something that indeed converges in the
physical $s$-channel region. It is, of course, not immediately clear
why the rewriting inherent in the Sommerfeld-Watson transform is
useful. However, the power of having a complex integral representation
(eq.~(\ref{Sommerfeld})) instead of a discrete sum
(eq.~(\ref{Aplusminus})) is that we can deform the contour of
integration to get an $s$-channel representation of the scattering
amplitude that has better convergence properties. Put another way,
deforming the contour of our Sommerfeld-Watson transform amounts to
carefully analytically continuing the amplitude from the physical
$t$-channel region to the $s$-channel region, so that we can safely
study its properties in the Regge limit of eq.~(\ref{Reggelim}).

The original contour $C$ in figure~\ref{fig:Sommerfeld}(a) encloses
the real axis (n.b. it is assumed to be closed at positive infinity on
the real $l$ axis). Let us deform it, as in
figure~\ref{fig:Sommerfeld}(b), to a vertical contour $C'$ at
$Re(l)=-1/2$, which is then closed to the right at infinity. In doing
this, we will end up moving the contour past any other singularities
that might be present in the complex plane, and that are not
associated with the simple poles along the real axis from the
denominator in eq.~(\ref{Sommerfeld}). Instead, they would be
associated with the partial wave functions $a^\pm(l,t)$ sitting in
eq.~(\ref{Sommerfeld}).  For now, we will assume that only simple
poles are present. Given that they are contained in the partial wave
functions, we must include the possibility that the location of these
additional poles in the complex angular momentum plane can depend on
$t$. We will thus label the positions of the poles by
$\{\alpha_i(t)\}$, as exemplified by figure~\ref{fig:Sommerfeld}(b).

By the usual tricks of complex analysis, moving the contour $C$ past a
simple pole means that we pick up its residue. In total, then,
deforming the contour $C\rightarrow C'$ means that we pick up the
residues of all poles $\{\alpha_i(t)\}$ lying between $C$ and
$C'$. Assuming the amplitude vanishes at infinity (so that there is no
contribution from the contour integral there) eq.~(\ref{Sommerfeld})
becomes, after the contour deformation, 
\begin{align}
{\cal A}^\pm(s,t)&=8\pi^2\sum_i\frac{(2\alpha^\pm_i(t)+1)\beta_i^\pm(t)}
{\sin[\pi\alpha_i^\pm(t)]}\left[P_{\alpha^\pm_i}(-z_t)\pm P_{\alpha_i^\pm}
(z_t)\right]\notag\\
&\quad-4\pi i\int_{-1/2-i\infty}^{1/2+i\infty}dl
(2l+1)a^\pm(l,t)\frac{[P_l(-z_t)\pm P_l(z_t)]}{\sin(\pi l)}.
\label{Sommerfeld2}
\end{align}
Here the first line arises from collecting the residues of all the
poles $\{\alpha_i(t)\}$, where $\beta_i^\pm (t)$ is a residue
factor. We have no way of knowing what the latter is without knowing
the partial wave amplitudes (which we are claiming to know nothing
about). However, the residue factor can depend only on $t$, which
still amounts to non-zero information. The second line in
eq.~(\ref{Sommerfeld2}) corresponds to the remaining complex integral
over the contour $C'$. Whilst it may not seem like it at first glance,
we have indeed finally put the $s$-channel amplitude in a form that
allows us to safely examine the Regge limit. Furthermore, it
simplifies considerably!

Let us first consider the first line of eq.~(\ref{Sommerfeld2}),
namely the pole contributions. Recalling eq.~(\ref{costhetat}), we
find in the Regge limit of eq.~(\ref{Reggelim})
that~\footnote{Equation~(\ref{ztlim}) would be different had we
  defined the Regge limit such that the momentum transfer was much
  less than the particle masses.}
\begin{equation}
z_t\rightarrow \frac{2s}{t}.
\label{ztlim}
\end{equation}
Given $t<0$ in the physical $s$-channel region, we see that $z_t$ is
large and negative. We may utilise the known relation for Legendre
polynomials~\footnote{Equation~(\ref{Legendrereflect2}) differs from
  eq.~(\ref{Legendrereflect}) in that it applies for arbitrary complex
  $\alpha$.}
\begin{equation}
P_\alpha(z_t)=e^{-i\pi\alpha} P_\alpha(-z_t)
\label{Legendrereflect2}
\end{equation}
to write
\begin{equation}
P_\alpha(-z_t)\pm P_\alpha(z_t)=[1\pm e^{-i\pi\alpha}]
P_\alpha(-z_t).
\label{Psum}
\end{equation}
Then, using the asymptotic behaviour
\begin{equation}
P_\alpha(x)\xrightarrow{x\gg 1} \frac{\Gamma(2\alpha+1)}
{\Gamma^2(1+\alpha)}\left(\frac{x}{2}\right)^\alpha,\quad
{\rm Re}(\alpha)\geq \frac12,
\label{Pasymp}
\end{equation}
where $\Gamma(z)$ is the Euler gamma function, the contribution in the
first line of eq.~(\ref{Sommerfeld2}) becomes
\begin{equation}
8\pi^2\sum_i\frac{(2\alpha_i^\pm+1)\beta_i^\pm(t)}
{\sin[\pi\alpha_i^\pm(t)]}\left(
1\pm e^{-i\pi\alpha^\pm_i(t)}\right)\frac{\Gamma(1+2\alpha_i^\pm)}
{\Gamma^2(1+\alpha_i^\pm)}\left(\frac{s}{-t}\right)^{\alpha_i^\pm}.
\label{poles1}
\end{equation}
We may tidy this up further using {\it Euler's reflection
  formula}~\footnote{Equation~(\ref{Euler}) is usually written with
  $z\rightarrow -z$.}
\begin{equation}
\Gamma(1+z)\Gamma(-z)=-\frac{\pi}{\sin(\pi z)},
\label{Euler}
\end{equation}
so that eq.~(\ref{poles1}) becomes
\begin{equation}
\sum_i \tilde{\beta}^\pm_i(t)\Gamma[-\alpha_i^\pm(t)]
\left(1\pm e^{-i\pi\alpha_i^\pm(t)}\right)
\left(\frac{s}{-t}\right)^{\alpha_i^\pm}, 
\label{poles2}
\end{equation}
where we have absorbed some factors in the residue factor (hence the
tilde). We must also worry about the remaining complex integral in the
second line of eq.~(\ref{Sommerfeld2}), which is known as the {\it
  background integral} in old literature on this subject. However,
Mandelstam has shown~\cite{MANDELSTAM1962254} that this can be made
arbitrarily small, at the expense of including extra poles
$\alpha_i^\pm(t)$ in the above sum. Thus, the form of the result is
unchanged, and our complete result for the $s$-channel physical
amplitude in the Regge limit is
\begin{equation}
{\cal A}^\pm(s,t)=\sum_i \tilde{\beta}_i^\pm(t)\Gamma[-\alpha_i(t)]
\left(1\pm e^{-i\pi \alpha_i^\pm(t)}\right)
\left(\frac{s}{-t}\right)^{\alpha_i^\pm(t)}. 
\label{highenergyamp}
\end{equation}
We have obtained a remarkable result. Based purely on well-motivated
assumptions for how the S-matrix should behave, we have found that the
behaviour of the $2\rightarrow 2$ scattering amplitude in the high
energy (Regge) limit of eq.~(\ref{Reggelim}) consists of a series of
power-like terms in $s$, with $t$-dependent prefactors. Each power
corresponds to a pole in the complex angular momentum plane, that
moves about as a function of $t$. In fact, we can simplify this even
further: if all we care about is the {\it leading} behaviour as
$s\rightarrow\infty$, we can keep only the term in
eq.~(\ref{highenergyamp}) whose power has the largest real part. In
other words, instead of summing over all poles in
eq.~(\ref{highenergyamp}), we can keep only the right-most pole
$\alpha(t)$ in the complex $l$ plane. This might occur in either the
even or the odd amplitude, in which case it will dominate the result
for the total amplitude of eq.~(\ref{Asum}). Let us introduce a factor
$\eta=\pm1$, called the {\it signature} of the right-most pole, where
the upper (lower) sign corresponds to its being in the even (odd)
amplitude. Then the total amplitude will be given by
\begin{equation}
{\cal A}(s,t)\xrightarrow{s\gg -t}\tilde{\beta}(t)
\Gamma[-\alpha(t)]\left(1+\eta e^{-i\pi\alpha}\right)
\left(\frac{s}{-t}\right)^{\alpha(t)}.
\label{ARegge2}
\end{equation}
You might be wondering, given the incredibly convoluted series of
mathematical steps that it took to get this result, whether there is
any hope of understanding the physics of what is going on. Indeed
there is! We saw already above that power-like growth with $s$ in the
$s$-channel can be identified with the exchange of single particle
states in the $t$-channel. We saw this for the case of a {\it single}
particle being exchanged, but in fact eq.~(\ref{ARegge2}) corresponds
to infinitely many states being exchanged in the $t$-channel. To see
why, note that the Euler gamma function has poles at all non-positive
integers $z\in\{0,-1,-2,-3,\ldots\}$. Thus, the factor of
$\Gamma[-\alpha(t)]$ in eq.~(\ref{ARegge2}) implies that the amplitude
${\cal A}(s,t)$ has poles in $t$, whenever $t$ is such that
$\alpha(t)$ is a physical angular momentum. But, as we derived in
section~\ref{sec:Smatrix}, poles in any Lorentz invariant represent
the exchange of bound states!  Thus there is an infinite family of
bound states that can be exchanged in the $t$-channel, and it is the
exchange of this entire family that gives the power-like growth in $s$
in the Regge limit.

We can really think of the entire family of bound states as being
contained in the function $\alpha(t)$: this parametrises a curve in
the complex angular momentum plane, such that whenever the curve
crosses a physical value of ${\rm Re}(l)$
(i.e. $l\in\{0,1,2\ldots\}$), the corresponding value of $t$ is the
squared mass of a state of spin $l$. Returning to the case where we
consider many different simple poles, each function $\alpha_i(t)$ is
called a {\it Regge trajectory}. A commonly-performed exercise in the
1960s was to figure out which known mesons (bound states) lay on the
same Regge trajectory, and to use the measured values of their spin
and masses to construct the function $\alpha(t)$ for that particular
trajectory. For mesons, the functions were found to be linear to a
very good approximation, which happens to be exactly the relation one
would expect if mesons were described by rotating strings. This is how
string theory was born: see chapter 1 of ref.~\cite{Green:1987sp} for
a nice historical overview.

Above, we only admitted the possibility of single poles in the complex
angular momentum plane. The arguments are relatively straightforward
to repeat, however, in the presence of other types of singularity. It
can be shown~\cite{Eden:1966dnq,Collins:1977jy}, for example, that
higher-order poles dress the above behaviour according to
\begin{equation}
{\cal A}(s,t)\sim \log^n (s) s^{\alpha(t)},
\label{higherorderpole}
\end{equation}
for some integer $n$, and that cuts lead to the behaviour
\begin{equation}
{\cal A}(s,t)\sim s^{\alpha_c(t)}(\log s)^{-\gamma(t)},
\label{cuts}
\end{equation}
where $\alpha_c(t)$ is the location of the branch point on the complex
$l$ plane, and $\gamma(t)$ is related to the discontinuity across the
cut. 

In this section, we have shown that generic behaviour arises in the
Regge limit of scattering amplitudes, namely power-like growth in the
centre of mass energy (up to possible logarithmic corrections). This
behaviour is traced to the exchange of particle states in the
$t$-channel, and thus will only occur if such states exist in the
theory. Note that, in the absence of an underlying theory, we cannot
say what the singularities in the complex angular momentum plane will
be. One can then ``guess'' a possible structure of singularities,
before fitting the results to data. This approach is still used today
for fitting certain observables in strong interaction physics which
lack a perturbative description e.g. total cross-section data at the
LHC~\cite{Donnachie:1998gm,Donnachie:2013xia}.

Given the above behaviour is completely generic, it must also be the
case that any particular underlying theory - such as a perturbative
quantum field theory - must reproduce the power-like growth in $s$ in
the high energy limit. In the following section, we will show examples
of this in both QCD and gravity, and examine the relationship between
them.

\section{Examples in QCD and gravity}
\label{sec:examples}

In this section, we will examine the high energy behaviour of
scattering amplitudes in two widely studied theories, namely QCD
(relevant for collider physics) and gravity (relevant for astrophysics
and cosmology). Before doing so, we should of course apologetically
point out that we don't necessarily expect the ideas of the previous
section to be at all relevant: both the gluon and graviton are
massless, and yet we have assumed thus far that all particle states in
our theory of interest are massive~\footnote{As noted above, QCD does
  actually have a mass gap, but this affects larger distances, where
  the strong interaction will be mediated by mesons rather than the
  gluon.}. This does not pose a problem for the exercise we are about
to carry out, which is simply to quote the known form of amplitudes in
the Regge limit, and examine their structure. However, we should keep
in mind that when we see results that correlate with the generic Regge
behaviour described above (as indeed we will), we should perhaps be
surprised!

\subsection{High energy scattering in QCD}
\label{sec:QCD}

\begin{figure}
\begin{center}
\scalebox{0.6}{\includegraphics{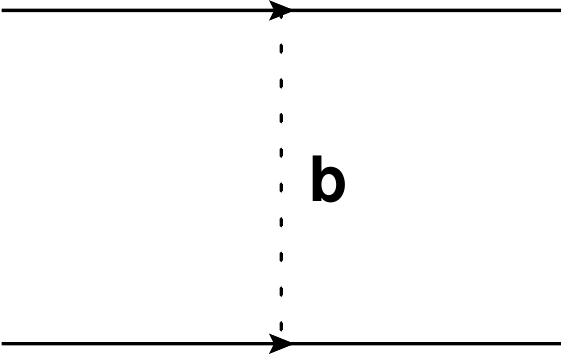}}
\caption{In the Regge limit of $2\rightarrow 2$ scattering, the
  incoming particles barely glance off each other, thus follow
  straight-line trajectories separated by an {\it impact parameter}
  $\vec{b}$.}
\label{fig:impact}
\end{center}
\end{figure}

\begin{figure}
\begin{center}
\scalebox{0.6}{\includegraphics{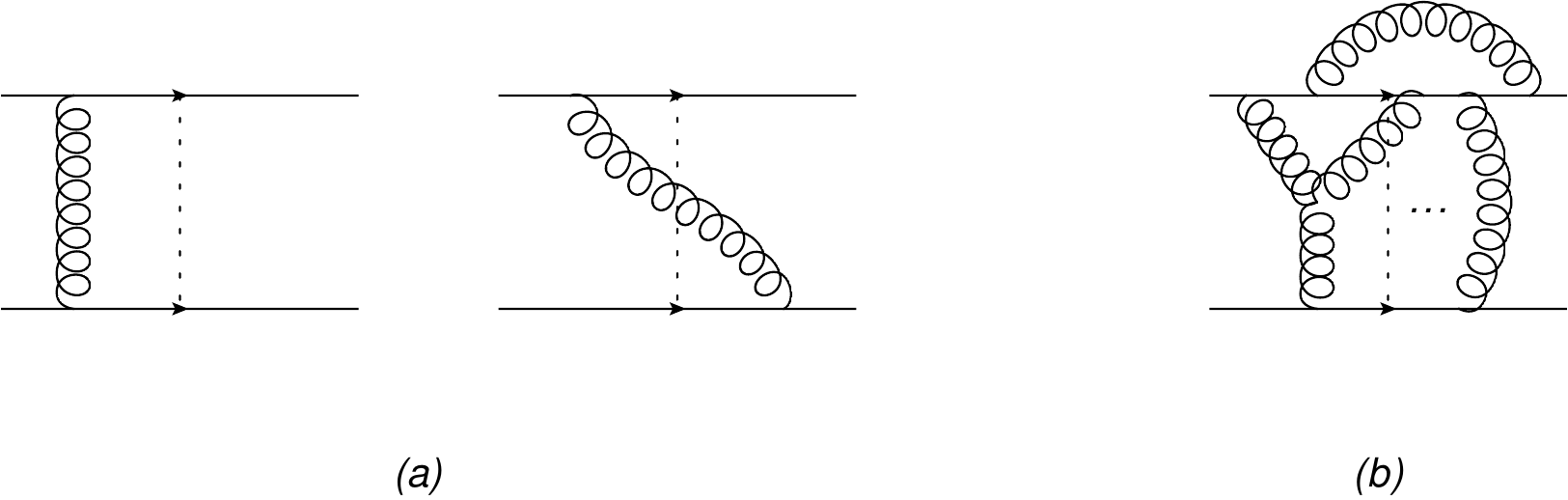}}
\caption{Example virtual gluon radiation at (a) 1-loop order; (b)
  arbitrary loop order.}
\label{fig:gluons}
\end{center}
\end{figure}

Let us consider $2\rightarrow2$ scattering of scalar particles that
carry a colour charge in a non-abelian gauge theory, where we label
momenta according to figure~\ref{fig:2to2}(a). If we consider the
Regge limit of eq.~(\ref{Reggelim}), the momentum transfer between
particles 1 and 2 is much smaller than the centre of mass energy. This
has a simple physical interpretation, in that it corresponds to highly
forward scattering, such that the incoming and outgoing particles
suffer a tiny deflection. Each incoming particle does not change
species, but instead follows a straight-line trajectory such that the
incoming particles (1,2) become collinear with the outgoing particles
(3,4) respectively~\footnote{This is not the usual {\it timelike}
  collinear limit associated with two final state particles becoming
  mutually collinear. The Regge limit is instead a {\it spacelike}
  collinear limit.}. We show this situation in
figure~\ref{fig:impact}, where we note that the two particle
trajectories are separated by a transverse distance $\vec{b}$ in
general. This is referred to as the {\it impact parameter}, and would
correspond to the distance of closest approach if we were not in the
Regge limit.
If the (scalar) particles of figure~\ref{fig:impact} carry colour
charge, they can emit gluon radiation, so that we must dress the
diagram with all possible virtual gluon emissions, as shown in
figure~\ref{fig:gluons}. 
At each order in the strong coupling $g_s$, we can then ask what the
dominant terms are in the Regge limit. Remarkably, this can be
successfully answered at {\it all} orders in perturbation theory!
Furthermore, the leading terms can be summed up exactly to give a
complete function of the
coupling~\cite{Korchemskaya:1994qp,Melville:2013qca} (for related work
on this subject, see refs.~\cite{Mandelstam:1965zz,Abers:1967zz,
  McCoy:1976ff,Frolov:1970ij,Grisaru:1973vw,Grisaru:1973ku,Grisaru:1974cf,
  Gribov:1970ik,Cheng:1969bf,Balitsky:1979ap,Bogdan:2006af,Tyburski:1975mr,
  Lipatov:1976zz,Mason:1976fr,Cheng:1977gt,Fadin:1975cb,
  Kuraev:1977fs,Kuraev:1976ge,Mason:1976ky,
  Sen:1982xv,Fadin:1977jr,Fadin:1995xg,
  Fadin:1996tb,Fadin:1995km,Blumlein:1998ib,DelDuca:2001gu,Bogdan:2002sr,
  Fadin:2006bj,Caron-Huot:2013fea,DelDuca:2013ara,DelDuca:2013dsa,DelDuca:2014cya,Caron-Huot:2017fxr,Caron-Huot:2017zfo,Vernazza:2018gyb}):
\begin{equation}
{\cal A}(s,t)\xrightarrow{s\gg -t}\exp\left\{K\left[i\pi {\bf T}_s^2
+\log\left(\frac{s}{-t}\right){\bf T}_t^2\right]+\ldots\right\}
{\cal A}_{\rm LO}(s,t),
\label{Aexact}
\end{equation}
where $K$ is a coupling-dependent parameter that we will define
shortly, and ${\cal A}_{\rm LO}$ the leading order amplitude for the
$2\rightarrow 2$ scattering to take place. The quantities ${\bf
  T}_s^2$ and ${\bf T}_t^2$ are operators that act on the colour
charge information carried by the LO amplitude. A full definition of
them is given in e.g. refs.~\cite{Bret:2011xm,DelDuca:2011ae}, but we
can try to explain the main idea in more qualitative terms
here. Imagine that the LO amplitude is dominated by the exchange of a
coloured particle (e.g. a gluon) in the $s$-channel. Such amplitudes
are eigenstates of the operator ${\bf T}_s^2$, and the eigenvalue is
just the squared colour charge of the exchanged particle (more
formally, the {\it quadratic Casimir invariant} in the appropriate
representation). We can draw this schematically as
\begin{equation}
{\bf T}_s^2\left[\quad 
\raisebox{-14pt}{\scalebox{0.35}{\includegraphics{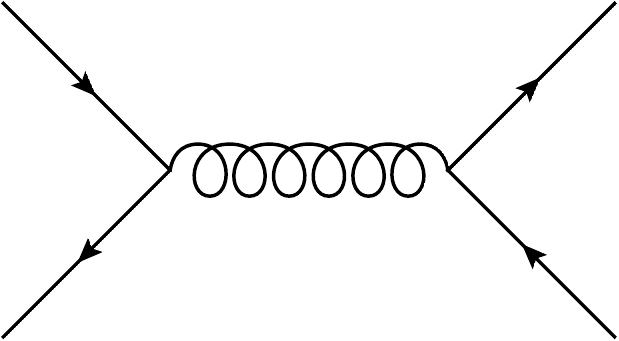}}}
\quad\right]
=C_A \quad \raisebox{-14pt}{\scalebox{0.35}{\includegraphics{Ts.pdf}}}\quad,
\label{Ts}
\end{equation}
where $C_A$ is indeed the appropriate quadratic Casimir for the gluon.
Likewise, eigenstates of the operator ${\bf T}_t^2$ are definite
exchanges in the $t$-channel, so that we may write
\begin{equation}
{\bf T}_t^2 \left[\quad
\raisebox{-9pt}{\scalebox{0.35}{\includegraphics{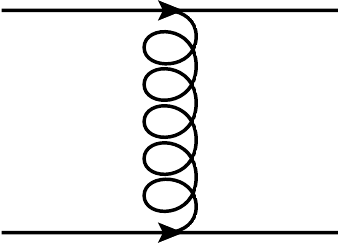}}}
\quad\right]=C_A \quad
\raisebox{-9pt}{\scalebox{0.35}{\includegraphics{Tt.pdf}}}
\label{Tt}
\end{equation}
Imagine that the LO amplitude contains the $t$-channel exchange of a
particle $X$, of spin $J$. We already argued in
section~\ref{sec:Regge} that this will dominate the physical
$s$-channel amplitude in the Regge limit, so that we have
\begin{equation}
{\cal A}_{\rm LO}(s,t)\sim 
\raisebox{-12pt}{\scalebox{0.35}{\includegraphics{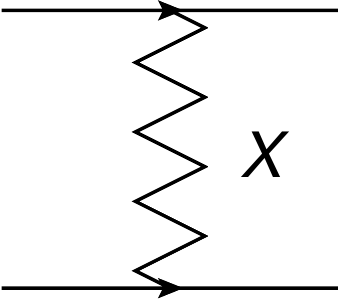}}}
\sim \left(\frac{s}{-t}\right)^J.
\label{Xexchange}
\end{equation}
Thus, in the Regge limit, the LO amplitude is indeed an eigenstate of
the colour operator ${\bf T}_t^2$. Furthermore, the exponent in
eq.~(\ref{Aexact}) becomes dominated by the second term, which is
logarithmically enhanced for $s\gg|t|$. From eqs.~(\ref{Aexact},
\ref{Tt}) we then have
\begin{align}
{\cal A}(s,t)&\rightarrow\exp\left\{K\log\left(\frac{s}{-t}\right){\bf T}_t^2
\right\}{\cal A}_{\rm LO}\notag\\
&=\exp\left\{K\log\left(\frac{s}{-t}\right)C_X
\right\}{\cal A}_{\rm LO}\notag\\
&=\left(\frac{s}{-t}\right)^{KC_X+J},
\label{Aexact2}
\end{align}
where $C_X$ is the squared colour charge of $X$. The overall effect of
all the virtual radiation is as if the propagator for the $X$ particle
has been dressed by an overall power of $s/|t|$. For example, if $X$
is a gluon, we could reproduce the result of eq.~(\ref{Aexact2}) by
modifying the gluon propagator (e.g. in the Feynman gauge) as follows:
\begin{equation}
-\frac{\eta_{\mu\nu}}{q^2}\rightarrow -\frac{\eta_{\mu\nu}}{q^2}
\left(\frac{s}{-t}\right)^{KC_A},
\label{Reggeisation}
\end{equation}
where $q$ is the exchanged momentum. This is called {\it
  Reggeisation}, and the quantity $KC_A$ is known as the {\it Regge
  trajectory} of the gluon. Indeed, we have precisely reproduced the
expectations of Regge theory, namely that the physical $s$-channel
amplitude in the Regge limit should contain a power-like growth in
$s$, where this is associated with particle exchange in the
$t$-channel! We could go further than this and show that the Reggeised
gluon indeed corresponds to a pole in the complex angular momentum
plane, but will not do so. 

\begin{figure}
\begin{center}
\scalebox{0.6}{\includegraphics{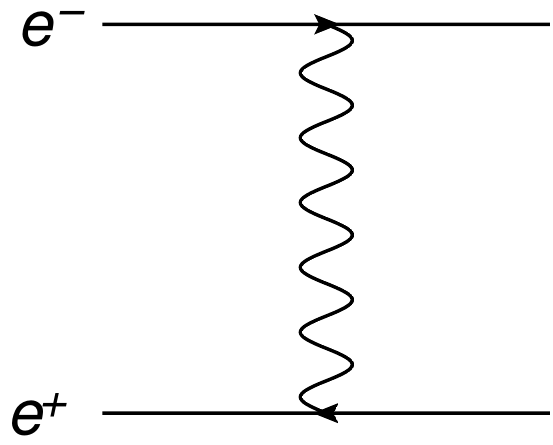}}
\caption{Electron-positron scattering in the high energy limit.}
\label{fig:QED}
\end{center}
\end{figure}

Note that there are gauge theories, and scattering examples within
those theories, such that the squared charge of $t$-channel particles
is zero. Consider, for example, the QED scattering process of
figure~\ref{fig:QED}, consisting of electron-positron scattering via
photon exchange~\footnote{In figure~\ref{fig:QED}, we can ignore the
  $s$-channel diagram, which will be kinematically subleading in the
  Regge limit, at least if we use the Feynman gauge.}. In this case,
the squared charge of the exchanged particle (the photon) is zero, and
one can show that the correct abelian form of eq.~(\ref{Aexact}) is
\begin{equation}
{\cal A}(s,t)\rightarrow e^{i\pi K_{\rm QED}}
{\cal A}_{\rm LO}(s,t),
\label{QED}
\end{equation}
for some parameter $K_{\rm QED}$ that depends upon the electron
charge. Thus, the LO amplitude gets dressed by an overall phase, which
is known as the {\it eikonal phase} in the literature. It is there
also in QCD (i.e. as the first term in the exponent of
eq.~(\ref{Aexact})), but in that case is kinematically subleading, in
that the Reggeisation term dominates due to the logarithmic
enhancement in energy.
What is the value of $K$ in eq.~(\ref{Aexact})? In fact, it turns out
to be infinite! To see what has gone wrong, recall that scattering
amplitudes contain {\it infrared (IR) singularities} associated with
the emission of virtual radiation which is ``soft'' (i.e. whose
momentum goes to zero). The Regge limit says that the total momentum
exchange between the scattering particles must be negligible compared
to their energy, which is precisely the statement that the exchanged
virtual gauge bosons be soft! We can make $K$ finite by introducing a
regulator, and ref.~\cite{Melville:2013qca} used dimensional
regularisation in $d=4-2\epsilon$ dimensions to obtain
\begin{equation}
K=\frac{g_s^2\Gamma(1-\epsilon)}{4\pi^{2-\epsilon}}
\frac{(\mu^2\vec{b}^2)^\epsilon}{2\epsilon},
\label{K}
\end{equation}
where $\mu$ is the dimensional regularisation scale. This depends on
the strong coupling $g_s$, as stated above, and also on the impact
parameter $\vec{b}$, which combines with $\mu$ to make a dimensionless
combination. Indeed we see a singularity as $\epsilon\rightarrow 0$,
corresponding to the IR singularity alluded to above. 

Infrared singularities are ultimately a consequence of the fact that
interactions involving the exchange of massless particles (e.g. the
photon or gluon) are not short-range~\footnote{Strong interactions
  {\it are} short-range, but this is due to the confinement of the
  gluon within mesons, which is a non-perturbative effect.}. Thus, it
is not correct to assume that the physical incoming and outgoing
states are single particles: they should be dressed with a virtual
cloud of gauge bosons. It is possible to set up QFT in this manner in
both gauge theories~\cite{Kulish1970,CATANI1986588} and
gravity~\cite{Ware:2013zja}, although the procedure is rather
cumbersome. Furthermore, in practice one can proceed without this
complication. Above, we included only {\it virtual} radiation. If we
included real emissions as well, all IR singularities would
cancel~\footnote{In QCD processes involving incoming quarks / gluons,
  we would also have to include {\it parton distribution functions}
  describing how these are emitted from the incoming hadrons in our
  experiment. These can then be used to absorb initial state collinear
  singularities.}. We would then find IR-finite Regge pole effects in
the total cross-section. The leading Regge pole with the same quantum
numbers as the vacuum is called the {\it Pomeron}, and has been widely
studied (see e.g. refs.~\cite{Forshaw:1997dc,DelDuca:1995hf} for a
review). We started this section by saying that we should perhaps be
surprised if the Regge theory analysis of section~\ref{sec:Regge}
(which assumed a mass gap in the theory) turned out to be relevant for
theories containing massless particles. We have now seen that it is in
fact relevant, provided one can accept the presence of IR
singularities. How to interpret the S-matrix in arbitrary massless
theories remains a subject of ongoing research.

Interestingly, in the QED case of eq.~(\ref{QED}), we can Fourier
transform the amplitude from impact parameter space to momentum space
exactly, and obtain the result~\cite{Melville:2013qca}
\begin{equation}
\tilde{A}\sim \left(-\frac{t}{\mu^2}\right)^{-i\alpha}
\frac{\Gamma(1+i\alpha)}{\Gamma(1-i\alpha)},\quad
\alpha=\frac{e^2}{4\pi}\frac{s-2m^2}{\sqrt{s(s-4m^2)}},
\label{Atilde}
\end{equation}
where $e$ is the electron charge. The gamma function in the numerator
has an infinite series of poles, located at
\begin{equation}
s=2m^2\left[1-\left(1+\frac{e^4}{16\pi^2 N^2}\right)^{-1/2}\right],
\quad N\in{\mathbb Z}^+.
\label{spoles}
\end{equation}
In section~\ref{sec:Smatrix}, we learnt that poles in the complex $s$
plane should represent bound states, so what are they? The answer is
that we have derived (at least in some perturbative approximation) the
spectrum of {\it positronium}, and the above result can be expanded
about a non-relativistic limit to obtain known results for the energy
levels~\cite{Brezin:1970zr}~\footnote{Equation~(\ref{spoles}) has been
  derived for spinless particles, rather than spin-1/2 electrons and
  positrons. However, one may show that spin effects are subleading in
  the Regge limit.}. 

\subsection{High energy scattering in gravity}
\label{sec:gravity}

We can repeat the QCD-like calculation of the previous section in
General Relativity (GR), and the answer for the $2\rightarrow 2$
scattering amplitude in the leading Regge limit, dressed with
arbitrary amounts of virtual graviton radiation, turns out to
be~\cite{Melville:2013qca}
\begin{equation}
{\cal M}(s,\vec{b})\rightarrow\exp\left\{K_g\left[i\pi s
+t\log\left(\frac{s}{-t}\right)\right]\right\}{\cal M}_{\rm LO},
\label{MRegge}
\end{equation}
where ${\cal M}_{\rm LO}$ is the leading-order gravitational
amplitude, and
\begin{equation}
K_g=\left(\frac{\kappa}{2}\right)^2\frac{\Gamma(1-\epsilon)}
{4\pi^{2-\epsilon}}\frac{(\mu^2\vec{b}^2)^\epsilon}{2\epsilon}.
\label{Kg}
\end{equation}
Here $\kappa^2=32\pi G_N$, with $G_N$ the Newton constant.  Note that
eq.~(\ref{MRegge}) is identical in form to the QCD result of
eq.~(\ref{Aexact}), and can indeed can be obtained from the latter
with the replacements
\begin{equation}
g_s\rightarrow\frac{\kappa}{2},\quad {\bf T}_s^2\rightarrow s,\quad
{\bf T}_t^2\rightarrow t.
\label{replace}
\end{equation}
These replacements in fact make perfect sense. The first tells us that
the strength of the strong force must be replaced by the appropriate
strength of the force in gravity. Furthermore, the operators
representing the squared colour charge exchanged in the $s-$ and
$t-$channels are replaced with the squared 4-momentum exchanged in
these channels. However, in GR the ``charge'' of a particle (i.e. the
thing that makes it couple to the graviton) is its 4-momentum, so that
$s$ and $t$ are indeed the squared charges! You may also recognise the
replacements of eq.~(\ref{replace}) as being consistent with the BCJ
double copy of refs.~\cite{Bern:2008qj,Bern:2010ue,Bern:2010yg}. Its
appearance here can be understood from the above-mentioned fact that
the Regge limit involves the exchange of soft gravitons, and the
leading soft structure of QCD and gravity amplitudes is known to
double copy~\cite{Oxburgh:2012zr,Vera:2014tda} (see
refs.~\cite{Johansson:2013nsa,Vera:2012ds,Saotome:2012vy} for other
studies of the Regge limit in a double copy context).

In eq.~(\ref{MRegge}), we again see the presence of an eikonal phase
term~\footnote{See refs.~\cite{Amati:1987wq,
    Muzinich:1987in,Kabat:1992tb} for earlier work on the
  gravitational eikonal phase.}, and a Reggeisation term, exactly
mirroring the structure in the QCD result of
eq.~(\ref{Aexact}). However, whereas the Reggeisation term dominated
in the QCD case, here the eikonal phase dominates, as it is
power-enhanced in $s/|t|$ with respect to the Reggeisation term. Put
another way, the Regge trajectory of the graviton is proportional to
the squared charge of the graviton, which is $t$ given that it is
being exchanged in the $t$-channel. This then means that Reggeisation
is a subleading effect compared to the phase, which explains why high
energy QCD people talk mainly about Reggeised gluons, whereas high
energy gravity people talk mostly about eikonal phases. Both effects
are present in both theories, but the physics ends up being very
different! There was also some confusion for many years about whether
the graviton Reggeises at all (see
refs.~\cite{Lipatov:1982it,Lipatov:1982vv,Schnitzer:2007kh,Grisaru:1981ra,Grisaru:1982bi,Naculich:2007ub,Schnitzer:2007rn}
for earlier work), and the above results explain why this confusion
persisted for so long: Reggeisation is there, but hidden (by being
power-suppressed) in the very limit in which it is expected to
occur. As in the QED case, we can Fourier transform to momentum space,
and find a spectrum of bound states, as originally discussed in
ref.~\cite{Kabat:1992tb}. They are bound states associated with the
linearised $(1/r)$ part of the gravitational potential.

Throughout these notes, we have limited ourselves to the case of
$2\rightarrow 2$ scattering only. However, Regge theory can also be
applied to multiparticle production: see e.g. the classic review of
ref.~\cite{Collins:1977jy}. Indeed, the so-called {\it multi-Regge
  limit} has continued to be an important playground for constraining
scattering amplitudes in a variety of different field
theories~\footnote{The generalisation of the results of this section
  to many particle final states was considered in
  ref.~\cite{Melville:2013qca}.}. In recent years, a number of studies
have examined corrections to the leading Regge
limit~\cite{Akhoury:2013yua,Luna:2016idw,Collado:2018isu}, namely the
possibility that the exchanged radiation is no longer strictly
soft. People have also applied knowledge from the Regge limit to check
and constrain higher-order computations in
supergravity~\cite{Bartels:2012ra,SabioVera:2019edr,DiVecchia:2019myk,BoucherVeronneau:2011qv,Henn:2019rgj}. There
are doubtless many future insights to be gained not only by comparing
such calculations with the similar situation in gauge theories (QCD),
but also with the vintage physics of the 1960s, which does not assume
a particular theory at all! It is hoped that these notes will prove
useful in this regard.

\section*{Acknowledgements}
I am very grateful to Andreas Brandhuber, Marcel Hughes, Arnau Koemans
Collado, Rodolfo Russo and Gabriele Travaglini for useful advice when
preparing these lectures. Furthermore, I thank Ben Maybee and Alex
Ochirov for encouraging me to make these notes more widely available,
and / or for detailed comments on the manuscript.


\paragraph{Funding information}
This work was supported by the UK Science and Technology Facilities
Council (STFC) Consolidated Grant ST/P000754/1 ``String theory, gauge
theory and duality'', and by the European Union Horizon 2020 research
and innovation programme under the Marie Ck\l{}odowska-Curie grant
agreement No. 764850 ``SAGEX''.






\bibliography{refs.bib}

\begin{thebibliography}{100}
\providecommand{\url}[1]{\texttt{#1}}
\providecommand{\urlprefix}{URL }
\expandafter\ifx\csname urlstyle\endcsname\relax
  \providecommand{\doi}[1]{doi:\discretionary{}{}{}#1}\else
  \providecommand{\doi}{doi:\discretionary{}{}{}\begingroup
  \urlstyle{rm}\Url}\fi
\providecommand{\eprint}[2][]{\url{#2}}

\bibitem{Elvang:2013cua}
H.~Elvang and Y.-t. Huang,
\newblock \emph{{Scattering Amplitudes}}  (2013),
\newblock \eprint{1308.1697}.

\bibitem{Henn:2014yza}
J.~M. Henn and J.~C. Plefka,
\newblock \emph{{Scattering Amplitudes in Gauge Theories}},
\newblock Lect. Notes Phys. \textbf{883}, pp.1 (2014),
\newblock \doi{10.1007/978-3-642-54022-6}.

\bibitem{Duhr:2014woa}
C.~Duhr,
\newblock \emph{{Mathematical aspects of scattering amplitudes}},
\newblock In \emph{{Proceedings, Theoretical Advanced Study Institute in
  Elementary Particle Physics: Journeys Through the Precision Frontier:
  Amplitudes for Colliders (TASI 2014): Boulder, Colorado, June 2-27, 2014}},
  pp. 419--476,
\newblock \doi{10.1142/9789814678766_0010} (2015), \eprint{1411.7538}.

\bibitem{Caron-Huot:2016owq}
S.~Caron-Huot, L.~J. Dixon, A.~McLeod and M.~von Hippel,
\newblock \emph{{Bootstrapping a Five-Loop Amplitude Using Steinmann
  Relations}},
\newblock Phys. Rev. Lett. \textbf{117}(24), 241601 (2016),
\newblock \doi{10.1103/PhysRevLett.117.241601},
\newblock \eprint{1609.00669}.

\bibitem{Dixon:2016nkn}
L.~J. Dixon, J.~Drummond, T.~Harrington, A.~J. McLeod, G.~Papathanasiou and
  M.~Spradlin,
\newblock \emph{{Heptagons from the Steinmann Cluster Bootstrap}},
\newblock JHEP \textbf{02}, 137 (2017),
\newblock \doi{10.1007/JHEP02(2017)137},
\newblock \eprint{1612.08976}.

\bibitem{Almelid:2017qju}
{\O}.~Almelid, C.~Duhr, E.~Gardi, A.~McLeod and C.~D. White,
\newblock \emph{{Bootstrapping the QCD soft anomalous dimension}},
\newblock JHEP \textbf{09}, 073 (2017),
\newblock \doi{10.1007/JHEP09(2017)073},
\newblock \eprint{1706.10162}.

\bibitem{Chicherin:2017dob}
D.~Chicherin, J.~Henn and V.~Mitev,
\newblock \emph{{Bootstrapping pentagon functions}},
\newblock JHEP \textbf{05}, 164 (2018),
\newblock \doi{10.1007/JHEP05(2018)164},
\newblock \eprint{1712.09610}.

\bibitem{Henn:2018cdp}
J.~Henn, E.~Herrmann and J.~Parra-Martinez,
\newblock \emph{{Bootstrapping two-loop Feynman integrals for planar $
  \mathcal{N}=4 $ sYM}},
\newblock JHEP \textbf{10}, 059 (2018),
\newblock \doi{10.1007/JHEP10(2018)059},
\newblock \eprint{1806.06072}.

\bibitem{Dixon:2014xca}
L.~J. Dixon, J.~M. Drummond, C.~Duhr, M.~von Hippel and J.~Pennington,
\newblock \emph{{Bootstrapping six-gluon scattering in planar N=4
  super-Yang-Mills theory}},
\newblock PoS \textbf{LL2014}, 077 (2014),
\newblock \doi{10.22323/1.211.0077},
\newblock \eprint{1407.4724}.

\bibitem{Luisoni:2015xha}
G.~Luisoni and S.~Marzani,
\newblock \emph{{QCD resummation for hadronic final states}},
\newblock J. Phys. \textbf{G42}(10), 103101 (2015),
\newblock \doi{10.1088/0954-3899/42/10/103101},
\newblock \eprint{1505.04084}.

\bibitem{Becher:2014oda}
T.~Becher, A.~Broggio and A.~Ferroglia,
\newblock \emph{{Introduction to Soft-Collinear Effective Theory}},
\newblock Lect. Notes Phys. \textbf{896}, pp.1 (2015),
\newblock \doi{10.1007/978-3-319-14848-9},
\newblock \eprint{1410.1892}.

\bibitem{Maldacena:1997re}
J.~M. Maldacena,
\newblock \emph{{The Large N limit of superconformal field theories and
  supergravity}},
\newblock Int. J. Theor. Phys. \textbf{38}, 1113 (1999),
\newblock \doi{10.1023/A:1026654312961, 10.4310/ATMP.1998.v2.n2.a1},
\newblock [Adv. Theor. Math. Phys.2,231(1998)],
\newblock \eprint{hep-th/9711200}.

\bibitem{Mason:2005kn}
L.~J. Mason and D.~Skinner,
\newblock \emph{{An Ambitwistor Yang-Mills Lagrangian}},
\newblock Phys. Lett. \textbf{B636}, 60 (2006),
\newblock \doi{10.1016/j.physletb.2006.02.061},
\newblock \eprint{hep-th/0510262}.

\bibitem{Casali:2015vta}
E.~Casali, Y.~Geyer, L.~Mason, R.~Monteiro and K.~A. Roehrig,
\newblock \emph{{New Ambitwistor String Theories}},
\newblock JHEP \textbf{11}, 038 (2015),
\newblock \doi{10.1007/JHEP11(2015)038},
\newblock \eprint{1506.08771}.

\bibitem{Geyer:2014fka}
Y.~Geyer, A.~E. Lipstein and L.~J. Mason,
\newblock \emph{{Ambitwistor Strings in Four Dimensions}},
\newblock Phys. Rev. Lett. \textbf{113}(8), 081602 (2014),
\newblock \doi{10.1103/PhysRevLett.113.081602},
\newblock \eprint{1404.6219}.

\bibitem{Bern:2008qj}
Z.~Bern, J.~J.~M. Carrasco and H.~Johansson,
\newblock \emph{{New Relations for Gauge-Theory Amplitudes}},
\newblock Phys. Rev. \textbf{D78}, 085011 (2008),
\newblock \doi{10.1103/PhysRevD.78.085011},
\newblock \eprint{0805.3993}.

\bibitem{Bern:2010ue}
Z.~Bern, J.~J.~M. Carrasco and H.~Johansson,
\newblock \emph{{Perturbative Quantum Gravity as a Double Copy of Gauge
  Theory}},
\newblock Phys. Rev. Lett. \textbf{105}, 061602 (2010),
\newblock \doi{10.1103/PhysRevLett.105.061602},
\newblock \eprint{1004.0476}.

\bibitem{Bern:2010yg}
Z.~Bern, T.~Dennen, Y.-t. Huang and M.~Kiermaier,
\newblock \emph{{Gravity as the Square of Gauge Theory}},
\newblock Phys. Rev. \textbf{D82}, 065003 (2010),
\newblock \doi{10.1103/PhysRevD.82.065003},
\newblock \eprint{1004.0693}.

\bibitem{Bern:2019prr}
Z.~Bern, J.~J. Carrasco, M.~Chiodaroli, H.~Johansson and R.~Roiban,
\newblock \emph{{The Duality Between Color and Kinematics and its
  Applications}}  (2019),
\newblock \eprint{1909.01358}.

\bibitem{Monteiro:2014cda}
R.~Monteiro, D.~O'Connell and C.~D. White,
\newblock \emph{{Black holes and the double copy}},
\newblock JHEP \textbf{12}, 056 (2014),
\newblock \doi{10.1007/JHEP12(2014)056},
\newblock \eprint{1410.0239}.

\bibitem{Luna:2015paa}
A.~Luna, R.~Monteiro, D.~O'Connell and C.~D. White,
\newblock \emph{{The classical double copy for Taub–NUT spacetime}},
\newblock Phys. Lett. \textbf{B750}, 272 (2015),
\newblock \doi{10.1016/j.physletb.2015.09.021},
\newblock \eprint{1507.01869}.

\bibitem{Luna:2016due}
A.~Luna, R.~Monteiro, I.~Nicholson, D.~O'Connell and C.~D. White,
\newblock \emph{{The double copy: Bremsstrahlung and accelerating black
  holes}},
\newblock JHEP \textbf{06}, 023 (2016),
\newblock \doi{10.1007/JHEP06(2016)023},
\newblock \eprint{1603.05737}.

\bibitem{White:2016jzc}
C.~D. White,
\newblock \emph{{Exact solutions for the biadjoint scalar field}},
\newblock Phys. Lett. \textbf{B763}, 365 (2016),
\newblock \doi{10.1016/j.physletb.2016.10.052},
\newblock \eprint{1606.04724}.

\bibitem{Luna:2016hge}
A.~Luna, R.~Monteiro, I.~Nicholson, A.~Ochirov, D.~O'Connell, N.~Westerberg and
  C.~D. White,
\newblock \emph{{Perturbative spacetimes from Yang-Mills theory}},
\newblock JHEP \textbf{04}, 069 (2017),
\newblock \doi{10.1007/JHEP04(2017)069},
\newblock \eprint{1611.07508}.

\bibitem{DeSmet:2017rve}
P.-J. De~Smet and C.~D. White,
\newblock \emph{{Extended solutions for the biadjoint scalar field}},
\newblock Phys. Lett. \textbf{B775}, 163 (2017),
\newblock \doi{10.1016/j.physletb.2017.11.007},
\newblock \eprint{1708.01103}.

\bibitem{Bahjat-Abbas:2017htu}
N.~Bahjat-Abbas, A.~Luna and C.~D. White,
\newblock \emph{{The Kerr-Schild double copy in curved spacetime}},
\newblock JHEP \textbf{12}, 004 (2017),
\newblock \doi{10.1007/JHEP12(2017)004},
\newblock \eprint{1710.01953}.

\bibitem{Luna:2017dtq}
A.~Luna, I.~Nicholson, D.~O'Connell and C.~D. White,
\newblock \emph{{Inelastic Black Hole Scattering from Charged Scalar
  Amplitudes}},
\newblock JHEP \textbf{03}, 044 (2018),
\newblock \doi{10.1007/JHEP03(2018)044},
\newblock \eprint{1711.03901}.

\bibitem{Berman:2018hwd}
D.~S. Berman, E.~Chacón, A.~Luna and C.~D. White,
\newblock \emph{{The self-dual classical double copy, and the Eguchi-Hanson
  instanton}},
\newblock JHEP \textbf{01}, 107 (2019),
\newblock \doi{10.1007/JHEP01(2019)107},
\newblock \eprint{1809.04063}.

\bibitem{Bahjat-Abbas:2018vgo}
N.~Bahjat-Abbas, R.~Stark-Muchão and C.~D. White,
\newblock \emph{{Biadjoint wires}},
\newblock Phys. Lett. \textbf{B788}, 274 (2019),
\newblock \doi{10.1016/j.physletb.2018.11.026},
\newblock \eprint{1810.08118}.

\bibitem{CarrilloGonzalez:2019gof}
M.~Carrillo~González, B.~Melcher, K.~Ratliff, S.~Watson and C.~D. White,
\newblock \emph{{The classical double copy in three spacetime dimensions}},
\newblock JHEP \textbf{07}, 167 (2019),
\newblock \doi{10.1007/JHEP07(2019)167},
\newblock \eprint{1904.11001}.

\bibitem{Anastasiou:2018rdx}
A.~Anastasiou, L.~Borsten, M.~J. Duff, S.~Nagy and M.~Zoccali,
\newblock \emph{{Gravity as Gauge Theory Squared: A Ghost Story}},
\newblock Phys. Rev. Lett. \textbf{121}(21), 211601 (2018),
\newblock \doi{10.1103/PhysRevLett.121.211601},
\newblock \eprint{1807.02486}.

\bibitem{LopesCardoso:2018xes}
G.~Lopes~Cardoso, G.~Inverso, S.~Nagy and S.~Nampuri,
\newblock \emph{{Comments on the double copy construction for gravitational
  theories}},
\newblock PoS \textbf{CORFU2017}, 177 (2018),
\newblock \doi{10.22323/1.318.0177},
\newblock \eprint{1803.07670}.

\bibitem{Cardoso:2016ngt}
G.~L. Cardoso, S.~Nagy and S.~Nampuri,
\newblock \emph{{A double copy for $ \mathcal{N}=2 $ supergravity: a linearised
  tale told on-shell}},
\newblock JHEP \textbf{10}, 127 (2016),
\newblock \doi{10.1007/JHEP10(2016)127},
\newblock \eprint{1609.05022}.

\bibitem{Anastasiou:2014qba}
A.~Anastasiou, L.~Borsten, M.~J. Duff, L.~J. Hughes and S.~Nagy,
\newblock \emph{{Yang-Mills origin of gravitational symmetries}},
\newblock Phys. Rev. Lett. \textbf{113}(23), 231606 (2014),
\newblock \doi{10.1103/PhysRevLett.113.231606},
\newblock \eprint{1408.4434}.

\bibitem{Goldberger:2016iau}
W.~D. Goldberger and A.~K. Ridgway,
\newblock \emph{{Radiation and the classical double copy for color charges}},
\newblock Phys. Rev. \textbf{D95}(12), 125010 (2017),
\newblock \doi{10.1103/PhysRevD.95.125010},
\newblock \eprint{1611.03493}.

\bibitem{Goldberger:2017frp}
W.~D. Goldberger, S.~G. Prabhu and J.~O. Thompson,
\newblock \emph{{Classical gluon and graviton radiation from the bi-adjoint
  scalar double copy}},
\newblock Phys. Rev. \textbf{D96}(6), 065009 (2017),
\newblock \doi{10.1103/PhysRevD.96.065009},
\newblock \eprint{1705.09263}.

\bibitem{Goldberger:2017vcg}
W.~D. Goldberger and A.~K. Ridgway,
\newblock \emph{{Bound states and the classical double copy}},
\newblock Phys. Rev. \textbf{D97}(8), 085019 (2018),
\newblock \doi{10.1103/PhysRevD.97.085019},
\newblock \eprint{1711.09493}.

\bibitem{Goldberger:2017ogt}
W.~D. Goldberger, J.~Li and S.~G. Prabhu,
\newblock \emph{{Spinning particles, axion radiation, and the classical double
  copy}},
\newblock Phys. Rev. \textbf{D97}(10), 105018 (2018),
\newblock \doi{10.1103/PhysRevD.97.105018},
\newblock \eprint{1712.09250}.

\bibitem{Collins:1977jy}
P.~D.~B. Collins,
\newblock \emph{{An Introduction to Regge Theory and High-Energy Physics}},
\newblock Cambridge Monographs on Mathematical Physics. Cambridge Univ. Press,
  Cambridge, UK,
\newblock ISBN 9780521110358,
\newblock \doi{10.1017/CBO9780511897603} (2009).

\bibitem{Eden:1966dnq}
R.~J. Eden, P.~V. Landshoff, D.~I. Olive and J.~C. Polkinghorne,
\newblock \emph{{The analytic S-matrix}},
\newblock Cambridge Univ. Press, Cambridge (1966).

\bibitem{Forshaw:1997dc}
J.~R. Forshaw and D.~A. Ross,
\newblock \emph{{Quantum chromodynamics and the pomeron}},
\newblock Cambridge Lect. Notes Phys. \textbf{9}, 1 (1997).

\bibitem{DelDuca:2018nsu}
V.~Del~Duca and L.~Magnea,
\newblock \emph{{The long road from Regge poles to the LHC}} (2018),
  \eprint{1812.05829}.

\bibitem{Andersen:2019yzo}
J.~R. Andersen, T.~Hapola, M.~Heil, A.~Maier and J.~Smillie,
\newblock \emph{{HEJ 2: High Energy Resummation for Hadron Colliders}}  (2019),
\newblock \eprint{1902.08430}.

\bibitem{Andersen:2017sht}
J.~R. Andersen, H.~M. Brooks and L.~Lönnblad,
\newblock \emph{{Merging High Energy with Soft and Collinear Logarithms using
  HEJ and PYTHIA}},
\newblock JHEP \textbf{09}, 074 (2018),
\newblock \doi{10.1007/JHEP09(2018)074},
\newblock \eprint{1712.00178}.

\bibitem{Andersen:2017kfc}
J.~R. Andersen, T.~Hapola, A.~Maier and J.~M. Smillie,
\newblock \emph{{Higgs Boson Plus Dijets: Higher Order Corrections}},
\newblock JHEP \textbf{09}, 065 (2017),
\newblock \doi{10.1007/JHEP09(2017)065},
\newblock \eprint{1706.01002}.

\bibitem{Andersen:2012gk}
J.~R. Andersen, T.~Hapola and J.~M. Smillie,
\newblock \emph{{W Plus Multiple Jets at the LHC with High Energy Jets}},
\newblock JHEP \textbf{09}, 047 (2012),
\newblock \doi{10.1007/JHEP09(2012)047},
\newblock \eprint{1206.6763}.

\bibitem{Andersen:2011hs}
J.~R. Andersen and J.~M. Smillie,
\newblock \emph{{Multiple Jets at the LHC with High Energy Jets}},
\newblock JHEP \textbf{06}, 010 (2011),
\newblock \doi{10.1007/JHEP06(2011)010},
\newblock \eprint{1101.5394}.

\bibitem{Andersen:2011zd}
J.~R. Andersen, L.~Lonnblad and J.~M. Smillie,
\newblock \emph{{A Parton Shower for High Energy Jets}},
\newblock JHEP \textbf{07}, 110 (2011),
\newblock \doi{10.1007/JHEP07(2011)110},
\newblock \eprint{1104.1316}.

\bibitem{Andersen:2009he}
J.~R. Andersen and J.~M. Smillie,
\newblock \emph{{The Factorisation of the t-channel Pole in Quark-Gluon
  Scattering}},
\newblock Phys. Rev. \textbf{D81}, 114021 (2010),
\newblock \doi{10.1103/PhysRevD.81.114021},
\newblock \eprint{0910.5113}.

\bibitem{Andersen:2009nu}
J.~R. Andersen and J.~M. Smillie,
\newblock \emph{{Constructing All-Order Corrections to Multi-Jet Rates}},
\newblock JHEP \textbf{01}, 039 (2010),
\newblock \doi{10.1007/JHEP01(2010)039},
\newblock \eprint{0908.2786}.

\bibitem{Andersen:2008gc}
J.~R. Andersen, V.~Del~Duca and C.~D. White,
\newblock \emph{{Higgs Boson Production in Association with Multiple Hard
  Jets}},
\newblock JHEP \textbf{02}, 015 (2009),
\newblock \doi{10.1088/1126-6708/2009/02/015},
\newblock \eprint{0808.3696}.

\bibitem{Andersen:2008ue}
J.~R. Andersen and C.~D. White,
\newblock \emph{{A New Framework for Multijet Predictions and its application
  to Higgs Boson production at the LHC}},
\newblock Phys. Rev. \textbf{D78}, 051501 (2008),
\newblock \doi{10.1103/PhysRevD.78.051501},
\newblock \eprint{0802.2858}.

\bibitem{Jung:2010si}
H.~Jung \emph{et~al.},
\newblock \emph{{The CCFM Monte Carlo generator CASCADE version 2.2.03}},
\newblock Eur. Phys. J. \textbf{C70}, 1237 (2010),
\newblock \doi{10.1140/epjc/s10052-010-1507-z},
\newblock \eprint{1008.0152}.

\bibitem{White:2006yh}
C.~D. White and R.~S. Thorne,
\newblock \emph{{A Global Fit to Scattering Data with NLL BFKL Resummations}},
\newblock Phys. Rev. \textbf{D75}, 034005 (2007),
\newblock \doi{10.1103/PhysRevD.75.034005},
\newblock \eprint{hep-ph/0611204}.

\bibitem{White:2006xv}
C.~D. White and R.~S. Thorne,
\newblock \emph{{A Variable flavor number scheme for heavy quark production at
  small x}},
\newblock Phys. Rev. \textbf{D74}, 014002 (2006),
\newblock \doi{10.1103/PhysRevD.74.014002},
\newblock \eprint{hep-ph/0603030}.

\bibitem{Ball:2017otu}
R.~D. Ball, V.~Bertone, M.~Bonvini, S.~Marzani, J.~Rojo and L.~Rottoli,
\newblock \emph{{Parton distributions with small-x resummation: evidence for
  BFKL dynamics in HERA data}},
\newblock Eur. Phys. J. \textbf{C78}(4), 321 (2018),
\newblock \doi{10.1140/epjc/s10052-018-5774-4},
\newblock \eprint{1710.05935}.

\bibitem{Bonvini:2016wki}
M.~Bonvini, S.~Marzani and T.~Peraro,
\newblock \emph{{Small-$x$ resummation from HELL}},
\newblock Eur. Phys. J. \textbf{C76}(11), 597 (2016),
\newblock \doi{10.1140/epjc/s10052-016-4445-6},
\newblock \eprint{1607.02153}.

\bibitem{Altarelli:2008aj}
G.~Altarelli, R.~D. Ball and S.~Forte,
\newblock \emph{{Small x Resummation with Quarks: Deep-Inelastic Scattering}},
\newblock Nucl. Phys. \textbf{B799}, 199 (2008),
\newblock \doi{10.1016/j.nuclphysb.2008.03.003},
\newblock \eprint{0802.0032}.

\bibitem{Altarelli:2003hk}
G.~Altarelli, R.~D. Ball and S.~Forte,
\newblock \emph{{An Anomalous dimension for small x evolution}},
\newblock Nucl. Phys. \textbf{B674}, 459 (2003),
\newblock \doi{10.1016/j.nuclphysb.2003.09.040},
\newblock \eprint{hep-ph/0306156}.

\bibitem{Altarelli:2001ji}
G.~Altarelli, R.~D. Ball and S.~Forte,
\newblock \emph{Factorization and resummation of small x scaling violations
  with running coupling},
\newblock Nucl. Phys. \textbf{B621}, 359 (2002),
\newblock \eprint[http://arXiv.org/abs]{hep-ph/0109178}.

\bibitem{Altarelli:1999vw}
G.~Altarelli, R.~D. Ball and S.~Forte,
\newblock \emph{Resummation of singlet parton evolution at small x},
\newblock Nucl. Phys. \textbf{B575}, 313 (2000),
\newblock \eprint[http://arXiv.org/abs]{hep-ph/9911273}.

\bibitem{Ciafaloni:2007gf}
M.~Ciafaloni, D.~Colferai, G.~P. Salam and A.~M. Stasto,
\newblock \emph{{A Matrix formulation for small-x singlet evolution}},
\newblock JHEP \textbf{08}, 046 (2007),
\newblock \doi{10.1088/1126-6708/2007/08/046},
\newblock \eprint{0707.1453}.

\bibitem{Ciafaloni:2006yk}
M.~Ciafaloni, D.~Colferai, G.~P. Salam and A.~M. Stasto,
\newblock \emph{{Minimal subtraction vs. physical factorisation schemes in
  small-x QCD}},
\newblock Phys. Lett. \textbf{B635}, 320 (2006),
\newblock \doi{10.1016/j.physletb.2006.03.014},
\newblock \eprint{hep-ph/0601200}.

\bibitem{Ciafaloni:2003rd}
M.~Ciafaloni, D.~Colferai, G.~P. Salam and A.~M. Stasto,
\newblock \emph{{Renormalization group improved small x Green's function}},
\newblock Phys. Rev. \textbf{D68}, 114003 (2003),
\newblock \doi{10.1103/PhysRevD.68.114003},
\newblock \eprint{hep-ph/0307188}.

\bibitem{Giddings:2010pp}
S.~B. Giddings, M.~Schmidt-Sommerfeld and J.~R. Andersen,
\newblock \emph{{High energy scattering in gravity and supergravity}},
\newblock Phys. Rev. \textbf{D82}, 104022 (2010),
\newblock \doi{10.1103/PhysRevD.82.104022},
\newblock \eprint{1005.5408}.

\bibitem{Schwartz:2013pla}
M.~D. Schwartz,
\newblock \emph{{Quantum Field Theory and the Standard Model}},
\newblock Cambridge University Press,
\newblock ISBN 1107034736, 9781107034730 (2014).

\bibitem{PhysRev.135.B745}
D.~I. Olive,
\newblock \emph{Exploration of {S}-matrix theory},
\newblock Phys. Rev. \textbf{135}, B745 (1964),
\newblock \doi{10.1103/PhysRev.135.B745}.

\bibitem{Weinberg:1995mt}
S.~Weinberg,
\newblock \emph{{The Quantum theory of fields. Vol. 1: Foundations}},
\newblock Cambridge University Press,
\newblock ISBN 9780521670531, 9780511252044 (2005).

\bibitem{Veltman:1994wz}
M.~J.~G. Veltman,
\newblock \emph{{Diagrammatica: The Path to Feynman rules}},
\newblock Cambridge Lect. Notes Phys. \textbf{4}, 1 (1994).

\bibitem{PhysRev.135.B1255}
D.~Branson,
\newblock \emph{Time and the $s$ matrix},
\newblock Phys. Rev. \textbf{135}, B1255 (1964),
\newblock \doi{10.1103/PhysRev.135.B1255}.

\bibitem{Chachamis:2017vgr}
F.~Caporale, F.~G. Celiberto, D.~Gordo~Gomez, A.~Sabio~Vera and G.~Chachamis,
\newblock \emph{{Multi-jet production in the high energy limit at LHC}},
\newblock In \emph{{25th Low-x Meeting (Low-x 2017) Bari, Italy, June 13-17,
  2017}} (2017), \eprint{1801.00014}.

\bibitem{Chachamis:2015ico}
G.~Chachamis and A.~Sabio~Vera,
\newblock \emph{{The high-energy radiation pattern from BFKLex with double-log
  collinear contributions}},
\newblock JHEP \textbf{02}, 064 (2016),
\newblock \doi{10.1007/JHEP02(2016)064},
\newblock \eprint{1512.03603}.

\bibitem{Caporale:2015int}
F.~Caporale, F.~G. Celiberto, G.~Chachamis and A.~Sabio~Vera,
\newblock \emph{{Multi-Regge kinematics and azimuthal angle observables for
  inclusive four-jet production}},
\newblock Eur. Phys. J. \textbf{C76}(3), 165 (2016),
\newblock \doi{10.1140/epjc/s10052-016-3963-6},
\newblock \eprint{1512.03364}.

\bibitem{Caporale:2015vya}
F.~Caporale, G.~Chachamis, B.~Murdaca and A.~Sabio~Vera,
\newblock \emph{{Balitsky-Fadin-Kuraev-Lipatov Predictions for Inclusive Three
  Jet Production at the LHC}},
\newblock Phys. Rev. Lett. \textbf{116}(1), 012001 (2016),
\newblock \doi{10.1103/PhysRevLett.116.012001},
\newblock \eprint{1508.07711}.

\bibitem{Amati:1987wq}
D.~Amati, M.~Ciafaloni and G.~Veneziano,
\newblock \emph{{Superstring Collisions at Planckian Energies}},
\newblock Phys. Lett. \textbf{B197}, 81 (1987),
\newblock \doi{10.1016/0370-2693(87)90346-7}.

\bibitem{Muzinich:1987in}
I.~J. Muzinich and M.~Soldate,
\newblock \emph{{High-Energy Unitarity of Gravitation and Strings}},
\newblock Phys. Rev. \textbf{D37}, 359 (1988),
\newblock \doi{10.1103/PhysRevD.37.359}.

\bibitem{tHooft:1987vrq}
G.~'t~Hooft,
\newblock \emph{{Graviton Dominance in Ultrahigh-Energy Scattering}},
\newblock Phys. Lett. \textbf{B198}, 61 (1987),
\newblock \doi{10.1016/0370-2693(87)90159-6}.

\bibitem{Amati:1987uf}
D.~Amati, M.~Ciafaloni and G.~Veneziano,
\newblock \emph{{Classical and Quantum Gravity Effects from Planckian Energy
  Superstring Collisions}},
\newblock Int. J. Mod. Phys. \textbf{A3}, 1615 (1988),
\newblock \doi{10.1142/S0217751X88000710}.

\bibitem{Amati:1990xe}
D.~Amati, M.~Ciafaloni and G.~Veneziano,
\newblock \emph{{Higher Order Gravitational Deflection and Soft Bremsstrahlung
  in Planckian Energy Superstring Collisions}},
\newblock Nucl. Phys. \textbf{B347}, 550 (1990),
\newblock \doi{10.1016/0550-3213(90)90375-N}.

\bibitem{Verlinde:1991iu}
H.~L. Verlinde and E.~P. Verlinde,
\newblock \emph{{Scattering at Planckian energies}},
\newblock Nucl. Phys. \textbf{B371}, 246 (1992),
\newblock \doi{10.1016/0550-3213(92)90236-5},
\newblock \eprint{hep-th/9110017}.

\bibitem{Amati:1992zb}
D.~Amati, M.~Ciafaloni and G.~Veneziano,
\newblock \emph{{Planckian scattering beyond the semiclassical approximation}},
\newblock Phys. Lett. \textbf{B289}, 87 (1992),
\newblock \doi{10.1016/0370-2693(92)91366-H}.

\bibitem{Amati:1993tb}
D.~Amati, M.~Ciafaloni and G.~Veneziano,
\newblock \emph{{Effective action and all order gravitational eikonal at
  Planckian energies}},
\newblock Nucl. Phys. \textbf{B403}, 707 (1993),
\newblock \doi{10.1016/0550-3213(93)90367-X}.

\bibitem{Kabat:1992tb}
D.~N. Kabat and M.~Ortiz,
\newblock \emph{{Eikonal quantum gravity and Planckian scattering}},
\newblock Nucl. Phys. \textbf{B388}, 570 (1992),
\newblock \doi{10.1016/0550-3213(92)90627-N},
\newblock \eprint{hep-th/9203082}.

\bibitem{DAppollonio:2010krb}
G.~D'Appollonio, P.~Di~Vecchia, R.~Russo and G.~Veneziano,
\newblock \emph{{High-energy string-brane scattering: Leading eikonal and
  beyond}},
\newblock JHEP \textbf{11}, 100 (2010),
\newblock \doi{10.1007/JHEP11(2010)100},
\newblock \eprint{1008.4773}.

\bibitem{Akhoury:2013yua}
R.~Akhoury, R.~Saotome and G.~Sterman,
\newblock \emph{{High Energy Scattering in Perturbative Quantum Gravity at Next
  to Leading Power}}  (2013),
\newblock \eprint{1308.5204}.

\bibitem{Collado:2018isu}
A.~K. Collado, P.~Di~Vecchia, R.~Russo and S.~Thomas,
\newblock \emph{{The subleading eikonal in supergravity theories}},
\newblock JHEP \textbf{10}, 038 (2018),
\newblock \doi{10.1007/JHEP10(2018)038},
\newblock \eprint{1807.04588}.

\bibitem{KoemansCollado:2019lnh}
A.~Koemans~Collado and S.~Thomas,
\newblock \emph{{Eikonal Scattering in Kaluza-Klein Gravity}},
\newblock JHEP \textbf{04}, 171 (2019),
\newblock \doi{10.1007/JHEP04(2019)171},
\newblock \eprint{1901.05869}.

\bibitem{KoemansCollado:2019ggb}
A.~Koemans~Collado, P.~Di~Vecchia and R.~Russo,
\newblock \emph{{Revisiting the 2PM eikonal and the dynamics of binary black
  holes}}  (2019),
\newblock \eprint{1904.02667}.

\bibitem{Regge:1959mz}
T.~Regge,
\newblock \emph{{Introduction to complex orbital momenta}},
\newblock Nuovo Cim. \textbf{14}, 951 (1959),
\newblock \doi{10.1007/BF02728177}.

\bibitem{MANDELSTAM1962254}
S.~Mandelstam,
\newblock \emph{An extension of the {R}egge formula},
\newblock Annals of Physics \textbf{19}(2), 254  (1962),
\newblock \doi{https://doi.org/10.1016/0003-4916(62)90218-X}.

\bibitem{Green:1987sp}
M.~B. Green, J.~H. Schwarz and E.~Witten,
\newblock \emph{{SUPERSTRING THEORY. VOL. 1: INTRODUCTION}},
\newblock Cambridge Monographs on Mathematical Physics,
\newblock ISBN 9780521357524 (1988).

\bibitem{Donnachie:1998gm}
A.~Donnachie and P.~V. Landshoff,
\newblock \emph{{Small x: Two pomerons!}},
\newblock Phys. Lett. \textbf{B437}, 408 (1998),
\newblock \doi{10.1016/S0370-2693(98)00899-5},
\newblock \eprint{hep-ph/9806344}.

\bibitem{Donnachie:2013xia}
A.~Donnachie and P.~V. Landshoff,
\newblock \emph{{$pp$ and $\bar pp$ total cross sections and elastic
  scattering}},
\newblock Phys. Lett. \textbf{B727}, 500 (2013),
\newblock \doi{10.1016/j.physletb.2015.09.017, 10.1016/j.physletb.2013.10.068},
\newblock [Erratum: Phys. Lett.B750,669(2015)],
\newblock \eprint{1309.1292}.

\bibitem{Korchemskaya:1994qp}
I.~Korchemskaya and G.~Korchemsky,
\newblock \emph{{High-energy scattering in QCD and cross singularities of
  Wilson loops}},
\newblock Nucl.Phys. \textbf{B437}, 127 (1995),
\newblock \doi{10.1016/0550-3213(94)00553-Q},
\newblock \eprint{hep-ph/9409446}.

\bibitem{Melville:2013qca}
S.~Melville, S.~G. Naculich, H.~J. Schnitzer and C.~D. White,
\newblock \emph{{Wilson line approach to gravity in the high energy limit}},
\newblock Phys. Rev. \textbf{D89}(2), 025009 (2014),
\newblock \doi{10.1103/PhysRevD.89.025009},
\newblock \eprint{1306.6019}.

\bibitem{Mandelstam:1965zz}
S.~Mandelstam,
\newblock \emph{{Non-Regge Terms in the Vector-Spinor Theory}},
\newblock Phys. Rev. \textbf{137}, B949 (1965),
\newblock \doi{10.1103/PhysRev.137.B949}.

\bibitem{Abers:1967zz}
E.~Abers and V.~L. Teplitz,
\newblock \emph{{Kinematic Constraints, Crossing, and the Reggeization of
  Scattering Amplitudes}},
\newblock Phys. Rev. \textbf{158}, 1365 (1967),
\newblock \doi{10.1103/PhysRev.158.1365}.

\bibitem{McCoy:1976ff}
B.~M. McCoy and T.~T. Wu,
\newblock \emph{{Theory of Fermion Exchange in Massive Quantum Electrodynamics
  at High-Energy. 1.}},
\newblock Phys. Rev. \textbf{D13}, 369 (1976),
\newblock \doi{10.1103/PhysRevD.13.369}.

\bibitem{Frolov:1970ij}
G.~V. Frolov, V.~N. Gribov and L.~N. Lipatov,
\newblock \emph{{On Regge poles in quantum electrodynamics}},
\newblock Phys. Lett. \textbf{31B}, 34 (1970),
\newblock \doi{10.1016/0370-2693(70)90013-4}.

\bibitem{Grisaru:1973vw}
M.~T. Grisaru, H.~J. Schnitzer and H.-S. Tsao,
\newblock \emph{{Reggeization of yang-mills gauge mesons in theories with a
  spontaneously broken symmetry}},
\newblock Phys. Rev. Lett. \textbf{30}, 811 (1973),
\newblock \doi{10.1103/PhysRevLett.30.811}.

\bibitem{Grisaru:1973ku}
M.~T. Grisaru, H.~J. Schnitzer and H.-S. Tsao,
\newblock \emph{{THE REGGEIZATION OF ELEMENTARY PARTICLES IN RENORMALIZABLE
  GAUGE THEORIES: SCALARS}},
\newblock Phys. Rev. \textbf{D9}, 2864 (1974),
\newblock \doi{10.1103/PhysRevD.9.2864}.

\bibitem{Grisaru:1974cf}
M.~T. Grisaru, H.~J. Schnitzer and H.-S. Tsao,
\newblock \emph{{Reggeization of elementary particles in renormalizable gauge
  theories - vectors and spinors}},
\newblock Phys. Rev. \textbf{D8}, 4498 (1973),
\newblock \doi{10.1103/PhysRevD.8.4498}.

\bibitem{Gribov:1970ik}
V.~N. Gribov, L.~N. Lipatov and G.~V. Frolov,
\newblock \emph{{The leading singularity in the j plane in quantum
  electrodynamics}},
\newblock Sov. J. Nucl. Phys. \textbf{12}, 543 (1971),
\newblock [Yad. Fiz.12,994(1970)].

\bibitem{Cheng:1969bf}
H.~Cheng and T.~T. Wu,
\newblock \emph{{High-energy collision processes in quantum electrodynamics.
  i}},
\newblock Phys. Rev. \textbf{182}, 1852 (1969),
\newblock \doi{10.1103/PhysRev.182.1852}.

\bibitem{Balitsky:1979ap}
I.~I. Balitsky, L.~N. Lipatov and V.~S. Fadin,
\newblock \emph{{REGGE PROCESSES IN NONABELIAN GAUGE THEORIES. (IN RUSSIAN)}}
  (1979).

\bibitem{Bogdan:2006af}
A.~V. Bogdan and V.~S. Fadin,
\newblock \emph{{A Proof of the reggeized form of amplitudes with quark
  exchanges}},
\newblock Nucl. Phys. \textbf{B740}, 36 (2006),
\newblock \doi{10.1016/j.nuclphysb.2006.01.033},
\newblock \eprint{hep-ph/0601117}.

\bibitem{Tyburski:1975mr}
L.~Tyburski,
\newblock \emph{{Reggeization of the Fermion-Fermion Scattering Amplitude in
  Nonabelian Gauge Theories}},
\newblock Phys. Rev. \textbf{D13}, 1107 (1976),
\newblock \doi{10.1103/PhysRevD.13.1107}.

\bibitem{Lipatov:1976zz}
L.~N. Lipatov,
\newblock \emph{Reggeization of the vector meson and the vacuum singularity in
  nonabelian gauge theories},
\newblock Sov. J. Nucl. Phys. \textbf{23}, 338 (1976).

\bibitem{Mason:1976fr}
A.~L. Mason,
\newblock \emph{{Radiation Gauge Calculation of High-Energy Scattering
  Amplitudes}},
\newblock Nucl. Phys. \textbf{B120}, 275 (1977),
\newblock \doi{10.1016/0550-3213(77)90044-X}.

\bibitem{Cheng:1977gt}
H.~Cheng and C.~Y. Lo,
\newblock \emph{{High-Energy Amplitudes of Yang-Mills Theory in Arbitrary
  Perturbative Orders. 1.}},
\newblock Phys. Rev. \textbf{D15}, 2959 (1977),
\newblock \doi{10.1103/PhysRevD.15.2959}.

\bibitem{Fadin:1975cb}
V.~S. Fadin, E.~A. Kuraev and L.~N. Lipatov,
\newblock \emph{On the pomeranchuk singularity in asymptotically free
  theories},
\newblock Phys. Lett. \textbf{B60}, 50 (1975).

\bibitem{Kuraev:1977fs}
E.~A. Kuraev, L.~N. Lipatov and V.~S. Fadin,
\newblock \emph{{The Pomeranchuk Singularity in Nonabelian Gauge Theories}},
\newblock Sov. Phys. JETP \textbf{45}, 199 (1977),
\newblock [Zh. Eksp. Teor. Fiz.72,377(1977)].

\bibitem{Kuraev:1976ge}
E.~A. Kuraev, L.~N. Lipatov and V.~S. Fadin,
\newblock \emph{Multi - reggeon processes in the yang-mills theory},
\newblock Sov. Phys. JETP \textbf{44}, 443 (1976).

\bibitem{Mason:1976ky}
A.~L. Mason,
\newblock \emph{{Factorization and Hence Reggeization in Yang-Mills Theories}},
\newblock Nucl. Phys. \textbf{B117}, 493 (1976),
\newblock \doi{10.1016/0550-3213(76)90411-9}.

\bibitem{Sen:1982xv}
A.~Sen,
\newblock \emph{{Asymptotic Behavior of the Fermion and Gluon Exchange
  Amplitudes in Massive Quantum Electrodynamics in the Regge Limit}},
\newblock Phys. Rev. \textbf{D27}, 2997 (1983),
\newblock \doi{10.1103/PhysRevD.27.2997}.

\bibitem{Fadin:1977jr}
V.~S. Fadin and V.~E. Sherman,
\newblock \emph{{Processes Involving Fermion Exchange in Nonabelian Gauge
  Theories}},
\newblock Zh. Eksp. Teor. Fiz. \textbf{72}, 1640 (1977).

\bibitem{Fadin:1995xg}
V.~S. Fadin, M.~I. Kotsky and R.~Fiore,
\newblock \emph{{Gluon Reggeization in QCD in the next-to-leading order}},
\newblock Phys. Lett. \textbf{B359}, 181 (1995),
\newblock \doi{10.1016/0370-2693(95)01016-J}.

\bibitem{Fadin:1996tb}
V.~S. Fadin, R.~Fiore and M.~I. Kotsky,
\newblock \emph{{Gluon Regge trajectory in the two loop approximation}},
\newblock Phys. Lett. \textbf{B387}, 593 (1996),
\newblock \doi{10.1016/0370-2693(96)01054-4},
\newblock \eprint{hep-ph/9605357}.

\bibitem{Fadin:1995km}
V.~S. Fadin, R.~Fiore and A.~Quartarolo,
\newblock \emph{{Reggeization of quark quark scattering amplitude in QCD}},
\newblock Phys. Rev. \textbf{D53}, 2729 (1996),
\newblock \doi{10.1103/PhysRevD.53.2729},
\newblock \eprint{hep-ph/9506432}.

\bibitem{Blumlein:1998ib}
J.~Blumlein, V.~Ravindran and W.~L. van Neerven,
\newblock \emph{{On the gluon Regge trajectory in O alpha-s**2}},
\newblock Phys. Rev. \textbf{D58}, 091502 (1998),
\newblock \doi{10.1103/PhysRevD.58.091502},
\newblock \eprint{hep-ph/9806357}.

\bibitem{DelDuca:2001gu}
V.~Del~Duca and E.~W.~N. Glover,
\newblock \emph{{The High-energy limit of QCD at two loops}},
\newblock JHEP \textbf{10}, 035 (2001),
\newblock \doi{10.1088/1126-6708/2001/10/035},
\newblock \eprint{hep-ph/0109028}.

\bibitem{Bogdan:2002sr}
A.~V. Bogdan, V.~Del~Duca, V.~S. Fadin and E.~W.~N. Glover,
\newblock \emph{{The Quark Regge trajectory at two loops}},
\newblock JHEP \textbf{03}, 032 (2002),
\newblock \doi{10.1088/1126-6708/2002/03/032},
\newblock \eprint{hep-ph/0201240}.

\bibitem{Fadin:2006bj}
V.~S. Fadin, R.~Fiore, M.~G. Kozlov and A.~V. Reznichenko,
\newblock \emph{{Proof of the multi-Regge form of QCD amplitudes with gluon
  exchanges in the NLA}},
\newblock Phys. Lett. \textbf{B639}, 74 (2006),
\newblock \doi{10.1016/j.physletb.2006.03.031},
\newblock \eprint{hep-ph/0602006}.

\bibitem{Caron-Huot:2013fea}
S.~Caron-Huot,
\newblock \emph{{When does the gluon reggeize?}},
\newblock JHEP \textbf{05}, 093 (2015),
\newblock \doi{10.1007/JHEP05(2015)093},
\newblock \eprint{1309.6521}.

\bibitem{DelDuca:2013ara}
V.~Del~Duca, G.~Falcioni, L.~Magnea and L.~Vernazza,
\newblock \emph{{High-energy QCD amplitudes at two loops and beyond}},
\newblock Phys. Lett. \textbf{B732}, 233 (2014),
\newblock \doi{10.1016/j.physletb.2014.03.033},
\newblock \eprint{1311.0304}.

\bibitem{DelDuca:2013dsa}
V.~Del~Duca, G.~Falcioni, L.~Magnea and L.~Vernazza,
\newblock \emph{{Beyond Reggeization for two- and three-loop QCD amplitudes}},
\newblock PoS \textbf{RADCOR2013}, 046 (2013),
\newblock \doi{10.22323/1.197.0046},
\newblock \eprint{1312.5098}.

\bibitem{DelDuca:2014cya}
V.~Del~Duca, G.~Falcioni, L.~Magnea and L.~Vernazza,
\newblock \emph{{Analyzing high-energy factorization beyond next-to-leading
  logarithmic accuracy}},
\newblock JHEP \textbf{02}, 029 (2015),
\newblock \doi{10.1007/JHEP02(2015)029},
\newblock \eprint{1409.8330}.

\bibitem{Caron-Huot:2017fxr}
S.~Caron-Huot, E.~Gardi and L.~Vernazza,
\newblock \emph{{Two-parton scattering in the high-energy limit}},
\newblock JHEP \textbf{06}, 016 (2017),
\newblock \doi{10.1007/JHEP06(2017)016},
\newblock \eprint{1701.05241}.

\bibitem{Caron-Huot:2017zfo}
S.~Caron-Huot, E.~Gardi, J.~Reichel and L.~Vernazza,
\newblock \emph{{Infrared singularities of QCD scattering amplitudes in the
  Regge limit to all orders}},
\newblock JHEP \textbf{03}, 098 (2018),
\newblock \doi{10.1007/JHEP03(2018)098},
\newblock \eprint{1711.04850}.

\bibitem{Vernazza:2018gyb}
L.~Vernazza, S.~Caron-Huot, E.~Gardi and J.~Reichel,
\newblock \emph{{The Regge Limit and infrared singularities of QCD scattering
  amplitudes to all orders}},
\newblock PoS \textbf{LL2018}, 038 (2018),
\newblock \doi{10.22323/1.303.0038}.

\bibitem{Bret:2011xm}
V.~Del~Duca, C.~Duhr, E.~Gardi, L.~Magnea and C.~D. White,
\newblock \emph{{An infrared approach to Reggeization}},
\newblock Phys.Rev. \textbf{D85}, 071104 (2012),
\newblock \doi{10.1103/PhysRevD.85.071104},
\newblock \eprint{1108.5947}.

\bibitem{DelDuca:2011ae}
V.~Del~Duca, C.~Duhr, E.~Gardi, L.~Magnea and C.~D. White,
\newblock \emph{{The Infrared structure of gauge theory amplitudes in the
  high-energy limit}},
\newblock JHEP \textbf{1112}, 021 (2011),
\newblock \doi{10.1007/JHEP12(2011)021},
\newblock \eprint{1109.3581}.

\bibitem{Kulish1970}
P.~P. Kulish and L.~D. Faddeev,
\newblock \emph{Asymptotic conditions and infrared divergences in quantum
  electrodynamics},
\newblock Theoretical and Mathematical Physics \textbf{4}(2), 745 (1970),
\newblock \doi{10.1007/BF01066485}.

\bibitem{CATANI1986588}
S.~Catani, M.~Ciafaloni and G.~Marchesini,
\newblock \emph{Non-cancelling infrared divergences in qcd coherent states},
\newblock Nuclear Physics B \textbf{264}, 588  (1986),
\newblock \doi{https://doi.org/10.1016/0550-3213(86)90500-6}.

\bibitem{Ware:2013zja}
J.~Ware, R.~Saotome and R.~Akhoury,
\newblock \emph{{Construction of an asymptotic S matrix for perturbative
  quantum gravity}},
\newblock JHEP \textbf{10}, 159 (2013),
\newblock \doi{10.1007/JHEP10(2013)159},
\newblock \eprint{1308.6285}.

\bibitem{DelDuca:1995hf}
V.~Del~Duca,
\newblock \emph{{An introduction to the perturbative QCD pomeron and to jet
  physics at large rapidities}}  (1995),
\newblock \eprint{hep-ph/9503226}.

\bibitem{Brezin:1970zr}
E.~Brezin, C.~Itzykson and J.~Zinn-Justin,
\newblock \emph{{Relativistic balmer formula including recoil effects}},
\newblock Phys. Rev. \textbf{D1}, 2349 (1970),
\newblock \doi{10.1103/PhysRevD.1.2349}.

\bibitem{Oxburgh:2012zr}
S.~Oxburgh and C.~D. White,
\newblock \emph{{BCJ duality and the double copy in the soft limit}},
\newblock JHEP \textbf{02}, 127 (2013),
\newblock \doi{10.1007/JHEP02(2013)127},
\newblock \eprint{1210.1110}.

\bibitem{Vera:2014tda}
A.~Sabio~Vera and M.~A. Vazquez-Mozo,
\newblock \emph{{The Double Copy Structure of Soft Gravitons}},
\newblock JHEP \textbf{03}, 070 (2015),
\newblock \doi{10.1007/JHEP03(2015)070},
\newblock \eprint{1412.3699}.

\bibitem{Johansson:2013nsa}
H.~Johansson, A.~Sabio~Vera, E.~Serna~Campillo and M.~A. V\'{a}zquez-Mozo,
\newblock \emph{{Color-Kinematics Duality in Multi-Regge Kinematics and
  Dimensional Reduction}},
\newblock JHEP \textbf{10}, 215 (2013),
\newblock \doi{10.1007/JHEP10(2013)215},
\newblock \eprint{1307.3106}.

\bibitem{Vera:2012ds}
A.~Sabio~Vera, E.~Serna~Campillo and M.~A. Vazquez-Mozo,
\newblock \emph{{Color-Kinematics Duality and the Regge Limit of Inelastic
  Amplitudes}},
\newblock JHEP \textbf{04}, 086 (2013),
\newblock \doi{10.1007/JHEP04(2013)086},
\newblock \eprint{1212.5103}.

\bibitem{Saotome:2012vy}
R.~Saotome and R.~Akhoury,
\newblock \emph{{Relationship Between Gravity and Gauge Scattering in the High
  Energy Limit}},
\newblock JHEP \textbf{01}, 123 (2013),
\newblock \doi{10.1007/JHEP01(2013)123},
\newblock \eprint{1210.8111}.

\bibitem{Lipatov:1982it}
L.~N. Lipatov,
\newblock \emph{{Multi - Regge Processes in Gravitation}},
\newblock Sov. Phys. JETP \textbf{55}, 582 (1982),
\newblock [Zh. Eksp. Teor. Fiz.82,991(1982)].

\bibitem{Lipatov:1982vv}
L.~N. Lipatov,
\newblock \emph{{GRAVITON REGGEIZATION}},
\newblock Phys. Lett. \textbf{116B}, 411 (1982),
\newblock \doi{10.1016/0370-2693(82)90156-3}.

\bibitem{Schnitzer:2007kh}
H.~J. Schnitzer,
\newblock \emph{{Reggeization of N=8 supergravity and N=4 Yang-Mills theory}}
  (2007),
\newblock \eprint{hep-th/0701217}.

\bibitem{Grisaru:1981ra}
M.~T. Grisaru and H.~J. Schnitzer,
\newblock \emph{{Dynamical Calculation of Bound State Supermultiplets in $N=8$
  Supergravity}},
\newblock Phys. Lett. \textbf{107B}, 196 (1981),
\newblock \doi{10.1016/0370-2693(81)90811-X}.

\bibitem{Grisaru:1982bi}
M.~T. Grisaru and H.~J. Schnitzer,
\newblock \emph{{Bound States in $N=8$ Supergravity and $N=4$ Supersymmetric
  {Yang-Mills} Theories}},
\newblock Nucl. Phys. \textbf{B204}, 267 (1982),
\newblock \doi{10.1016/0550-3213(82)90148-1}.

\bibitem{Naculich:2007ub}
S.~G. Naculich and H.~J. Schnitzer,
\newblock \emph{{Regge behavior of gluon scattering amplitudes in N=4 SYM
  theory}},
\newblock Nucl. Phys. \textbf{B794}, 189 (2008),
\newblock \doi{10.1016/j.nuclphysb.2007.10.026},
\newblock \eprint{0708.3069}.

\bibitem{Schnitzer:2007rn}
H.~J. Schnitzer,
\newblock \emph{{Reggeization of N=8 supergravity and N=4 Yang-Mills theory.
  II.}}  (2007),
\newblock \eprint{0706.0917}.

\bibitem{Luna:2016idw}
A.~Luna, S.~Melville, S.~G. Naculich and C.~D. White,
\newblock \emph{{Next-to-soft corrections to high energy scattering in QCD and
  gravity}},
\newblock JHEP \textbf{01}, 052 (2017),
\newblock \doi{10.1007/JHEP01(2017)052},
\newblock \eprint{1611.02172}.

\bibitem{Bartels:2012ra}
J.~Bartels, L.~N. Lipatov and A.~Sabio~Vera,
\newblock \emph{{Double-logarithms in Einstein-Hilbert gravity and
  supergravity}},
\newblock JHEP \textbf{07}, 056 (2014),
\newblock \doi{10.1007/JHEP07(2014)056},
\newblock \eprint{1208.3423}.

\bibitem{SabioVera:2019edr}
A.~Sabio~Vera,
\newblock \emph{{Double-logarithms in $ \mathcal{N} $ = 8 supergravity: impact
  parameter description \& mapping to 1-rooted ribbon graphs}},
\newblock JHEP \textbf{07}, 080 (2019),
\newblock \doi{10.1007/JHEP07(2019)080},
\newblock \eprint{1904.13372}.

\bibitem{DiVecchia:2019myk}
P.~Di~Vecchia, A.~Luna, S.~G. Naculich, R.~Russo, G.~Veneziano and C.~D. White,
\newblock \emph{{A tale of two exponentiations in ${\cal N}=8$ supergravity}}
  (2019),
\newblock \eprint{1908.05603}.

\bibitem{BoucherVeronneau:2011qv}
C.~Boucher-Veronneau and L.~J. Dixon,
\newblock \emph{{N >- 4 Supergravity Amplitudes from Gauge Theory at Two
  Loops}},
\newblock JHEP \textbf{12}, 046 (2011),
\newblock \doi{10.1007/JHEP12(2011)046},
\newblock \eprint{1110.1132}.

\bibitem{Henn:2019rgj}
J.~M. Henn and B.~Mistlberger,
\newblock \emph{{Four-graviton scattering to three loops in $ \mathcal{N}=8 $
  supergravity}},
\newblock JHEP \textbf{05}, 023 (2019),
\newblock \doi{10.1007/JHEP05(2019)023},
\newblock \eprint{1902.07221}.

\end{thebibliography}

\nolinenumbers

\end{document}